\documentclass[%
    aip,
    jmp,%
    amsmath,amssymb,floatfix,
    reprint,%
    onecolumn,
    longbibliography,
    ]{revtex4-1}
    
    \usepackage{amsmath,amsthm,latexsym,amssymb,amsfonts,epsfig}

    \usepackage[english]{babel}
    \usepackage[utf8]{inputenc}			
    \usepackage{graphicx, texdraw}   %for figures
    \usepackage{bm}

    \usepackage{lmodern}
    \usepackage{mathtools}
    \usepackage{mathrsfs} 

    \usepackage[left= 3cm, right = 3cm, top= 3cm, bottom = 3cm]{geometry}    % preset margins

    \usepackage{enumerate}

    \usepackage[colorlinks=true,citecolor=cyan, linkcolor=blue]{hyperref}

    \usepackage{csquotes} 

    \usepackage[usenames, dvipsnames]{color}

    \usepackage{amsmath}
    \usepackage{amsfonts}
    \usepackage{amssymb}
    \usepackage{braket}

    \usepackage[normalem]{ulem} %to strike the words 
\newcommand{\vect}[1]{\boldsymbol{#1}}

\newcommand{\boldNabla}{\boldsymbol{\nabla}}

\newcommand{\kS}{\kappa_\mathrm{S}}

\newcommand{\kB}{\kappa_\mathrm{B}}

\newcommand{\alphaB}{\alpha_\mathrm{B}}

\newcommand{\betaB}{\beta_\mathrm{B}}

\newcommand{\R}{\vect{r}}
\newcommand{\Intd}{\mathrm{d }}

\newcommand{\F}{\vect{F}}

\newcommand{\muS}{\mu^{\mathrm{S}}}
\newcommand{\muP}{\mu^{\mathrm{P}}}

\newcommand{\Faxen}{Fax\'{e}n }

\newcommand{\eR}{\vect{e}_r}
\newcommand{\ePhi}{\vect{e}_{\phi}}

\newcommand{\eZ}{\vect{e}_{z}}

\newcommand{\gOne}{\vect{g}_1}
\newcommand{\gTwo}{\vect{g}_2}

\newcommand{\JS}{J_{\mathrm{S}}}

\newcommand{\vStok}{\vect{v}^{\mathrm{S}}}
\newcommand{\pStok}{p^{\mathrm{S}}}
\newcommand{\vStokcom}{v^{\mathrm{S}}}

\newcommand{\vIm}{\vect{v}^{*}}
\newcommand{\pIm}{p^{*}}
\newcommand{\vImcom}{v^{*}}

\newcommand{\EB}{E_\mathrm{B}}

\newcommand{\MSpara}{{M_\parallel}_\mathrm{S}}
\newcommand{\LSpara}{{L_\parallel}_\mathrm{S}}
\newcommand{\NSpara}{{N_\parallel}_\mathrm{S}}

\newcommand{\MBpara}{{M_\parallel}_\mathrm{B}}
\newcommand{\LBpara}{{L_\parallel}_\mathrm{B}}
\newcommand{\NBpara}{{N_\parallel}_\mathrm{B}}

\newcommand{\new}[1]{#1}

\newcommand{\phanSp}{\phantom{cccc}}

    \usepackage{epstopdf}
    
    \let\mathbf\undefined
    \usepackage{mathptmx} % APS-like font (Times New Roman)

\begin{document}
    %\title{Hydrodynamic mobility of a sphere moving along the centerline of a cylindrical elastic tube}
    \title{Hydrodynamic mobility of a sphere moving on the centerline of an elastic tube}

    \author{Abdallah Daddi-Moussa-Ider}
    \email{ider@thphy.uni-duesseldorf.de}
    \affiliation
    {Biofluid Simulation and Modeling, Fachbereich Physik, Universit\"at Bayreuth, Universit\"{a}tsstra{\ss}e 30, Bayreuth 95440, Germany}

    \affiliation
    {Institut f\"{u}r Theoretische Physik II: Weiche Materie, Heinrich-Heine-Universit\"{a}t D\"{u}sseldorf, Universit\"{a}tsstra\ss e 1, D\"{u}sseldorf 40225, Germany}

    \author{Maciej Lisicki}

    \affiliation
    {Department of Applied Mathematics and Theoretical Physics, Wilberforce Rd, Cambridge CB3 0WA, United Kingdom}

    \affiliation
    {Institute of Theoretical Physics, Faculty of Physics, University of Warsaw, Pasteura 5, 02-093 Warsaw, Poland }

    \author{Stephan Gekle}
    \email{stephan.gekle@uni-bayreuth.de}
    \affiliation
    {Biofluid Simulation and Modeling, Fachbereich Physik, Universit\"at Bayreuth, Universit\"{a}tsstra{\ss}e 30, Bayreuth 95440, Germany}

    \date{\today}

    \begin{abstract}
    Elastic channels are an important component of many soft matter systems, in which hydrodynamic interactions with confining membranes determine the behavior of particles in flow. 
   	In this work, we derive analytical expressions for the Green's functions associated to a point-force (Stokeslet) directed parallel or perpendicular to the axis of an elastic cylindrical channel exhibiting resistance against shear and bending.
   	We then compute the leading order self- and pair mobility functions of particles on the cylinder axis, finding that the mobilities are primarily determined by membrane shear and that bending does not play a significant role.
   	In the quasi-steady limit of vanishing frequency, the particle self- and pair mobilities near a no-slip hard cylinder are recovered only if the membrane possesses a non-vanishing shear rigidity.
   	We further compute the membrane deformation, finding that deformation is generally more pronounced in the axial (radial) directions, for the motion along (perpendicular) to the cylinder centerline, respectively. 
   	Our analytical calculations \new{for the Green's functions in an elastic cylinder can serve as a fundamental building block for future studies} and are verified by fully resolved boundary integral simulations where a very good agreement is obtained.

    \end{abstract}
    \maketitle

    \section{Introduction}

    Many biological and industrial microscale processes occur in geometric confinement, which is known to strongly affect the diffusional dynamics in a viscous fluid~\cite{brady88, bleibel14}. Hydrodynamic interactions with boundaries play a key role in such systems by determining their transport properties~\cite{wei00, lutz04, Janssen_2012, lele11, gross14}. 
    Tubular confinement is of particular interest, since flow in living organisms often involves channel-like structures, such as arteries in the cardiovascular system~\cite{Frey1952}.
    % or phloem tissues containing latex in {\it Hevea} trees~\cite{Southorn1969}. 
    A common feature of these complex networks of channels is the elasticity of their building material. 
    Arteries and capillaries of the blood system involve a large number of collagen and elastin filaments, which gives them the ability to stretch in response to changing pressure~\cite{Shadwick1999,Caro2011}. 
    Elastic deformation has been further utilized to control and direct fluid flow within flexible microfluidic devices~\cite{stone04, holmes13, tavakol14}.

    The motion of a small sphere in a viscous fluid filling a rigid cylinder is a well studied problem. 
    A review of most analytical developments can be found in the  classic book of Happel and Brenner~\cite{happel12}. In particular, axial motion has been studied using the method of reflections by \Faxen~\cite{faxen59, faxen22}, Wakiya~\cite{wakiya53}, Bohlin~\cite{bohlin60} and Zimmerman~\cite{zimmerman04}, to name a few, expressing the mobility in power series of the ratio of particle to cylinder diameter. 
    These works have been extended to finite-sized spheres~\cite{Leichtberg_1976, kkedzierski10}, pair interactions~\cite{cui02, cui02b,misiunas15} and recently to non-spherical particles~\cite{Yeh_2013}. 
    For an arbitrarily positioned particle, and in the presence of an external Poiseuille flow, the procedure has been generalized to yield expressions in terms of the particle and channel radius, and the eccentricity of the position of the particle, as derived e.g. in the works of Happel and collaborators~\cite{happel54, brenner58, greenstein68, bungay73} and Liron and Shahar~\cite{liron78}. 
    The slow motion of two spherical particles symmetrically placed about the axis of a cylinder in a direction perpendicular to their line of centers has later been studied by Greenstein and Happel~\cite{greenstein70}.
    Experimental verification of these results has been performed e.g. by the use of laser interferometry by Lecoq~\textit{et al.}~\cite{Lecoq1993} or using digital video microscopy measurements by Cui \textit{et al.}~\cite{cui02}.
    Theoretical developments have been supplemented by numerical computations of the resistance functions for spheres, bubbles and drops in cylindrical tubes~\cite{higdon95, pozrikidis05, keh07, Bhattacharya_2010, imperio11, Navardi_2015}. Other works include motion perpendicular to the axis~\cite{hasimoto76}, finite length of the tube~\cite{sano87} and the flow around a line of equispaced spheres moving at a prescribed velocity along the axis of a circular tube~\cite{sheard07}.
    Transient effects have also been taken into account in the works of Felderhof, both in the case of an incompressible~\cite{felderhof09} and compressible fluid~\cite{felderhof10, felderhof10b, felderhof11}.

    For elastic cylinders, most previous work has focused on the flow itself which is driven through a deformable elastic channel~\cite{Rubinow1972, fung13} where various physiological phenomena related to the cardiovascular and respiratory systems have been observed, including the generation of instabilities~\cite{bertram89}, small-amplitude wave propagation~\cite{grotberg01, grotberg04}, hysteresis behavior of arterial walls~\cite{canic06}, \new{peristaltic pumping \cite{Takagi2011}} and anomalous bubble propagation~\cite{heap08, heap09}.
    Further work has been devoted to investigate the influence of elastic tube deformation on flow behavior of a shear-thinning fluid~\cite{zheng91, nahar12, nahar13}, the steady flow in thick-walled flexible elastic tubes~\cite{mikelic07, marzo05} or the tensile instability under an axial load~\cite{alexander71, shan06}.
    Moreover, the lateral mobility of membrane inclusions in a cylindrical biological membrane has been studied using computer simulations~\cite{rahimi13, rahimi13thesis}.
\new{Regarding particles in elastic tubes, some works \cite{Lighthill_1968, Tozeren_1982} considered a closely fitting particle in an elastic cylinder which in some sense represents the opposite limit of the present problem.}

%    More recently, the slow rotational motion of a solid particle inside a cylindrical elastic tube has been theoretically investigated~\cite{daddi17d}. 

    The translational mobility of a particle inside an elastic cylinder has not been studied so far \new{(rotational mobility has recently been investigated in our related work \cite{daddi17d}).}
    Motivated by this knowledge gap, we turn our attention to the problem of hydrodynamic mobility of a small spherical particle slowly moving in a viscous fluid filling a circular cylindrical elastic tube. 
    In many situations such as blood flow through small capillaries, the Reynolds number is typically very small allowing us to adopt the framework of creeping (Stokes) flow~\cite{leal80}.
    It is known from previous works on systems bounded by elastic surfaces~\cite{tan12} that their deformations introduce memory into the system, which may lead to transient anomalous subdiffusion~\cite{daddi16, daddi16b} or a change of sign of pair hydrodynamic interactions~\cite{daddi16c}. 
    Here we determine \new{an analytical expression for the Green's function in a cylindrical membrane of given elastic shear modulus and bending rigidity filled with and surrounded by a Newtonian fluid as the fundamental solution governing Stokes flow in that particular geometry.} From this, we compute the frequency-dependent mobility of a small \new{massless point} particle inside the cylinder.
    The solution is obtained by directly solving the Stokes equations in cylindrical geometry by the use of Fourier-Bessel expansion to represent the fluid velocity and pressure.

    The remainder of the paper is organized as follows. 
    In section~\ref{theoretical}, we formulate the problem of axial and radial motions of a small colloid inside an elastic tube in terms of the Stokes equations supplemented by appropriate boundary conditions. 
    We then present the method of solving these equations and use the obtained results in section~\ref{mobility} to derive explicit expressions for the frequency-dependent self- and pair mobility functions for colloids moving along or perpendicular to the centerline of the tube.
    Further, we calculate the deformation of the membrane for a given actuation. 
    In section~\ref{comparison}, we compare our theoretical developments to boundary integral numerical simulations for a chosen set of parameters for particles moving under a harmonic or a steady constant external force. 
    \new{We conclude the paper in section~\ref{conclusion} and relegate technical details to the appendices.    
    In appendix~\ref{appendix:membraneMech}, we derive in cylindrical coordinates the traction jumps across a membrane endowed with shear and bending resistances, which serve as boundary conditions for the calculation of the relevant Stokes flow.
    Appendices~\ref{appendix:determinationConstantsAxial} and~\ref{appendix:determinationConstantsRadial} provide explicit analytical solutions for axial and radial motions, respectively, for the two limiting cases of a membrane resisting either only to shear or only to bending. 
    The solution combining the two can be derived in the same way.}

    %%%%%%%%%%%%%%%%%%%%%%%%%%%%%%%%%%%%%%%%%%%%%%%%%%%%%%%%%%%%%%%%%%%%

    \section{Theoretical description}\label{theoretical}

    %\subsection{Problem formulation}

    \begin{figure}
    \begin{center}
      %\scalebox{0.9}{\input{Pics/cylinderIllustration}}
            \includegraphics[scale=0.7]{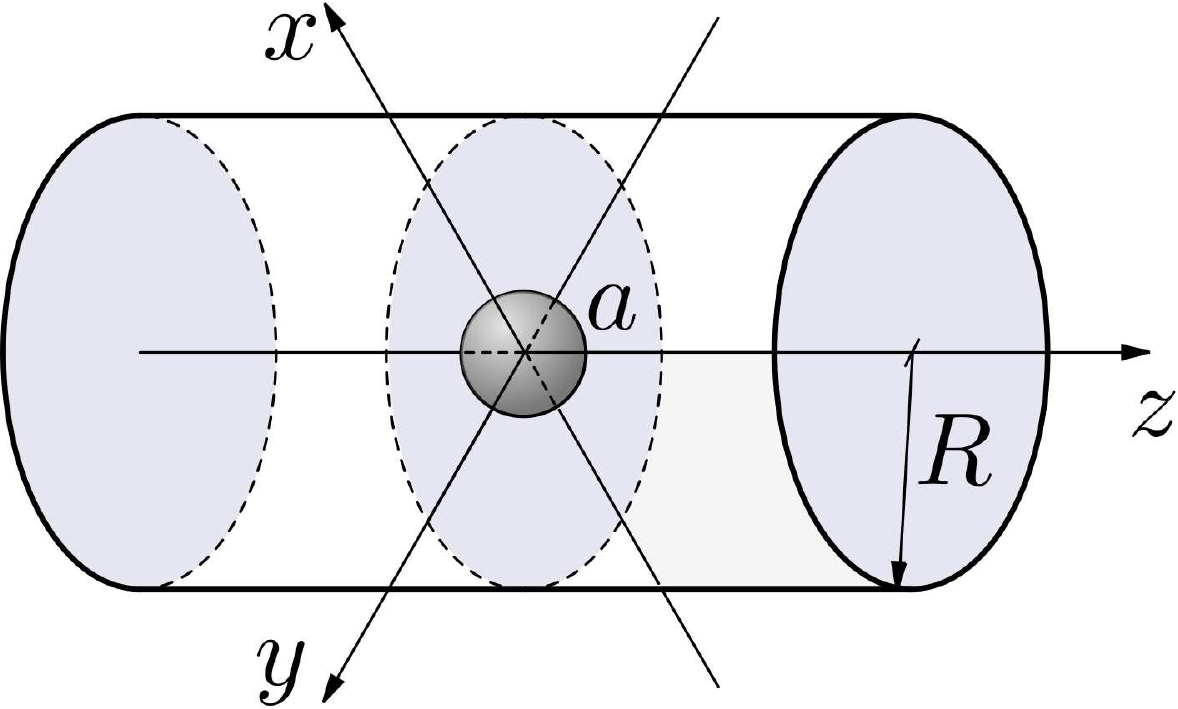}
    \caption{Illustration of the system setup. A small spherical solid particle of radius~$a$ located at the origin moving on the centerline of a deformable elastic tube of radius~$R$.} 
    \label{cylinderIllustration}
    \end{center}
    \end{figure}

    We consider a small spherical particle of radius~$a$ fully immersed in a Newtonian fluid and moving on the axis of a cylindrical elastic tube of initial (undeformed) radius $R \gg a$.
    The tube membrane exhibits a resistance against shear and bending.
    We choose the cylindrical coordinate system $(r,\phi,z)$ where the $z$ coordinate is directed along the cylinder axis with the origin located at the center of the particle (see figure~\ref{cylinderIllustration} for an illustration of the system setup). The regions inside and outside the cylinder are labeled 1 and~2, respectively.

    We proceed by computing the Green's functions which are solutions of the Stokes equations
    \begin{subequations}
    \begin{align}
    \eta \boldNabla^2 \vect{v}_1 - \boldNabla p_1 + \vect{F} (t) \, \delta (\R)  &= 0 \, , \\
    \boldNabla \cdot \vect{v}_1 &= 0 \, , 
    \end{align}
    \label{equationMotion_fluid1}
    \end{subequations}
    inside the tube (for $r<R$) and 
    \begin{subequations}
    \begin{align}
    \eta \boldNabla^2 \vect{v}_2 - \boldNabla p_2 &= 0 \, , \\
    \boldNabla \cdot \vect{v}_2 &= 0 \, , 
    \end{align}
    \label{equationMotion_fluid2}
    \end{subequations}
    outside (for $r>R$). 
    Here $\eta$ denotes the fluid shear viscosity, assumed to be the same everywhere.
    $\vect{F} (t)$ is an arbitrary time-dependent point-force acting at the particle position.
    \new{The Reynolds number $\operatorname{Re} = \rho V a / \eta$ as well as the Strouhal number $\operatorname{St} = \omega a / V $ are defined via the velocity amplitude of the oscillating particle $V$ with $\omega$ being the corresponding oscillation frequency.
    In the linear response regime, the latter is a small quantity thus making Re and St also small quantities such that the non-linear and the time-dependent terms of the Navier-Stokes equations can safely be neglected and the use of the steady Stokes equations is justified.}

    We therefore need to solve Eqs.~\eqref{equationMotion_fluid1} and \eqref{equationMotion_fluid2} subject to the regularity conditions 
    \begin{align}
    |\vect{v}_1| < \infty \,  &\text{ for } |\R| = 0 \, , \label{regularity_Inside} \\
    \vect{v}_1 \to \mathbf{0} \, &\text{ for } z \to \infty \, , \label{regularity_Inside_2} \\
    \vect{v}_2 \to \mathbf{0} \, &\text{ for } |\R| \to \infty \, , \label{regularity_Outside}
    \end{align}
    together with the boundary conditions imposed at the surface $r=R$ by assuming small deformations, 
    namely the natural continuity of fluid velocity
    \begin{align}
      [v_r] &= 0 \, , \label{BC:v_r} \\
      [v_\phi] &= 0 \, , \label{BC:v_phi} \\
      [v_z] &= 0 \, , \label{BC:v_z} 
    \end{align}
    and the traction jumps stemming from membrane elastic deformation
    \begin{align}
      [\sigma_{z r}] &= \Delta f_z^{\mathrm{S}} \, , \label{BC:sigma_r_z} \\
      [\sigma_{\phi r}] &= \Delta f_\phi^{\mathrm{S}} \, , \label{BC:sigma_r_phi} \\
      [\sigma_{rr}] &= \Delta f_r^{\mathrm{S}} + \Delta f_r^{\mathrm{B}} \, , \label{BC:sigma_r_r}
    \end{align}
    where the notation $[w] := w(r = R^{+}) - w(r = R^{-})$ stands for the jump of a given quantity $w$ across the cylindrical elastic membrane. 
    These linearized traction jumps can be decomposed into two contributions due to shear (superscript S) and bending (superscript B). 
    %    The shear part is accounted for using the new-Hookean model~\cite{ramanujan98}.
    The membrane is modeled by combining the neo-Hookean model for shear~\cite{ramanujan98, lac04, Freund_2014, BarthesBiesel_2016}, and the Helfrich model~\cite{helfrich73, Guckenberger_preprint} for bending of its surface. 
    As derived in appendix~\ref{appendix:membraneMech}, the linearized traction jumps due to shear are written as
    \begin{subequations}\label{shearTractions}
    \begin{align}
    \Delta f_\phi^{\mathrm{S}} &= -\frac{\kS}{3} \left( u_{\phi, zz} + \frac{3 u_{z, \phi z}}{R} + \frac{4 (u_{r,\phi} + u_{\phi, \phi\phi})}{R^2} \right)  \, , \\
    \Delta f_{z}^{\mathrm{S}}    &= -\frac{\kS}{3} \left( 4 u_{z,zz} + \frac{2 u_{r,z} + 3 u_{\phi, z\phi}}{R} + \frac{u_{z, \phi\phi}}{R^2} \right) \, , \\
    \Delta f_{r}^{\mathrm{S}}    &= \frac{2\kS}{3} \left( \frac{2 (u_r + u_{\phi, \phi}) }{R^2} + \frac{u_{z,z}}{R}  \right) \, ,
    \end{align}
    \end{subequations}
    where $\kS$ is the surface shear modulus (expressed in N/m).
    Here $\vect{u} (\phi, z) = u_r(\phi, z) \eR + u_\phi(\phi,z) \ePhi + u_z(\phi, z) \eZ$ is the membrane deformation field.
    The comma in indices denotes a partial spatial derivative. 
    % Note that due to the axial symmetry of the system, there is no shear deformation in the membrane and deformations in the azimuthal direction $u_\phi$ do not appear.

    %    Bending of the membrane is described following the well-established Helfrich model~\cite{helfrich73} which relates the normal traction jump to the radial deformation as (cf. Appendix)
    For bending, only a normal traction jump appears
    \begin{equation}\label{bendingTraction}
      %\begin{split}
      \Delta f_r^{\mathrm{B}} = \kB \Big( R^3 u_{r,zzzz} + 2R(u_{r,zz} + u_{r,zz\phi\phi}) \\
		  + \frac{u_r+2u_{r,\phi\phi}+u_{r, \phi\phi\phi\phi}}{R} \Big)  \, ,
      %\end{split}
    \end{equation}
    where $\kB$ is the bending modulus (expressed in Nm). 
    Note that Helfrich bending does not introduce a discontinuity in the tangential traction jumps~\cite{Guckenberger_preprint}.

    The effect of these two elastic modes, given the characteristic frequency of actuation~$\omega$, is determined by two dimensionless quantities, the shear coefficient $\alpha$ and the bending coefficient $\alphaB$, defined as
    \begin{equation}
    \alpha:=\frac{2 \kS}{3\eta R\omega}\, , 
      \qquad\qquad 
    \alphaB := \frac{1}{R} \left(\frac{\kB}{\eta\omega}\right)^{1/3}. 
    \end{equation}
    Note that this definition is slightly different than in our earlier work~\cite{daddi16}. 
    
    In cylindrical coordinates, the components of the fluid stress tensor are expressed in the usual way as~\cite{kim13}
    \begin{align}
      \sigma_{\phi r} &= \eta \left( v_{\phi ,r} - \frac{v_\phi + v_{r, \phi}}{r} \right) \, , \notag \\
      \sigma_{zr}     &= \eta (v_{z,r} + v_{r,z}) \, , \notag  \\
      \sigma_{rr}     &= -p + 2\eta v_{r,r} \, . \notag 
    \end{align}

    A direct relationship between velocity and displacement at the undisplaced membrane $r=R$ can be obtained from the no-slip boundary condition, $\vect{v}=\partial_t \vect{u}$. 
    Transforming to the temporal Fourier space, we have~\cite{bickel06}
    \begin{equation}
    u_\alpha (\phi, z) = \left. \frac{v_{\alpha} (r,\phi, z)}{i\omega} \right|_{r = R} \, , \qquad \alpha \in \{r, \phi, z\} \, . \label{no-slip-relation}
    \end{equation}

    It is worth mentioning that when the finite amplitude of deformation is important, it becomes necessary to apply the boundary conditions at the deformed membrane rather than at undisplaced membrane~\cite{weekley06, salez15, saintyves16, rallabandi17}

    We then solve the equations of motion by expanding them in the form of Fourier integrals in two distinct regions (inside and outside the cylindrical membrane). The solution can be written in terms of  integrals of harmonic functions with unknown coefficients, which we then determine from the boundary conditions of 
    (a) continuity of radial, azimuthal and axial velocities, and 
    (b) surface traction jumps deriving from the elastic properties of the membrane.
    We present the full analytic solutions for two limiting models of the membrane susceptible only to shear or bending deformations.

    %\subsection{Method of solution} 
    We begin by expressing the solution of  Eqs.~\eqref{equationMotion_fluid1} inside the cylinder as a sum of a point-force flow field and the flow reflected from the interface~\cite{fuentes88, fuentes89}
    \begin{align}
    \vect{v}_1 &= \vStok + \vIm \, , \notag  \\
    p_1        &= \pStok + \pIm \, , \notag
    \end{align}
    where $\vStok$ and $\pStok$ are the Stokeslet solution in an infinite (unbounded) medium 
    % given by the Oseen tensor 
    and $\vIm$ and $\pIm$ are the solutions of the homogenous (force-free) Stokes equations
    \begin{subequations}\label{equationMotion_fluid1_Image}
      \begin{align}
    \eta \boldNabla^2 \vIm - \boldNabla \pIm &= 0 \, , \\
    \boldNabla \cdot \vIm &= 0 \, ,  
    \end{align}
    \end{subequations}
    required such that the full flow field satisfies the regularity and boundary conditions.
    In the following, we shall consider the cases of particle motion parallel or perpendicular to the cylinder centerline separately.
\new{We note that the particle mobility for motion in an arbitrary direction cannot be obtained exactly by a simple weighted superposition of these two fundamental mobilities. 
This is due to the elastic nature of the boundary and in contrast to a hard cylinder.
By comparing with boundary-integral simulations further below, we show however that superposition does yield a pretty good approximation and therefore seems a reasonable approach.}

    %%%%%%%%%%%%%%%%%%%%%%%%%%%%%%%%%%%%%%%%%%%%%%%%%%%%%%%%%%

    \subsection{Axial motion}

    The Stokeslet solution for a point-force located at the origin and directed along the cylinder axis reads~\cite{happel12}
    %\SG{incorrect reference?}\AI{citing Happel and Brenner here makes more sens}
    \begin{equation}
    \vStokcom_r = \frac{F_z}{8\pi\eta} \frac{zr}{d^3} \, , \quad \vStokcom_z = \frac{F_z}{8\pi\eta} \left( \frac{1}{d} + \frac{z^2}{d^3} \right) \, , \quad \pStok = \frac{F_z}{4\pi} \frac{z}{d^3} \, , \notag 
    \end{equation}
    where $d := \sqrt{r^2+z^2}$ is the distance from the singularity position. We now rewrite the Stokeslet solution  in the form of a Fourier integral expansion noting that 
    \begin{equation}
    \frac{rz}{d^3} = -\frac{\partial}{\partial r} \frac{z}{d} \, , \qquad \frac{1}{d}+\frac{z^2}{d^3} = \frac{2}{d} - \frac{\partial}{\partial z} \frac{z}{d} \, , \label{firstFormula}
    \end{equation}
    and making use of the integral relations~\cite{watson95, brenner58}
    \begin{subequations}  \label{secondFormula}
    \begin{align}
    \frac{1}{d} &= \frac{2}{\pi} \int_0^\infty K_0(q r) \cos q z \, \Intd q \, , \\
    \frac{z}{d} &= \frac{2}{\pi} \,  r \int_0^\infty K_1(q r) \sin q z \, \Intd q \, ,
      \end{align} 
      \end{subequations}
    wherein $K_\alpha$ is the $\alpha$th order modified Bessel function of the second kind~\cite{abramowitz72}. 
    We thus express the axisymmetric Stokeslet solution in the integral form with the wavenumber $q$ as
    \begin{subequations} \label{stokeslet}
    \begin{align}
    \vStokcom_r (r,z) &= \frac{F_z}{4\pi^2\eta}  \int_0^\infty r q K_0(q r)  \sin q z \, \Intd q \, ,   \\
    \vStokcom_z (r,z) &= \frac{F_z}{4\pi^2\eta} \int_0^\infty \Big( 2K_0(q r) - q r K_1(q r) \Big) \cos q z \, \Intd q \, , \\
    \pStok      (r,z) &= \frac{F_z}{2\pi^2}  \int_0^\infty q K_0(q r) \sin q z \, \Intd q \, ,
    \end{align}
    \end{subequations}
    using the relation $\partial K_1(qr)/\partial r = -q K_0(qr) - K_1(qr)/r \, .$

    The reflected flow can also be represented in a similar way by noting that the homogenous Stokes equations~\eqref{equationMotion_fluid1_Image} for axisymmetric motion have a general solution expressed in terms of two harmonic functions $\Psi_\parallel$ and $\Phi_\parallel$ as~\cite[p. 77]{happel12}
    \begin{subequations}\label{vstar}
    \begin{align}
    \vImcom_r &= {\Psi_\parallel}_{,r} + r \, {\Phi_\parallel}_{,rr} \, , \label{v_r_prime} \\
    \vImcom_z &= {\Psi_\parallel}_{,z} + r \, {\Phi_\parallel}_{,rz} + {\Phi_\parallel}_{,z} \, , \label{v_z_prime} \\
    \pIm   &= -2\eta \, {\Phi_\parallel}_{,zz} \, . \label{pressure_prime}
    \end{align}
    \end{subequations}
    The two functions $\Psi_\parallel$ and $\Phi_\parallel$ are solutions to the axisymmetric Laplace equation which can be written in an integral form as
    \begin{subequations}\label{PhiPsi}
    \begin{align}
    \Phi_\parallel &= \frac{F_z}{4\pi^2\eta} \int_0^\infty \varphi_\parallel(q) f_\parallel(q r) \sin(q z) \, \Intd q \, ,  \\
    \Psi_\parallel &= \frac{F_z}{4\pi^2\eta} \int_0^\infty \psi_\parallel(q) f_\parallel(q r) \sin(q z) \, \Intd q  \, , 
    \end{align}
    \end{subequations}
    where $\varphi_\parallel$ and $\psi_\parallel$ are to be determined from the boundary conditions. 
    At this point, the arbitrary prefactor outside the integral is chosen such that the resulting velocity and pressure fields will in the end have a similar representation as the Stokeslet solution given by Eq.~\eqref{stokeslet}.
    %Moreover, $f(x)$ satisfies the differential equation~\cite{abramowitz72}
    %\begin{equation}
    % \frac{\Intd}{\Intd x} \left( x \, \frac{\Intd f(x)}{\Intd x} \right) - x f(x) = 0 \, . \label{modifiedBesselEquation}
    %\end{equation}
    For $\Psi_\parallel$ and $\Phi_\parallel$ to be solutions to the axisymmetric Laplace equation, the function $f_\parallel$ has to satisfy the zeroth order modified Bessel equation~\cite{abramowitz72}.
    Since the image solution inside the cylinder has to be regular at the origin owing to Eq.~\eqref{regularity_Inside}, we take $f_\parallel \equiv I_0 $ in the inner solution. 
    Combining Eqs.~\eqref{vstar} and \eqref{PhiPsi} together, the solution of Eq.~\eqref{equationMotion_fluid1_Image} reads
    \begin{subequations} \label{reflected}
    \begin{align}
    \vImcom_r (r,z) &= \frac{F_z}{4\pi^2\eta} \int_0^\infty q \Big(  \big( rq I_0(qr) - I_1 (q r) \big) \varphi^*_\parallel (q) % \notag \\
	      + I_1(q r) \psi^*_\parallel(q) \Big) \sin q z \, \Intd q \, , \label{v_r_star_final} \\
    \vImcom_z (r,z) &= \frac{F_z}{4\pi^2\eta} \int_0^\infty q \Big(  \big( rq I_1(qr) + I_0 (q r) \big) \varphi^*_\parallel(q) % \notag \\
	      + I_0(q r) \psi^*_\parallel(q) \Big) \cos q z \, \Intd q \, , \label{v_z_star_final} \\
    \pIm      (r,z) &=  \frac{F_z}{2\pi^2}       \int_0^\infty q^2 \varphi^*_\parallel(q) I_0(q r) \sin q z \, \Intd q \, . \label{pressure_star_final}
    \end{align}
    \end{subequations}

    Thus the Green's function inside the elastic cylindrical channel for the axial point-force is given explicitly by summing up the Stokeslet contribution \eqref{stokeslet} and the reflected flow \eqref{reflected}.

    The outer solution for the force-free Stokes equations~\eqref{equationMotion_fluid2} has an analogous structure with the only difference that the flow has to decay at infinity by virtue of Eq.~\eqref{regularity_Outside} and we therefore take $f_\parallel \equiv K_0$ leading to
    \begin{subequations} \label{flowOut}
    \begin{align}
    {v_2}_r (r,z) &= \frac{F_z}{4\pi^2\eta} \int_0^\infty q \Big( \big( rq K_0(qr) + K_1 (q r) \big) {\varphi_2}_\parallel (q)  % \notag \\
	    -K_1(q r) {\psi_2}_\parallel (q) \Big) \sin q z \, \Intd q \, , \label{v_r_2_final} \\
    {v_2}_z (r,z) &= \frac{F_z}{4\pi^2\eta} \int_0^\infty q \Big(  \big( K_0 (q r) - rq K_1(qr) \big) {\varphi_2}_\parallel (q) % \notag \\
	    + K_0(q r) {\psi_2}_\parallel (q) \Big) \cos q z \, \Intd q \, , \label{v_z_2_final} \\
    p_2     (r,z) &= \frac{F_z}{2\pi^2}  \int_0^\infty q^2 {\varphi_2}_\parallel (q) K_0(q r) \sin q z \, \Intd q \, , \label{pressure_2_final}
    \end{align}
    \end{subequations}
    after making use of the relations $\partial I_0(qr)/\partial r=q I_1(qr)$, $\partial I_1(qr)/\partial r = qI_0(qr)-I_1(qr)/r$ and $\partial K_0(qr)/\partial r = -q K_1(qr) \, .$
    The unknown functions $\psi^*_\parallel$, $\varphi^*_\parallel$, ${\psi_2}_\parallel$ and ${\varphi_2}_\parallel$ remain to be determined from the boundary conditions of continuous velocity and prescribed traction jumps at the membrane.

    The continuity of radial and axial velocity components across the membrane expressed by Eqs.~\eqref{BC:v_r} and \eqref{BC:v_z} leads to the expression of the functions ${\psi_2}_\parallel$ and ${\varphi_2}_\parallel$ in terms of $\psi^*_\parallel$ and $\varphi^*_\parallel$ as
    %\begin{subequations}
    %\begin{align}
     % -I_1\psi^*_\parallel + \big( I_1-s I_0 \big) \varphi^*_\parallel - K_1 {\psi_2}_\parallel + \big( K_1 + s K_0 \big) {\varphi_2}_\parallel &=  R K_0 \, , \notag  \\
     % -s I_0 \psi^*_\parallel -s \big( I_0+s I_1 \big) \varphi^*_\parallel  + s K_0 {\psi_2}_\parallel  + s \big( K_0-s K_1 \big) {\varphi_2}_\parallel  &=   R \left( 2K_0 - s K_1 \right)  \, , \notag 
    %\end{align}
    %\end{subequations}
    \begin{subequations}\label{psi_2_phi_2_from_psi_1_phi_1}
    \begin{align}
    {\psi_2}_\parallel &=  \frac{G_\parallel \psi^*_\parallel  + (1+s^2) S_\parallel \varphi^*_\parallel }{D_\parallel} +  \frac{  R }{s} \, , \label{psi_2_from_psi_1} \\
    {\varphi_2}_\parallel &= \frac{S_\parallel \psi^*_\parallel  + G_\parallel \varphi^*_\parallel }{D_\parallel} +  \frac{ R }{s} \, , \label{phi_2_from_phi_1}
    \end{align}
    \end{subequations}
    where $s := q R$ is a dimensionless wavenumber and
    %\begin{subequations}
    \begin{align}
    S_\parallel &= K_1 I_0 + K_0 I_1 \, , \notag \\
    G_\parallel &= \big( sK_1-K_0 \big) I_1 + \big( s K_0 + K_1 \big) I_0  \, , \notag \\
    D_\parallel &= s K_0^2 - s K_1^2 + 2 K_0 K_1 \, . \notag 
    \end{align}
    %\end{subequations} 
    The modified Bessel functions have the argument $s$ which is dropped for brevity.

    The form of $\psi^*_\parallel$ and $\varphi^*_\parallel$ may be determined given the constitutive model of the membrane. 
    In appendix~\ref{appendix:determinationConstantsAxial}, we provide explicit analytical expressions for $\psi^*_\parallel$ and $\varphi^*_\parallel$ by considering independently a shear-only or a bending-only membrane.
    An analogous resolution procedure can be employed by considering simultaneously shear and bending resistances.
    % The calculations can readily be performed by computer algebra software but analytical expressions are not listed here due to their complexity and lengthiness.
    
    For future reference, we shall express the solution near a membrane with both shear and bending rigidities as
    \begin{equation}
      \psi^*_\parallel  =  R \, \frac{M_\parallel}{N_\parallel} \, , \quad \varphi^*_\parallel  =   R \, \frac{L_\parallel}{N_\parallel} \, .  \label{psi_1_AND_phi_1_both}
    \end{equation}
    We note that the steady solution near a hard cylinder as first computed by Liron and Shahar~\cite{liron78} stated by Eq.~\eqref{LironandShahar} is recovered in the vanishing frequency limit.
    In the following, the solution for a point-force acting perpendicular to the cylinder axis will be derived.

    %%%%%%%%%%%%%%%%%%%%%%%%%%%%%%%%%%%%%%%%%

    \subsection{Radial motion}

    Without loss of generality, we shall consider that the point force is located at the origin and that its motion is directed along the $x$ direction in Cartesian coordinates.
    % corresponding to the $\phi=0$ direction in cylindrical coordinates.
    The induced velocity field reads~\cite{happel12}
    %\SG{This ref seems incorrect here}\AI{Changed}
    \begin{equation}
    \vStokcom_x = \frac{F_x}{8\pi\eta} \left( \frac{1}{d} + \frac{x^2}{d^3} \right) \, , \quad 
    \vStokcom_y = \frac{F_x}{8\pi\eta} \frac{xy}{d^3} \, , \quad 
    \vStokcom_z = \frac{F_x}{8\pi\eta} \frac{xz}{d^3} \, , \notag
    \end{equation}
    and the pressure
    \begin{equation}
      \pStok = \frac{F_x}{4\pi} \frac{x}{d^3} \, . \notag
    \end{equation}
    % where again $d := \sqrt{r^2+z^2}$. 
    Setting $x=r \cos\phi$ and $y=r \sin\phi$, the radial and tangential velocities read
    \begin{equation}
    \vStokcom_r = \frac{F_x}{8\pi\eta}  \left( \frac{1}{d} + \frac{r^2}{d^3} \right) \cos \phi \, , \qquad
    \vStokcom_\phi = -\frac{F_x}{8\pi\eta} \frac{\sin \phi}{d} \, . \notag
    \end{equation}

    By making use of Eqs.~\eqref{firstFormula} and \eqref{secondFormula}, the Stokeslet solution can thus be written in the form of a Fourier-Bessel integral expansion as
    \begin{subequations} \label{stokeslet_radial}
    \begin{align}
    \vStokcom_r (r,\phi, z) &= \frac{F_x}{4\pi^2\eta} \, \cos \phi \int_0^\infty \left( K_0(q r) +qr K_1(qr) \right)  \cos q z \, \Intd q \, ,  \\
    \vStokcom_\phi (r,\phi, z) &= -\frac{F_x}{4\pi^2\eta} \, \sin  \phi \int_0^\infty K_0(q r)   \cos q z \, \Intd q \, ,  \\
    \vStokcom_z (r,\phi, z) &= \frac{F_x}{4\pi^2\eta} \, \cos\phi  \int_0^\infty qr K_0(qr)  \sin q z \, \Intd q \, ,  \\
    \pStok      (r,\phi, z) &= \frac{F_x}{2\pi^2} \, \cos \phi \int_0^\infty q K_1(qr)  \cos qz \, \Intd q \, . 
    \end{align}
    \end{subequations}
    % using the fact that $\partial K_1(qr)/\partial r = -q K_0(qr) - K_1(qr)/r$.

    Similar, the reflected flow can also be represented by noting that the force-free Stokes equations~\eqref{equationMotion_fluid1_Image} have a general solution expressed in terms of three harmonic functions $\Psi_\perp$, $\Phi_\perp$ and $\Gamma_\perp$ as~\cite[p.~77]{happel12}
    \begin{subequations}\label{vstar_radial}
    \begin{align}
    \vImcom_r &= {\Psi_\perp}_{,r} + \frac{{\Gamma_\perp}_{,\phi}}{r} + r \, {\Phi_\perp}_{,rr} \, , \label{v_r_prime_radial} \\
    \vImcom_\phi &= \frac{{\Psi_\perp}_{,\phi}}{r} - {\Gamma_\perp}_{,r} - \frac{{\Phi_\perp}_{,\phi}}{r} + {\Phi_\perp}_{,\phi r} \, , \label{v_phi_prime_radial} \\ 
    \vImcom_z &= {\Psi_\perp}_{,z} + r \, {\Phi_\perp}_{,rz} + {\Phi_\perp}_{,z} \, , \label{v_z_prime_radial} \\
    \pIm   &= -2\eta \, {\Phi_\perp}_{,zz} \, . \label{pressure_prime_radial}
    \end{align}
    \end{subequations}

    The functions $\Psi_\perp$, $\Phi_\perp$ and $\Gamma_\perp$ are solutions to the asymmetric Laplace equation which can be written in an integral form as
    \begin{subequations}\label{PhiPsi_radial}
    \begin{align}
    \Phi_\perp &=   \frac{F_x}{4\pi^2\eta} \, \cos \phi \int_0^\infty \varphi_\perp(q) f_\perp(q r)  \cos(q z) \, \Intd q \, ,  \\
    \Psi_\perp &=   \frac{F_x}{4\pi^2\eta} \, \cos \phi \int_0^\infty \psi_\perp(q) f_\perp(q r)  \cos(q z) \, \Intd q  \, ,  \\
    \Gamma_\perp &= \frac{F_x}{4\pi^2\eta} \,  \sin \phi \int_0^\infty \gamma_\perp(q) f_\perp(q r) \cos(q z) \, \Intd q  \, , 
    \end{align}
    \end{subequations}
    where $\varphi_\perp$, $\psi_\perp$ and $\gamma_\perp$ are wavenumber-dependent quantities to be determined from the prescribed boundary conditions at the membrane.

    For $\Psi_\perp$, $\Phi_\perp$ and $\Gamma_\perp$ to be solutions to the Laplace equation, the function $f_\perp$ should be solution to the first order modified Bessel equation~\cite{abramowitz72}.
    In order to satisfy the regularity of the image solution inside the elastic cylinder as stated by Eq.~\eqref{regularity_Inside}, we take $f_\perp \equiv I_1 $ in the inner solution. 
    Upon combination of Eqs.~\eqref{vstar_radial} and \eqref{PhiPsi_radial} together, the solution of Eq.~\eqref{equationMotion_fluid1_Image} for a radial Stokeslet reads
    %\begin{widetext}
    %\begin{table*}
    \begin{subequations} \label{reflected_radial}
    \begin{align}
    \vImcom_r (r,\phi, z) &= \frac{F_x}{4\pi^2\eta} \frac{\cos \phi}{r} \int_0^\infty  \big( \left( (2+q^2 r^2)I_1(qr)-qr I_0(qr) \right)\varphi^*_\perp (q)   \notag \\
	      &+ \left( qr I_0(qr)-I_1(qr) \right) \psi^*_\perp(q) + I_1(qr) \, \gamma^*_\perp (q) \big) \cos qz      \, \Intd q \, , \label{v_r_star_final_radial} \\
    \vImcom_\phi (r,\phi, z) &= -\frac{F_x}{4\pi^2\eta} \frac{\sin\phi}{r} \int_0^\infty \big( \left( qrI_0(qr)-2I_1(qr) \right)\varphi^*_\perp (q)   \notag \\
		 &+ I_1(qr) \psi^*_\perp (q) + \left( qr I_0(qr)-I_1(qr) \right) \gamma^*_\perp (q) \big) \cos qz  \, \Intd q \, , \label{v_phi_star_final_radial} \\
    \vImcom_z (r,\phi, z) &= -\frac{F_x \cos\phi}{4\pi^2\eta}  \int_0^\infty q \big( qr I_0(qr) \varphi^*_\perp (q)
	      + I_1(qr) \psi^*_\perp (q) \big) \sin qz  \, \Intd q \, , \label{v_z_star_final_radial} \\
    \pIm   (r,\phi, z) &=  \frac{F_x \cos \phi}{2\pi^2}  \int_0^\infty q^2 I_1(qr) \varphi^*_\perp (q) \cos qz  \, \Intd q \, . \label{pressure_star_final_radial}
    \end{align}
    \end{subequations}

    The outer solution for the force-free Stokes equations~\eqref{equationMotion_fluid2} has to decay at infinity owing to Eq.~\eqref{regularity_Outside}, suggesting to take $f_\perp \equiv K_1$ leading to
    \begin{subequations} \label{flowOut_radial}
    \begin{align}
    {v_2}_r (r,\phi, z) &= \frac{F_x}{4\pi^2\eta} \frac{\cos \phi}{r} \int_0^\infty  \big( \left( (2+q^2 r^2)K_1(qr)+qr K_0(qr) \right) {\varphi_2}_\perp (q)   \notag \\
	    &- \left( qr K_0(qr) + K_1(qr) \right) {\psi_2}_\perp (q) + K_1(qr) \, {\gamma_2}_\perp (q) \big) \cos qz      \, \Intd q \, , \label{v_r_2_final_radial} \\
    {v_2}_\phi (r,\phi, z) &= \frac{F_x}{4\pi^2\eta} \frac{\sin\phi}{r} \int_0^\infty \big( \left( qrK_0(qr)+2K_1(qr) \right) {\varphi_2}_\perp (q)  \notag \\
	      &- K_1(qr) {\psi_2}_\perp (q) + \left( qr K_0(qr) + K_1(qr) \right) {\gamma_2}_\perp (q) \big) \cos qz  \, \Intd q \, , \label{v_phi_2_final_radial} \\
    {v_2}_z (r,\phi, z) &= \frac{F_x \cos\phi}{4\pi^2\eta}  \int_0^\infty q \left( qr K_0(qr) {\varphi_2}_\perp (q) - K_1(qr) {\psi_2}_\perp (q) \right) \sin qz  \, \Intd q \, , \label{v_z_2_final_radial} \\
    p_2  (r,\phi, z) &=  \frac{F_x \cos \phi}{2\pi^2} \,  \int_0^\infty q^2 K_1(qr) {\varphi_2}_\perp (q) \cos qz  \, \Intd q \, . \label{pressure_2_final_radial}
    \end{align}
    \end{subequations}

    The six unknown functions can thus be determined from the imposed boundary conditions, namely the continuity of fluid velocity and the traction jumps across the membrane.
    
    The continuity of the velocity field expressed by Eqs.~\eqref{BC:v_r} through \eqref{BC:v_z} leads to the expression of the unknown functions ${\varphi_2}_\perp$, ${\psi_2}_\perp$ and ${\gamma_2}_\perp$ outside the cylinder in terms of $\varphi^*_\perp$, $\psi^*_\perp$ and $\gamma^*_\perp$ on the inside as
    %\begin{align}
    %&\left( sI_0-(2+s^2)I_1 \right)\varphi^*_\perp +(I_1-sI_0)\psi^*_\perp -I_1\gamma^*_\perp +K_1 {\gamma_2}_\perp 
    %+  \left( s K_0+(2+s^2)K_1 \right) {\varphi_2}_\perp  \notag  \\ 
    %&\phanSp
    %-(K_1+sK_0) {\psi_2}_\perp = R(K_0+sK_1) \, , \notag  \\
    %&(sI_0-2I_1)\varphi^*_\perp +I_1 \psi^*_\perp +(sI_0-I_1)\gamma^*_\perp + (sK_0+2K_1) {\varphi_2}_\perp  
    %-K_1 {\psi_2}_\perp \notag \\
    %&\phanSp
    %+(K_1+sK_0) {\gamma_2}_\perp = -R K_0 \, , \notag \\
    %&s^2 I_0 \varphi^*_\perp + sI_1 \psi^*_\perp + s^2K_0 {\varphi_2}_\perp - sK_1 {\psi_2}_\perp = R s K_0 \, . \notag
    %\end{align}
    %\end{widetext}
    %\end{table*}
    \begin{align}
    {\varphi_2}_\perp   &= \frac{S_\perp \varphi^*_\perp + (K_1+sK_0)G_\perp \psi^*_\perp + K_1 G_\perp \gamma^*_\perp }{D_\perp} + \frac{R}{s} \, , \label{Eq_1_forPhi2} \\
    {\psi_2}_\perp   &= \frac{s \left((2+s^2)K_0+sK_1 \right) G_\perp \varphi^*_\perp + S_\perp \psi^*_\perp + s K_0 G_\perp \gamma^*_\perp }{D_\perp} \, , \label{Eq_2_forPsi2} \\
    {\gamma_2}_\perp &= \frac{\left( S_\perp-G_\perp \left( sK_0 + (2+s^2)K_1 \right) \right) \gamma^*_\perp }{D_\perp}  %\notag \\
	      -\frac{2sK_0 G_\perp \varphi^*_\perp - 2K_1 G_\perp \psi^*_\perp}{D_\perp} - \frac{2R}{s} \, , \label{Eq_3_forOmega2}
    \end{align}
    where we have defined
    \begin{align}
    S_\perp &= -sK_0K_1 \left( sI_0+(2+s^2)I_1 \right) 
	    - s^2 \left( sI_0 K_0^2+I_1K_1^2 \right) \, , \notag \\
    G_\perp &= -s \left( I_0 K_1 + I_1 K_0 \right) \, , \notag  \\
    D_\perp &= s \big( s^2 K_0^3 + sK_0^2 K_1 - sK_1^3 
	    - (2+s^2)K_0 K_1^2 \big) \, . \notag
    \end{align}

    In appendix~\ref{appendix:determinationConstantsRadial}, we provide explicitly the expressions of ${\psi}_\perp^*$, ${\phi}_\perp^*$ and ${\gamma}_\perp^*$ by considering independently membranes with pure shear or pure bending.

    %\subsubsection{Shear and bending}

    %The same resolution procedure can be adopted for the determination of the unknown coefficients when the membrane is endowed simultaneously with shear and bending resistances.
    %Explicit analytical expressions are lengthy and they will not be given here.
    For future reference, we shall express the solution for a membrane endowed with both shear and bending as
    \begin{equation}
      \psi^*_\perp  =  R \, \frac{M_\perp}{N_\perp} \, , \quad \varphi^*_\perp  =   R \, \frac{L_\perp}{N_\perp} \, , \quad \gamma^*_\perp  =   R \, \frac{K_\perp}{N_\perp} \, .  
      \label{psi_1_AND_phi_1_both_radial}
    \end{equation}

    We note here that for cylindrical membranes, shear and bending contributions do not add up linearly in the solution of the flow field, i.e.\@ in a similar way as previously observed between two parallel planar elastic membranes~\cite{daddi16b} or a spherical membrane~\cite{daddi17b, daddi17c} and in contrast to the case of a single planar membrane~\cite{daddi16}.

    %%%%%%%%%%%%%%%%%%%%%%%%%%%%%%%%%%%%%%%%%%%%%%%%%%%%%%%%%%%%%%%%%

    \section{Particle mobility and membrane deformation}\label{mobility}

    The exact results obtained in the previous section allow for the analysis of the effect of the membrane on the axial and radial motion of a colloidal particle, particularly for the calculation of leading-order self- and pair mobility functions~\cite{kim84} relevant to the transport of suspensions in a cylindrical channel. A more accurate description would be achievable by considering a distribution of point forces over the particle surface. Our simpler approximation nevertheless leads to a good agreement with numerical simulations performed with truly extended particles as will be shown below.

    \subsection{Axial mobility}

    We first compute the particle self-mobility correction due to the presence of the membrane for the axisymmetric motion parallel to the cylinder axis. At leading order, the self-mobility correction is calculated by evaluating the axial velocity component of the \emph{reflected} flow field at the Stokeslet position such that
    \begin{equation}
    \Delta \mu_{\parallel}^{\mathrm{S}} = F_z^{-1} \lim_{\R\to 0}  \vImcom_z \, , 
    \end{equation}
    where S appearing as superscript refers to \enquote{self}.
    By making use of Eq.~\eqref{v_z_star_final}, the latter equation can be written as
    \begin{equation}
    \Delta \mu_{\parallel}^{\mathrm{S}} = \frac{1}{4\pi^2\eta} \int_0^{\infty} q \big( \psi^*_\parallel + \varphi^*_\parallel \big) \, \Intd q \,  .
    \end{equation}
    Inserting $\psi^*_\parallel$ and $\varphi^*_\parallel$ from \eqref{psi_1_AND_phi_1_both}, the scaled self-mobility correction reads
    \begin{equation}
    \frac{\Delta \mu_{\parallel}^{\mathrm{S}}}{\mu_0} = \frac{3}{2\pi} \frac{a}{R}  \int_0^\infty \frac{M_\parallel + L_\parallel}{N_\parallel} \, s \, \Intd s  \, , \label{selfMobilityElastic}
    \end{equation}
    where $\mu_0 = 1/(6\pi\eta a)$ is the usual bulk mobility given by the Stokes law. 
    Notably, the correction vanishes for a very wide channel, as $R\to\infty$. 

    Considering a membrane with both shear and bending resistances, and by taking $\alpha$ to infinity, we recover the mobility correction near a hard cylinder with stick boundary conditions, namely
    \begin{equation}
      \lim_{\alpha\to\infty} \frac{\Delta \mu_{\parallel}^{\mathrm{S}}}{\mu_0} = -\frac{3}{2\pi} \frac{a}{R} 
	\int_0^\infty \frac{w_\parallel}{W_\parallel} \, \Intd s 
	\approx -2.10444\, \frac{a}{R} \, ,  \label{vanishingFreqLimShear}
    \end{equation}
    where numerical integration has been performed to obtain the latter estimate, which is in agreement with the result known in the literature~\cite{happel12, faxen22, wakiya53, bohlin60}.
    Moreover, 
    \begin{align}
    w_\parallel &= (I_0 K_1+I_1 K_0)s^2 - 2(I_0 K_0+I_1 K_1) s+4 I_1 K_0 \, , \notag \\
    W_\parallel &= s ( I_1^2 - I_0^2) + 2 I_0 I_1 \, . \notag 
    \end{align}
    The same result is obtained when considering a membrane with only shear rigidity.

    It is worth noting that a bending-only membrane produces a different correction to particle self-mobility when $\alphaB$ is taken to infinity, namely
    \begin{equation}
    \lim_{\alphaB\to\infty} \frac{\Delta \mu_{\parallel, \mathrm{B}}^{\mathrm{S}}}{\mu_0} = -\frac{3}{2\pi} \frac{a}{R} \int_0^\infty 
    \frac{{w_\parallel}_\mathrm{B}}{{W_\parallel}_\mathrm{B}} \, \Intd s \approx -1.80414 \, \frac{a}{R} \, ,   \label{vanishingFreqLimBending}
    \end{equation}
    where 
    \begin{align}
    {w_\parallel}_\mathrm{B} &= s K_0^2 \, , \notag \\ 
    {W_\parallel}_\mathrm{B} &= s(I_1 K_0 - I_0 K_1) + 2I_1 K_1 \, . \notag 
    \end{align}
      Clearly, Eq.~\eqref{vanishingFreqLimBending} does not coincide with the hard cylinder limit predicted by Eq.~\eqref{vanishingFreqLimShear}.
      The reason is the same as discussed in the appendix below Eq.~\eqref{eqn:flowHardBending}, namely that bending only restricts normal but not tangential motion.
    %    This is related to the nonlinear relation between shear and bending contributions to the total mobility.

    %\subsubsection{Particle pair mobility}

    We now turn our attention to hydrodynamic interactions between two particles positioned on the centerline of an elastic cylinder, with the second particle of the same radius~$a$ placed along the cylinder axis at $z=h$.
    For future reference, we shall denote by $\gamma$ the particle located at the origin and by $\lambda$ the particle at $z=h$. The leading order particle pair mobility parallel to the line of centers is readily obtained from the \emph{total} flow field evaluated at the position of the second particle,
    \begin{equation}
    {\muP_{\parallel}} = F_z^{-1} \lim_{\R\to \R_\lambda}  {v_1}_z \, , 
    \end{equation}
    where P appearing as superscript stands for \enquote{pair}.
    The latter equation can be written in a scaled form as
    \begin{equation}
    \frac{\muP_{\parallel}}{\mu_0} = \frac{3}{2} \frac{a}{h} + \frac{3}{2\pi} \frac{a}{R} \int_0^\infty \frac{M_\parallel + L_\parallel}{N_\parallel} \cos \left( \sigma s \right) s \, \Intd s \, , \label{pairMobilityElastic}
    \end{equation}
    where $\sigma:=h/R$.
    Note that $h>2a$ as overlap between the two particles should be avoided.
    The first term in Eq.~\eqref{pairMobilityElastic} is the leading-order bulk contribution to the pair mobility obtained from the Stokeslet solution~\cite{dufresne00, swan07, traenkle16}, 
    whereas the second term is the frequency-dependent correction to the particle pair mobility due to the presence of the elastic membrane.

    Similarly, for an infinite membrane shear modulus, the pair mobility near a hard cylinder limit is obtained,
    \begin{equation}
      \lim_{\alpha\to\infty} \frac{ \mu_{\parallel}^{\mathrm{P}}}{\mu_0} = \frac{3}{2} \frac{a}{h} -\frac{3}{2\pi} \frac{a}{R} 
      \int_0^\infty \frac{w_\parallel}{W_\parallel} \cos \left( \sigma s \right) \, \Intd s \, . \label{vanishingFreqLimShear_PairMobility}
    \end{equation}

    Interestingly, the latter result can also be expressed in terms of convergent infinite series as~\cite{blake79, cui02}
    \begin{equation}
      \lim_{\alpha\to\infty} \frac{\muP_{\parallel}}{\mu_0} = \frac{3}{4} \sum_{n=1}^{\infty} \big( a_n \cos(\beta_n \sigma) + b_n \sin (\beta_n \sigma) \big) e^{-\alpha_n \sigma} \, , \label{vanishingFreqLimShear_PairMobility_SERIES}
    \end{equation}
    where $u_n = \alpha_n + i \beta_n$ are the complex roots of the equation $u (J_0^2 (u_n) + J_1^2(u_n)) = 2 J_0(u_n) J_1 (u_n)$.
    Moreover, $a_n+i b_n = 2 \Big(\pi \big( 2J_1(u_n) Y_0(u_n) - u_n (J_0(u_n) Y_0(u_n) + J_1(u_n) Y_1(u_n)) \big) - u_n \Big)/J_1^2(u_n)$, where $J_\alpha$ and $Y_\alpha$ are the $\alpha$th order Bessel functions of the first and second kind, respectively.
    Although being different in form, our expressions (\ref{vanishingFreqLimShear_PairMobility}) and (\ref{vanishingFreqLimShear_PairMobility_SERIES}) give identical numerical values. 
    The pair mobility therefore has a sharp exponential decay as the interparticle distance becomes larger.
    For $\sigma  \gg 1$, the series in Eq.~\eqref{vanishingFreqLimShear_PairMobility_SERIES} can conveniently be truncated at the first term to give the estimate
    \begin{equation}
    \lim_{\alpha\to\infty} \frac{\muP_{\parallel}}{\mu_0} \simeq \frac{3}{4} \big( a_1 \cos(\beta_1 \sigma) + b_1 \sin (\beta_1 \sigma) \big) e^{-\alpha_1 \sigma} \, , 
    \end{equation}
    where $\alpha_1 \simeq 4.46630$, $\beta_1 \simeq 1.46747$, $a_1 \simeq -0.03698$ and $b_1 \simeq 13.80821$.
    We further mention that the pair mobility function inside a hard cylinder undergoes a sign reversal for $\sigma \gtrsim 2.14206$ before it vanishes as $\sigma$ goes to infinity~\cite{cui02}.

    \subsection{Radial mobility}

    We now compute the particle self-mobility correction caused by the presence of the membrane for the asymmetric motion perpendicular to the cylinder axis.
    At leading order in the ratio $a/R$, the mobility corrections are calculated by evaluating the reflected fluid velocity at the point-force position. Accordingly, 
    %\begin{align}
    %\Delta \muS_{rr} &= (F_x \cos \phi)^{-1} \lim_{\vect{r} \to 0} v_r^* \, ,  \\
    %\Delta \muS_{\phi\phi} &= (-F_x \sin \phi)^{-1} \lim_{\vect{r} \to 0} v_\phi^* \, , 
    %\end{align}
    \begin{align}
    \Delta \muS_{\perp} = F_r^{-1} \lim_{\vect{r} \to 0} v_r^* 
      \equiv F_\phi^{-1} \lim_{\vect{r} \to 0} v_\phi^* \, , 
    \end{align}
    where $F_r = F_x \cos \phi$ and $F_\phi = -F_x \sin \phi$. 
    Upon using Eq.~\eqref{v_r_star_final_radial}, we readily obtain 
    %\SG{Please fix reference to equation}\AI{Done}
    \begin{align}
    \Delta \mu_{\perp}^{\mathrm{S}} = \frac{1}{8\pi^2\eta} \int_0^{\infty} q \big( \psi^*_\perp + \gamma^*_\perp \big) \, \Intd q \,  .
    \end{align}
    %Note that $\Delta \muS_{\phi\phi} = \Delta \mu_{\perp}^{\mathrm{S}}$ for a particle located at the cylinder axis.
    Inserting $\psi^*_\perp$ and $\gamma^*_\perp$ from the general form given by \eqref{psi_1_AND_phi_1_both_radial}, and scaling by the bulk mobility $\mu_0$, we get
    \begin{equation}
    \frac{\Delta \mu_{\perp}^{\mathrm{S}}}{\mu_0} = \frac{3}{4\pi} \frac{a}{R}  \int_0^\infty \frac{M_\perp + K_\perp}{N_\perp} \, s \, \Intd s  \, . \label{selfMobilityElastic_radial}
    \end{equation}
    Similar, by taking $\alpha$ to infinity, we recover the mobility correction near a no-slip cylinder, namely
    \begin{equation}
      \lim_{\alpha\to\infty} \frac{\Delta \mu_{\perp}^{\mathrm{S}}}{\mu_0} = -\frac{3}{4\pi} \frac{a}{R} \int_0^\infty
		\frac{w_\perp}{W_\perp} \, \Intd s 
				      \approx -1.80436 \, \frac{a}{R} \, ,
    \end{equation}
    in full agreement with previous studies~\cite{hasimoto76, felderhof10b},  where we have defined
    \begin{align}
    w_\perp &= I_0(I_0 K_1+I_1K_0)s^3 + \left( (2I_0^2-3I_1^2)K_0-I_0I_1K_1 \right)s^2 % \notag \\
		  - 2I_1(I_0K_0+I_1K_1)s-4K_0 I_1^2 \, , \notag  \\
    W_\perp &= I_0(I_0^2-I_1^2)s^2 + I_1(I_1^2-I_0^2)s-2I_0 I_1^2 \, . \notag 
    \end{align}
    The same steady mobility is obtained when the membrane is endowed with pure shear.

    It is worth to note that for a bending-only membrane, however, the particle self-mobility in the limit when $\alphaB$ is taken to infinity reads
    \begin{equation}
      \lim_{\alphaB\to\infty} \frac{\Delta \mu_{\perp, \mathrm{B}}^{\mathrm{S}}}{\mu_0}  = -\frac{3}{4\pi} \frac{a}{R} \int_0^\infty
		\frac{{w_\perp}_\mathrm{B}}{{W_\perp}_\mathrm{B}} \, \Intd s 
				      \approx -1.55060	 \, \frac{a}{R} \, ,
    \end{equation}
    where we defined
    \begin{align}
    {w_\perp}_\mathrm{B} &= s^2 (s K_1+K_0)^2 \, , \notag \\
    {W_\perp}_\mathrm{B} &= s \left( (s^2+3)K_1+2sK_0 \right) I_0 - (s^2+3) (sK_0+2K_1) I_1 \, .  \notag 
    \end{align}

    Continuing, the particle pair mobility function is determined by evaluating the total velocity field at the nearby particle position leading to
    \begin{equation}
    {\muP_{\perp}} = F_r^{-1} \lim_{\R\to \R_\lambda}  {v_1}_r 
    \equiv F_\phi^{-1} \lim_{\R\to \R_\lambda}  {v_1}_\phi \, . \label{muP_rr}
    \end{equation}
    Eq.~\eqref{muP_rr} can be written in a scaled form as
    \begin{equation}
    \frac{\muP_{\perp}}{\mu_0} = \frac{3}{4} \frac{a}{h} + \frac{3}{4\pi} \frac{a}{R}  \int_0^\infty \frac{M_\perp + K_\perp}{N_\perp} \, \cos \left( \sigma s \right) \, s \, \Intd s \, . \label{pairMobilityElastic_radial}
    \end{equation}
    Similar, for an infinite membrane shear modulus, we recover the pair mobility near a hard cylinder,
    \begin{equation}
      \lim_{\alpha\to\infty} \frac{ \mu_{\perp}^{\mathrm{P}}}{\mu_0} = \frac{3}{4} \frac{a}{h} -\frac{3}{4\pi} \frac{a}{R} 
      \int_0^\infty \frac{w_\perp}{W_\perp} \cos \left( \sigma s \right) \, \Intd s \, . \label{vanishingFreqLimShear_PairMobility_radial}
    \end{equation}
    % \section{Membrane deformation}

% SG: no, it is not diagonal just as the single particle is not. The mobility is a scalar quantity specific for each direction. The concept of a mobility tensor does not work out here, as argued correctly by the referee.
%    \new{
%    It is worth mentioning that for two particles located on the centerline of an elastic tube the pair-mobility is a diagonal tensor. Therefore, the pair interaction is characterized by the parallel and perpendicular components that have been computed here. The same behavior is observed for motion in bulk fluid or inside a no-slip cylinder.
%    }

    % \subsection{Axial motion}

    \subsection{Startup motion}

    Here we will derive the mobility coefficients for a particle starting from rest and then moving under a constant external force (e.g.~gravity) exerted along or perpendicular to the cylinder axis.
    Mathematically, such force can be described by a Heaviside step function force $\vect{F}(t) = \vect{A} \, \theta(t)$ whose Fourier transform in the frequency domain reads~\cite{bracewell99}
    \begin{equation}
    \vect{F} (\omega) = \left( \pi \delta(\omega) - \frac{i}{\omega} \right) \vect{A} \, .
    \end{equation}
    Applying back Fourier transform, the time-dependent correction to the particle mobility for a startup motion reads
    \begin{equation}
    \Delta \mu(t) = \frac{\Delta \mu(0)}{2} + \frac{1}{2i\pi} \int_{-\infty}^{+\infty} \frac{\Delta \mu(\omega)}{\omega} \, e^{i\omega t} \, \Intd \omega \, . \label{mobilitySteadyForce}
    \end{equation}

    The second term in Eq.~\eqref{mobilitySteadyForce} is a real valued quantity which takes values between $-\Delta \mu(0)/2$ when $t\to 0$ and $+\Delta\mu(0)/2$ as $t \to \infty$. 
    Since the frequency-dependent mobility corrections are expressed as a Fourier-Bessel integral over the scaled wavenumber $s$, the computation of the time-dependent mobility requires a double integration procedure.
    For this purpose, we use the Cuba Divonne algorithm~\cite{hahn05, hahn16} for an accurate and fast numerical computation. 

    \subsection{Membrane deformation}

    Finally, our results can be used to compute the membrane deformation resulting from a time-dependent point-force acting along or perpendicular to the cylinder axis. 
    The membrane displacement field is readily obtained from the velocity at $r=R$ via the no-slip boundary condition stated by Eq.~\eqref{no-slip-relation}. We define the membrane frequency-dependent reaction tensor (again in an approximate sense if the direction is not perfectly radial or axial) as~\cite{bickel07}
    \begin{equation}
    u_\alpha (\phi,z, \omega) = R_{\alpha\beta} (\phi,z, \omega) F_\beta (\omega)\, , \label{VerformungVonDerKraft}
    \end{equation}
    bridging between the membrane displacement field and the force acting on the nearby particle.
    Restricting to a harmonic driving force $F_\alpha (t)=A_\alpha e^{i\omega_0 t}$, the membrane deformation in the temporal domain is calculated as 
    \begin{equation}
    u_\alpha (\phi,z,t) = R_{\alpha\beta} (\phi,z,\omega_0) A_\beta e^{i\omega_0 t} \, .
    \end{equation}
    Further, the physical displacement is obtained by taking the real part of the latter equation. 
    The radial-axial and axial-axial components of the reaction tensor are then computed from Eq.~\eqref{flowOut} as
    \begin{align}
    R_{rz}  &=  \Lambda \int_0^\infty s \Big(  \big( s K_0 + K_1  \big) {\varphi_2}_\parallel
		      -K_1 {\psi_2}_\parallel \Big) \sin \left( \frac{sz}{R} \right) \Intd s \, ,  \notag  \\
    R_{zz}  &=  \Lambda \int_0^\infty s \Big(  \big( K_0  - s K_1 \big) {\varphi_2}_\parallel 
		      + K_0 {\psi_2}_\parallel \Big) \cos \left( \frac{sz}{R} \right) \Intd s \, , \notag
    \end{align}
    with $\Lambda := 1/( {4} i\pi^2\eta\omega R^2)$, which give access to the radial and axial displacements after making use of Eq.~\eqref{VerformungVonDerKraft}.
    Moreover, $R_{\phi z} = 0$ due to axial symmetry.

    % \subsection{Radial motion} 
    For a point force directed perpendicular to the cylinder axis, the components of the reaction tensor can readily be computed from Eqs.~\eqref{flowOut_radial} to obtain
    % With the radial and tangential directions taken in the local coordinate frame of the observation point, we obtain
    \begin{subequations} 
    \begin{align}
    R_{rr}  &= \Lambda  \int_0^\infty  \big( \left( (2+s^2)K_1+s K_0 \right) {\varphi_2}_\perp % \notag \\
	    - \left( s K_0 + K_1 \right) {\psi_2}_\perp + K_1 {\gamma_2}_\perp \big) \cos \left( \frac{sz}{R} \right)  \Intd s \, , \notag \\
    R_{\phi \phi}  &= -\Lambda  \int_0^\infty \big( \left( s K_0+2K_1 \right) {\varphi_2}_\perp % \notag \\
	      - K_1 {\psi_2}_\perp + \left( s K_0 + K_1 \right) {\gamma_2}_\perp \big) \cos \left( \frac{sz}{R} \right)  \Intd s \, , \notag  \\
    R_{zr}  &= \Lambda \int_0^\infty s \left( s K_0 {\varphi_2}_\perp - K_1 {\psi_2}_\perp \right) \sin \left( \frac{sz}{R} \right)  \Intd s \, . \notag
    \end{align}
    \end{subequations}
    Additionally, we have $R_{r\phi} = R_{\phi r} = R_{z\phi} = 0$.

    \section{Comparison with Boundary Integral simulations}\label{comparison}

    \begin{figure}
    \begin{center}
       \scalebox{0.78}{\input{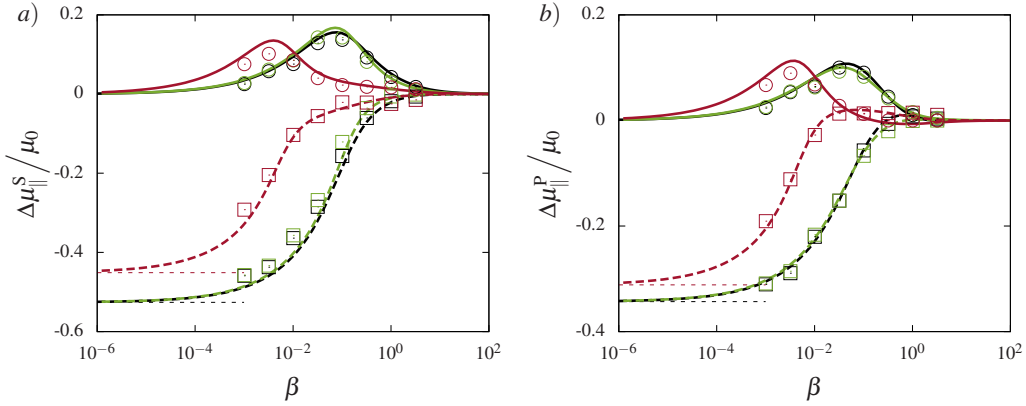}} 
    \caption{(Color online) $a)$ The axial component of the scaled frequency-dependent self-mobility correction versus the scaled frequency $\beta=1/\alpha$ nearby a cylindrical membrane endowed with only-shear (green or bright gray in a black and white printout), only-bending (red or dark gray in a black and white printout) and both rigidities (black).
    The particle is set on the centerline of an elastic cylinder of radius $R=4a$. Here we take a reduced bending modulus $\EB = 1/6$. The theoretical predictions are presented as dashed and solid lines for the real and imaginary parts, respectively. Boundary integral results are shown as squares for the real part and circles for the imaginary part. The horizontal dashed lines are the vanishing frequency limits given by Eqs.~\eqref{vanishingFreqLimShear} and \eqref{vanishingFreqLimBending}.
    $b)$ The parallel component of the scaled frequency-dependent pair mobility correction versus the scaled frequency $\beta$. The two particles are set a distance $h=R$ apart on the centerline of an elastic cylinder of radius $R=4a$. 
    } 
    \label{cylindricoSelfMobi}
    \end{center}
    \end{figure}

    The accuracy of the point-particle approximation employed throughout this work can be assessed by direct comparison with fully resolved numerical simulations. 
    To this end, we employ a completed double layer boundary integral method~\cite{phan93, kohr04, zhao11, zhao12} which has proven to be perfectly suited for simulating solid particles in the presence of deforming boundaries.
    Technical details concerning the algorithm and its numerical implementation have been reported by some of us elsewhere, e.g.~\onlinecite{daddi16b} and~\onlinecite{guckenberger16}. 
    The cylindrical membrane has a length of $200a$, meshed uniformly with 6550 triangles, and the spherical particle is meshed with 320 triangles obtained by consecutively refining an icosahedron.
    \new{We show in the Supporting Information that using finer or coarser meshes does not influence the results significantly  [URL will be inserted by the editor].}

    In order to determine the particle self- and pair mobilities numerically, a harmonic force ${F_\lambda}_\alpha (t) = {A_\lambda}_\alpha e^{i\omega_0 t}$ of amplitude ${A_\lambda}_\alpha$ and frequency $\omega_0$ is applied along the direction $\alpha$ at the surface of the particle labeled~$\lambda$. The force is directed \new{along the cylinder ($z$~direction), perpendicular to the axis ($x$~direction) or at angle $\theta$ to the axis.}
    After a brief transient evolution, both particles oscillate at the same frequency with different phases, i.e. ${V_{\lambda\alpha}} = {B_{\lambda\alpha}} e^{i\omega_0 t + \delta_\lambda}$ and ${V_{\gamma\alpha}} = {B_{\gamma\alpha}} e^{i\omega_0 t + \delta_\gamma}$.
    For the accurate determination of the velocity amplitudes and phase shifts, we use a nonlinear least-squares algorithm~\cite{marquardt63} based on the trust region method~\cite{conn00}.
    The particle self- and pair mobility functions can therefore be computed as
    \begin{equation}
    \mu_{\alpha\beta}^{\mathrm{S}} = \frac{{B_\lambda}_\alpha}{{A_\lambda}_\beta} \, e^{i\delta_\lambda} \, , \quad\quad  
    \mu_{\alpha\beta}^{\mathrm{P}} = \frac{{B_\gamma}_\alpha}{{A_\lambda}_\beta} \, e^{i\delta_\gamma} \, .
    \end{equation}

     \begin{figure}
    \begin{center}
      \scalebox{0.78}{\input{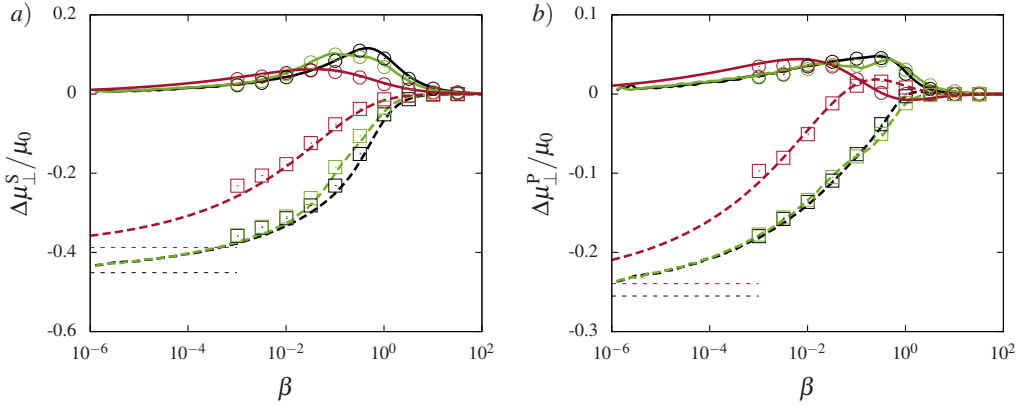}}
    \caption{(Color online) The radial component of the scaled frequency-dependent self $a)$ and pair $b)$ mobility corrections versus the scaled frequency $\beta$. The color code is the same as in figure~\ref{cylindricoSelfMobi}.
    }
    \label{cylindricoRad_TT_XX}
    \end{center}
    \end{figure}

    We now define the characteristic frequency for shear, $\beta := 1/\alpha = 3\eta \omega R/(2\kS)$, and for bending, $\betaB:=1/\alphaB^3 = \eta\omega R^3/\kB$. We also introduce the membrane reduced bending modulus as $\EB:=\kB/(\kS R^2)$ quantifying the nonlinear coupling between shear and bending~\cite{pozrikidis01jfm}.

    In figure~\ref{cylindricoSelfMobi}~$a)$, we show the correction to particle self-mobility versus the scaled frequency $\beta$ as predicted theoretically by Eq.~\eqref{selfMobilityElastic}.
    The particle is set on the centerline of an elastic cylinder of radius $R=4a$.
    For the simulation parameters, we take a reduced bending $\EB=1/6$ for which $\beta$ and $\betaB$ have about the same magnitude, and thus shear and bending manifest themselves equally.
    We observe that the real part is a monotonically increasing function of frequency whereas the imaginary part exhibits a bell-shape.
    \new{This form corresponds to the Debye shape often observed for complex linear response functions in systems with memory (the mathematical form in the present system is however much more complex than a simple Debye equation).}
    For small forcing frequencies, the real part of the mobility correction approaches that near a no-slip hard cylinder only if the membrane possesses resistance against shear.
    For large forcing frequencies, both the real and imaginary parts vanish, which corresponds to the bulk behavior.   
    It can clearly be seen that the mobility correction is primarily determined by shear resistance and bending does not play a significant role, similarly to what has been recently observed for spherical elastic membranes~\cite{daddi17b, daddi17c}.
    A good quantitative agreement is obtained between analytical predictions and numerical simulations over the whole range of applied frequencies.

    Analogous predictions for the pair mobility versus the scaled frequency $\beta$ are shown in figure~\ref{cylindricoSelfMobi}~$b)$. The two particles are set a distance $h=R$ apart along the axis of an elastic cylinder of radius $R=4a$. The overall shapes resemble those observed for the self-mobility, where again the effect of shear is more pronounced. 
    However, it can be seen that the real part for a bending-only membrane may undergo a change of sign at some intermediate frequencies in the same way as observed nearby planar membranes~\cite{daddi16c}. 
    Interestingly, we find that the correction to the pair mobility induced by the elastic membrane is almost as large as the bulk pair mobility itself.

    The frequency-dependent self- and pair mobility corrections for the motion perpendicular to the cylinder axis are shown in figure~\ref{cylindricoRad_TT_XX}.
    We observe that the total mobility corrections are primarily determined by membrane shear resistance as it has been observed for the axial motion along the cylinder axis.
    \new{
    This is somewhat surprising as for radial motion the particle ''pushes'' against the membrane and one may thus expect bending resistance to be more important than shear resistance. Indeed, for planar membranes \cite{daddi16} this is the case. 
    The surprisingly strong influence of shear resistance in the present system can thus be attributed to the cylindrical geometry.}
    Notably, the correction near a rigid cylinder is recovered only if the membrane possesses a finite resistance towards shear.
    
\begin{figure}
    \begin{center}
      \scalebox{0.78}{\input{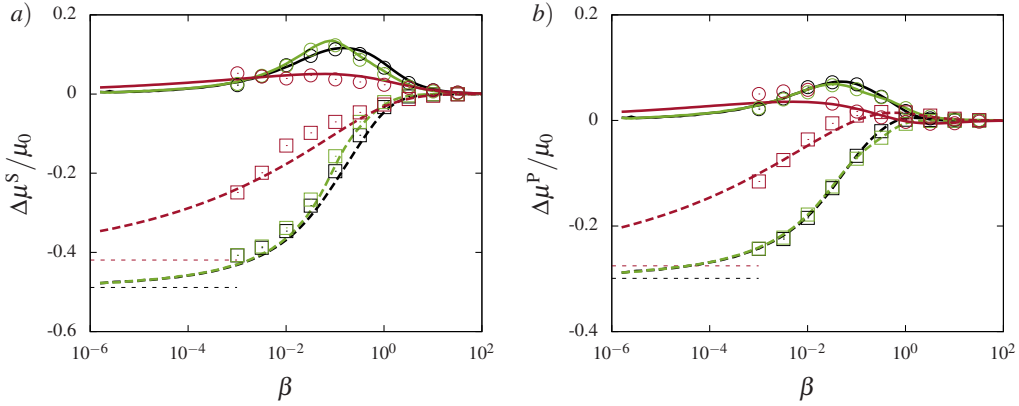}}
    \caption{(Color online) The scaled frequency-dependent self $a)$ and pair $b)$ mobility corrections versus the scaled frequency $\beta$ for a force acting at an angle $\theta=\pi/4$ with respect to the cylinder axis. The color code is the same as in figure~\ref{cylindricoSelfMobi}.
    }
    \label{cylindrico_oblik_TT_XX}
    \end{center}
    \end{figure}

    \new{
    Next, we address the general motion problem by considering an oblique force~$\F$ acting at an angle~$\theta$ relative to the cylinder axis.
    For a hard cylinder, this problem can straightforwardly be solved by decomposing the force into axial and radial components and linearly superposing axial and radial mobilities with the same relative weights.
    Due to the nature of the boundary conditions, for a deformable elastic boundary as in the present problem, this procedure is not strictly possible (the weighted linear sum of the axial and radial flow fields cannot be shown to satisfy the boundary conditions for oblique motion).
    Nevertheless, we will show by comparing with boundary-integral simulations below that this simple approach allows a surprisingly good approximation.
    We thus decompose the force into an axial component $F_z = F \cos\theta$ along the cylinder axis together with a radial component $F_r = F \sin\theta$.
    The particle velocity along the oblique direction $V=V_z \cos\theta + V_r \sin\theta$ with $V_z = \mu_\parallel^\mathrm{S} F \cos\theta$ and $V_r = \mu_\perp^\mathrm{S} F \sin\theta$.
    Accordingly, the particle self- and pair-mobility functions along the oblique direction can be estimated as 
    %\SG{why $\cos^2$?}\AI{It follows from the transformation equations. Please refer to Eqs.~(B37) and (B38) of our first pre paper.}
    %\SG{Please give one or two intermediate steps how to do this transformation.}
    \begin{equation}
     \mu^{\mathrm{S}, \mathrm{P}} = \mu_\parallel^{\mathrm{S}, \mathrm{P}} \cos^2\theta + \mu_\perp^{\mathrm{S}, \mathrm{P}} \sin^2\theta \, .
    \end{equation}    
    In figure~\ref{cylindrico_oblik_TT_XX}, we present the mobility corrections versus $\beta$ due to an oblique force acting at an angle $\theta= \pi/4$ with respect to the cylinder axis with $\phi=0$.
    In this particular situation, the particle mobility is the arithmetic mean of the mobilities parallel and perpendicular to the cylinder axis.
    We observe that the analytical predictions compare favourably with boundary integral simulations, where again the particle mobility is primly determined by membrane shearing resistance.
    Consequently, the present theory gives a good approximation of the mobilities associated with a general motion containing both radial and axial directions.}

    \begin{figure}
    \begin{center}
      \scalebox{0.78}{\input{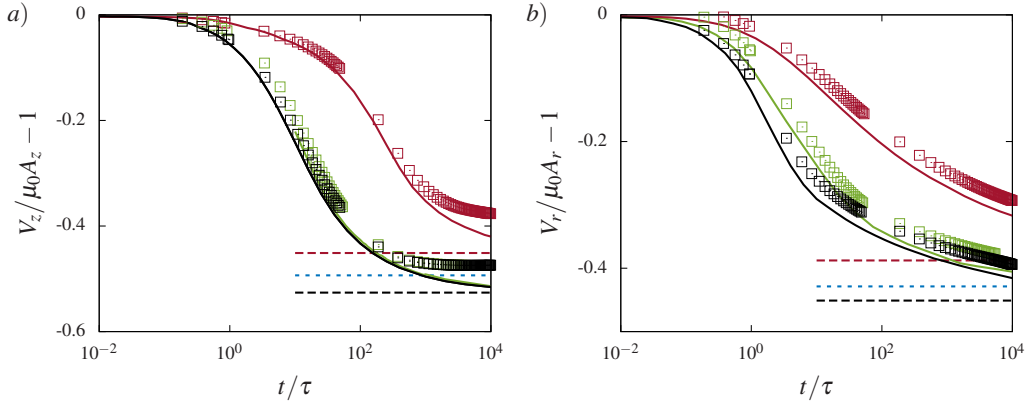}}
    \caption{Translational velocity of a particle starting from rest for $a)$ axial and $b)$ radial motion under the action of a constant external force, obtained using the same parameters as in figure~\ref{cylindricoSelfMobi} for a membrane with pure shear (green or bright gray in a black and white printout), pure bending (red or dark gray in a black and white printout) and both rigidities (black). Solid lines are the analytical predictions obtained from Eq.~\eqref{mobilitySteadyForce} and symbols are the boundary integral simulations. 
    Horizontal dashed lines are our theoretical predictions in the steady limit based on the point-particle approximation and the blue dotted lines are the higher order corrections given by Eqs.~\eqref{upTo3rd_order_Axial} and \eqref{upTo3rd_order_Radial} for the axial and radial motions, respectively. Here $\tau$ is a characteristic time scale defined as $\tau := \beta/\omega$. 
    }  
    \label{steadyMotion_Force}
    \end{center}
    \end{figure}

    In figure~\ref{steadyMotion_Force}, we show the time-dependent translational velocity of a particle starting from rest and subsequently moving under the action of a constant axial or radial force nearby a cylindrical membrane endowed with shear-only (green), bending-only (red) or both shear and bending resistances (black).
    The time is scaled by the characteristic time scale for shear $\tau := \beta/\omega = 3\eta R/(2\kS)$.
    At short time scales, we observe that the mobility correction amounts to a small value since the particle does not yet feel the presence of the elastic membrane.
    As the time increases, the membrane effect becomes more noticeable and the mobility curves bend down substantially to asymptotically approach the correction nearby a hard cylinder if the membrane possesses a non-vanishing resistance towards shear. Moreover, we observe that the steady state is more quickly achieved for the axial (parallel) motion than for the radial motion (perpendicular), i.e. in a way similar to what has been observed nearby planar elastic membranes~\cite{daddi16}. 
    At the end of the simulations, the particle position changes only by about 10~\% of its radius.

    Before continuing, we briefly comment on the importance of higher order terms. For this, we consider a hard cylinder for which the correction to the axial mobility can be obtained from Bohlin inverse series coefficients as~\cite[Tab.~2.1]{zimmerman04}
    \begin{equation}
      \lim_{\alpha\to\infty} \frac{\Delta \mu_{\parallel}^\mathrm{S}}{\mu_0} = -2.104443 \left( \frac{a}{R} \right) + 2.086694 \left( \frac{a}{R} \right)^3 + \cdots \, ,\label{upTo3rd_order_Axial}
    \end{equation}
    which has been truncated at the 3rd order here since higher order terms amount to an insignificant correction for $a\ll R$.
    For the radial motion, this reads 
    \begin{equation}
      \lim_{\alpha\to\infty} \frac{\Delta \mu_{\perp}^\mathrm{S}}{\mu_0} = -1.804360 \left( \frac{a}{R} \right) + 1.430590 \left( \frac{a}{R} \right)^3 + \cdots \, . \label{upTo3rd_order_Radial}
    \end{equation}

    Comparing the first and third order in the above equations for the present parameters, we find that the higher order terms lead to a correction of about 5~\%.

    \begin{figure}
    \begin{center}
      \scalebox{0.78}{\input{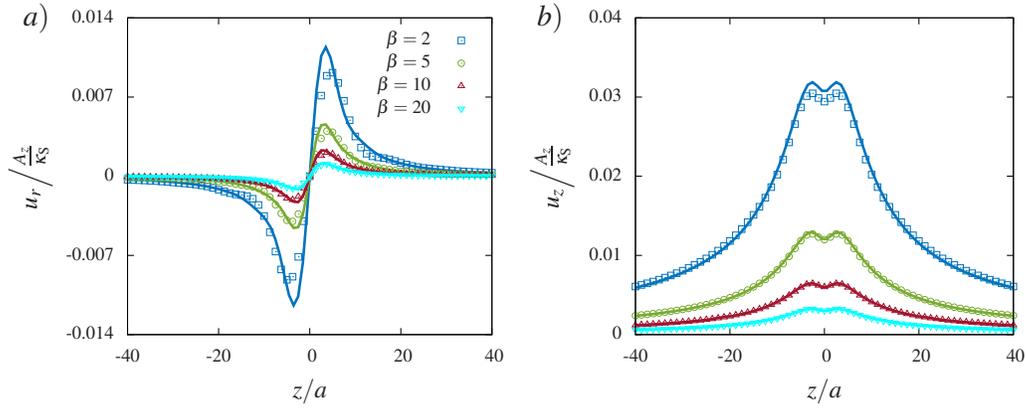}}
    \caption{(Color online) The scaled radial $a)$ and axial $b)$ membrane displacements versus $z/a$ at four different forcing frequencies calculated at quarter period, i.e. when $\omega_0 t=\pi/2$ and the particle reaches its maximal amplitude moving to the right along the $z$ axis. Solid lines refer to theoretical predictions and symbols are the boundary integral simulations.} 
    \label{deformation}
    \end{center}
    \end{figure}

    The membrane displacements induced by axial motion of the particle are illustrated in figure~\ref{deformation}, which includes the  theoretical predictions (solid lines) and boundary integral simulations (symbols) for four different forcing frequencies. The natural scale for the displacement, $A_z/\kS$ is set by the amplitude of forcing $A_z$ and the shear resistance $\kS$. Here we use the same parameters as in figure~\ref{cylindricoSelfMobi} for a membrane with both shear and bending rigidities. We plot the axial and radial displacement of the axial section (along~$z$) of the tube wall in the moment in which a particle moving harmonically with a very small amplitude reaches its maximal axial position. We observe that the radial displacement~$u_r$ is an odd function of~$z$ that vanishes at the origin and at infinity. The axial deformation $u_z$ shows a fundamentally different evolution with respect to~$z$, where the membrane is displaced along the direction of the force.
    Moreover, the maximum deformation reached in~$u_z$ is found to be about three times larger than that reached in~$u_r$. 
    Interestingly, the maximum in~$u_z$ is not attained at the particle position $z=0$, but slightly besides.
    By comparing the membrane deformation at various forcing frequencies, it can be seen that larger frequencies induce smaller deformations as the elastic membrane does not have enough time to react to the rapidly wiggling particle.

    \begin{figure}
    \begin{center}
      \scalebox{0.8}{\input{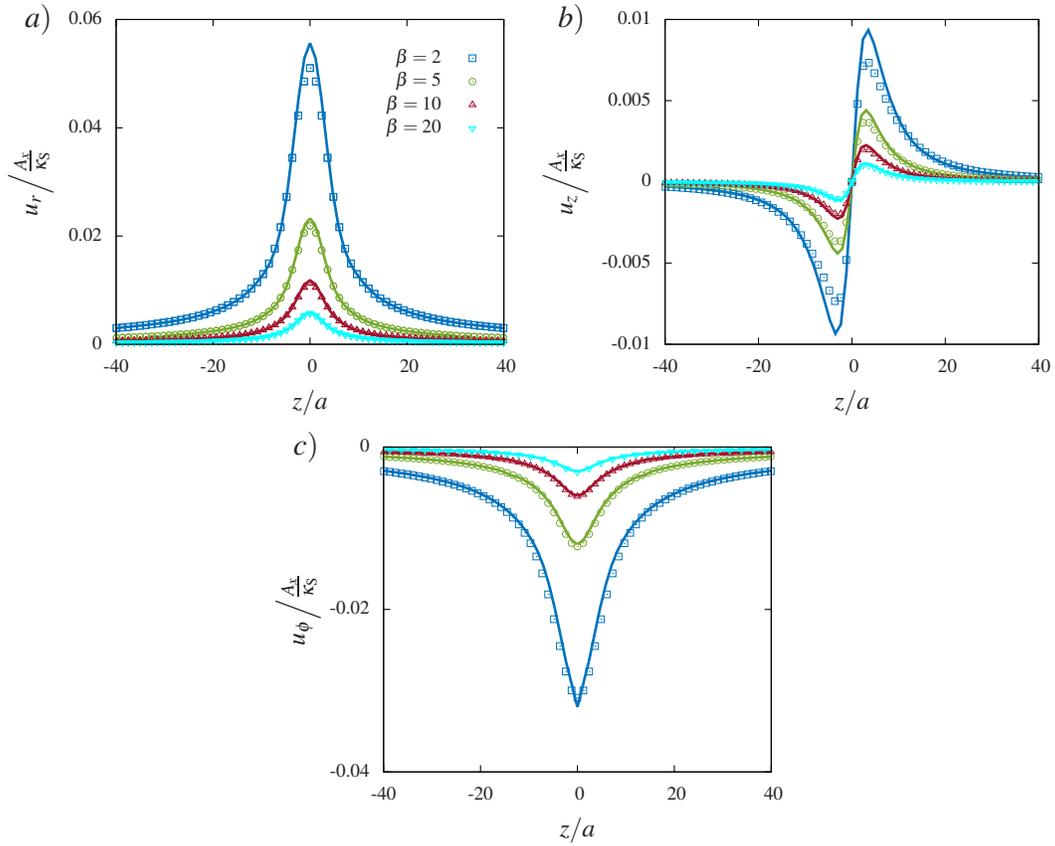}}
    \caption{(Color online) The scaled radial $a)$, azimuthal $b)$ and axial $c)$ membrane displacements versus $z/a$ at four forcing frequencies calculated at quarter period for $\omega_0 t=\pi/2$ when the particle reaches its maximal radial position. 
    Here deformations are shown in the plane of maximum deformation.
    Solid lines refer to theoretical predictions determined and symbols are the boundary integral simulations.} 
    \label{deformation_Radial}
    \end{center}
    \end{figure}

    In figure~\ref{deformation_Radial}, we show the scaled radial, axial and azimuthal displacement fields induced by the particle radial motion upon varying the forcing frequency.
    Deformations are plotted  when the oscillating particle reaches its maximal amplitude, in the plane of maximum deformation, i.e.\@ $\phi=0$ (or $y=0$) for $u_r$ and $u_z$, and $\phi=\pi/2$ (or $x=0$) for $u_\phi$ for a force directed along the $x$~direction.
    Not surprisingly, we observe that the membrane mainly undergoes radial deformation.
    The latter is found to be about twice as large as the azimuthal deformation and even six times larger that axial deformation.
    The numerical simulations are found to be in a very good agreement with analytical predictions, over the whole length of the deformed cylinder.

    For typical situations, the order of magnitude of the forces exerted by optical tweezers on suspended particles are of the order of 1~pN~\cite{cipparrone10}.
    For a cylinder radius of $10^{-6}~$m, a shear modulus of about $ 10^{-6}~$N/m  and a scaled forcing frequency $\beta = 2$, the membrane undergoes a maximal deformation of about 2~\% and 5~\% of its undeformed radius for the axial and radial motions, respectively.
    As a result, membrane deformations are generally small and deviations from cylindrical shape are indeed negligible.

%\SG{We have defined Re and St using the velocity amplitude. Which amplitude did you use here? I would remove this part, I think it is sufficient to say as we did above that the amplitude can be made arbitrarily small.}
%    As a final remark, we shall show that the range of frequencies employed throughout this work is consistent with the assumption of small Reynolds and Strouhal numbers.
%    In fact, by taking a fluid density $\rho = 10^3$~kg/m$^3$, a shear viscosity $\eta = 1.2 \times 10^{-3}$~Pas and a membrane bending modulus $\kB = 2 \times 10^{-19}$ as typical values~\cite{Freund_2014}, the condition $\operatorname{Re}  \operatorname{St} \ll 1 $ leads to
%    \begin{equation}
%    \beta \ll \frac{3 \eta^2}{2 \rho R \kS} \approx 2200 \, ,  \qquad  \betaB \ll \frac{R \eta^2}{\rho \kB} \approx 7200 \, .
%    \end{equation}
%    Clearly, both scaled frequencies satisfy these conditions in the frequency range considered in the present work.

    \section{Conclusions}\label{conclusion}

    In this paper, we derived explicit analytic expressions for the Green's functions, i.e., the flow field generated by a point particle (Stokeslet), acting either axially along or perpendicular to the centerline of an elastic cylindrical tube which exhibits resistance towards shear and bending. For this, we first derived the appropriate boundary conditions determining the surface traction jump across the membrane and then used a Fourier integral expansion to solve the Stokes equations.
    By examining the influence of shear and bending motion, we determined the full form of the solutions and discussed their behavior for the whole range of actuation frequencies for arbitrary elastic parameters of the membrane -- the bending rigidity $\kB$ and elastic modulus $\kS$. 

    The solution was then used to compute the leading order correction to the self- and pair mobility of particles moving axially or radially in the elastic tube, which are in good agreement with fully resolved boundary integral simulations performed for the particle radius being a quarter of the channel size. 
    \new{Additionally, we showed that our analytical theory can approximately be applied for the general motion due to an oblique force acting along both the radial and axial directions.}
    We also computed the deformation field of the membrane for an arbitrary time-dependent forcing and compared it with fully resolved numerical simulations. 

    The theoretical results prove that in this case the coupling between the effects of bending and shear of the membrane has a nonlinear nature, and the limit of a rigid tube is recovered only for non-zero shear resistance. 
    We further showed that the effects of shear are far more important for both axial and radial motions than bending and therefore determine the qualitative behavior of the elastically confined particle. 
    For two hydrodynamically interacting particles, the correction to pair mobility is found to be of the same order as the bulk pair mobility itself.
    % thus hinting at a possibly significant influence on particle agglomeration processes near elastic interfaces. 
    %\SG{Do you see attractive interactions for some parameters? Are the interactions for the hard cylinder attractive or repulsive or both? One referee asked for this.}
    %\AI{For two particles on the centerline of a cylinder, there are no off-diagonal components of the pair-mobility tensor. This is true for a hard-cylinder as well}

    \begin{acknowledgments}
    %\section*{Acknowledgments}
    ADMI and SG thank the Volkswagen Foundation for financial support and acknowledge the Gauss Center for Supercomputing e.V. for providing computing time on the GCS Supercomputer SuperMUC at Leibniz Supercomputing Center. 
    This work has been supported by the Ministry of Science and Higher Education of Poland via the Mobility Plus Fellowship awarded to ML. ML acknowledges funding from the Foundation for Polish Science within the START programme. 
    This article is based upon work from COST Action MP1305, supported by COST (European Cooperation in Science and Technology).
    We gratefully acknowledge support from the Elite Study Program Biological Physics.
    \end{acknowledgments}

    \appendix

    \section{Membrane mechanics} \label{appendix:membraneMech}

    In this appendix, we derive equations in cylindrical coordinates for the traction jump across a membrane endowed with shear and bending rigidities.
    We denote by $\vect{a} = R \eR + z \eZ$ the position vector of the points located at the undisplaced membrane, with $R$ being the undeformed membrane radius.
    Here $r$, $\phi$ and $z$ are used to refer to the radial, azimuthal and axial coordinates, respectively.
    After deformation, the vector position reads
    \begin{equation}
    \vect{r} = (R + u_r)\eR + u_\phi \ePhi + (z + u_z) \eZ \, ,
    \end{equation}
    where $\vect{u}$ denotes the displacement vector field.
    Hereafter, we shall use capital roman letters for the undeformed state and small roman letters for the deformed.
    The cylindrical membrane can be defined by the covariant base vectors $\gOne := \vect{r}_{,\phi}$ and $\gTwo := \vect{r}_{,z}$.
    % where commas in indices stands for a partial spatial derivative.
    The unit normal vector $\vect{n}$ is defined as
    \begin{equation}
    \vect{n} = \frac{\gOne \times \gTwo}{|\gOne \times \gTwo|}  \, .
    \end{equation}

    Hence, the covariant base vectors read
    \begin{align}
    \gOne &=  (u_{r,\phi} - u_\phi)\eR + (R + u_r + u_{\phi, \phi}) \ePhi  + u_{z,\phi} \eZ \, , \\
    \gTwo &=  u_{r,z} \eR + u_{\phi, z} \ePhi + (1 + u_{z,z} ) \eZ \, ,
    \end{align}
    and the unit normal vector at leading order in deformation reads
    \begin{equation}
    \vect{n} = \eR + \frac{u_\phi - u_{r, \phi}}{R} \, \ePhi - u_{r,z} \eZ \, . 
    \end{equation}

    Note that $\gOne$ has a length dimension while $\gTwo$ and $\vect{n}$ are dimensionless.
    The covariant components of the metric tensor are defined by the scalar product $g_{\alpha\beta} = \vect{g}_{\alpha} \cdot \vect{g}_{\beta}$.
    The contravariant tensor $g^{\alpha\beta}$ is the inverse of the metric tensor.
    In a linearized form, we obtain
    \begin{align}
    g_{\alpha\beta} &= \left(
		      \begin{array}{cc}
		      R^2 + 2R (u_r + u_{\phi, \phi}) & u_{z,\phi} + R u_{\phi, z} \\
		      u_{z,\phi} + R u_{\phi, z}  & 1+2u_{z,z} 
		      \end{array}
		      \right) \, ,
		      \\
    g^{\alpha\beta} &= \left(
		      \begin{array}{cc}
		      \frac{1}{R^2} - 2\frac{u_r + u_{\phi, \phi}}{R^3} & -\frac{u_{z,\phi} + R u_{\phi, z}}{R^2} \\
		      -\frac{u_{z,\phi} + R u_{\phi, z}}{R^2}  & 1-2u_{z,z} 
		      \end{array}
		      \right) \, .
			\label{cocontravariantTensor}
    \end{align}

    The covariant and contravariant tensors in the undeformed state $G_{\alpha\beta}$ and $G^{\alpha\beta}$ can immediately be obtained by considering a vanishing displacement field in Eq.~\eqref{cocontravariantTensor}.

    \subsection{Shear}

    In the following, we shall derive the traction jump equations across a cylindrical membrane endowed by an in-plane shear resistance.
    The two transformation invariants are given by Green and Adkins as~\cite{green60, zhu14, daddi-thesis}
    \begin{subequations}
    \begin{align}
      I_1 &= G^{\alpha\beta}  g_{\alpha\beta} - 2 \, , \\
      I_2 &= \det G^{\alpha\beta}  \det g_{\alpha\beta} - 1 \, .
    \end{align}
    \end{subequations}

    From the membrane constitutive relation, the contravariant components of the stress tensor $\tau^{\alpha\beta}$ can readily be obtained such that~\cite{lac04, zhu15}
    \begin{equation}
    \tau^{\alpha\beta} = \frac{2}{\JS} \frac{\partial W}{\partial I_1} \, G^{\alpha\beta} + 2\JS \frac{\partial W}{\partial I_2} \,  g^{\alpha\beta} \, ,
    \label{stressTensor}
    \end{equation}
    wherein $W$ is the areal strain energy functional and $\JS := \sqrt{1+I_2}$ is the Jacobian determinant.
    In the linear theory of elasticity, $\JS \simeq 1 + e$, where $e := (u_r+u_{\phi,\phi})/R + u_{z,z}$ being the dilatation function~\cite{sadd09}.
    In the present paper, we use the neo-Hookean model to describe the elastic properties of the membrane, whose areal strain energy reads~\cite{krueger12, lac05}
    \begin{equation}
    W(I_1, I_2) = \frac{\kS}{6} \left( I_1 - 1 + \frac{1}{1+I_2} \right) \, .
    \label{skalakEquation}
    \end{equation}

    By plugging Eq.~\eqref{skalakEquation} into Eq.~\eqref{stressTensor}, the linearized in-plane stress tensor reads
    \begin{equation}
    \tau^{\alpha\beta} = \frac{2 \kS}{3}
    \left(
    \begin{array}{cc}
      \frac{u_r + u_{\phi, \phi}}{R^3} + \frac{e}{R^2} & \frac{1}{2R} \left( u_{\phi, z} + \frac{u_{z, \phi}}{R} \right) \\
      \frac{1}{2R} \left( u_{\phi, z} + \frac{u_{z, \phi}}{R} \right) & u_{z,z} + e 
    \end{array}
    \right) \, .
    \end{equation}

    The equilibrium equations balancing the membrane elastic and external forces read
    \begin{subequations}
    \begin{align}
      \nabla_{\alpha} \tau^{\alpha\beta} + \Delta f^{\beta} &= 0 \, , \label{Equilibrium_Tangential} \\
      \tau^{\alpha\beta} b_{\alpha\beta} + \Delta f^{n} &= 0 \, , \label{Equilibrium_Normal}
    \end{align}
    \label{Equilibrium}
    \end{subequations}
    where $\Delta \vect{f} = \Delta f^{\beta} \vect{g}_{\beta} + \Delta f^{n} \vect{n} $ is the traction jump vector across the membrane.
    Here $\nabla_{\alpha}$ stands for the covariant derivative, which for a second-rank tensor is defined as~\cite{deserno15}
    \begin{equation}
    \nabla_{\alpha} \tau^{\alpha\beta} = \tau^{\alpha\beta}_{,\alpha} + \Gamma_{\alpha\eta}^{\alpha} \tau^{\eta\beta} + \Gamma_{\alpha\eta}^{\beta} \tau^{\alpha\eta} \, , \label{covariantDerivativeDef}
    \end{equation}
    with $\Gamma_{\alpha\beta}^{\lambda}$ being the Christoffel symbols of the second kind which read~\cite{synge69} 
    \begin{equation}
    \Gamma_{\alpha\beta}^{\lambda} = \frac{1}{2} g^{\lambda\eta} \left( g_{\alpha\eta,\beta} + g_{\eta\beta,\alpha} - g_{\alpha\beta, \eta} \right) \, .
    \end{equation}

    Moreover, $b_{\alpha\beta}$ is the curvature tensor defined by the dot product $b_{\alpha\beta} = \vect{g}_{\alpha,\beta} \cdot \vect{n}$.
    We obtain
    \begin{equation}
    b_{\alpha\beta} = 
    \left(
    \begin{array}{cc}
    u_{r,\phi\phi} - (R + u_r + 2 u_{\phi, \phi}) & u_{r, \phi z} - u_{\phi, z} \\
    u_{r, \phi z} - u_{\phi, z} & u_{r,zz}
    \end{array}
    \right) \, . \label{curvatureTensor}
    \end{equation}
    %\SG{I guess the last line should also be a $\phi$?}\AI{Yes, indeed, corrected}

    At leading order in deformation, only the partial derivative remains in Eq.~\eqref{covariantDerivativeDef}.
    After some algebra, we find that the traction jumps across the membrane given by Eqs.~\eqref{Equilibrium}  are written in the cylindrical coordinate basis as
    \begin{subequations}
    \begin{align}
    \frac{\kS}{3} \left( u_{\phi, zz} + \frac{3 u_{z, \phi z}}{R} + \frac{4 (u_{r,\phi} + u_{\phi, \phi\phi})}{R^2} \right) + \Delta f_\phi &=0 \, , \\
      \frac{\kS}{3} \left( 4 u_{z,zz} + \frac{2 u_{r,z} + 3 u_{\phi, z\phi}}{R} + \frac{u_{z, \phi\phi}}{R^2} \right) + \Delta f_{z} &= 0 \, , \\
      -\frac{2\kS}{3} \left( \frac{2 (u_r + u_{\phi, \phi}) }{R^2} + \frac{u_{z,z}}{R}  \right) + \Delta f_{r} &= 0 \, .
    \end{align}
    \label{Equilibrium_finalize}
    \end{subequations}

    % As an aside remark, a similar set of equations is obtained in the linear elasticity theory by taking $\mu = \kS/3$ and $\lambda = 2\mu$, where $\mu$ and $\lambda$ being the first and second Lam\'e coefficients respectively~\cite{sadd09} [Eqs. (3.7.6)].
    % However, we derive the equations here from scratch for the sake of completeness.

    Note that for curved membranes, the normal traction jump does not vanish in the plane stress formulation employed throughout this work as the zeroth order in the curvature tensor is not identically null. 
    For a planar elastic membrane however, the resistance to shear introduces a jump only in the tangential traction jumps~\cite{daddi16, daddi16b, daddi16c}.

    Continuing, the jump in the fluid stress tensor across the membrane reads
    \begin{equation}
    [ \sigma_{\beta r} ] = \Delta f_{\beta} \, , \quad \beta \in \{ r, \phi, z \} \, . \label{stressTensorTractionJump}
    \end{equation}

    %%%%
    Therefore, From Eqs.~\eqref{Equilibrium_finalize}, \eqref{stressTensorTractionJump} and \eqref{no-slip-relation}, it follows that 
    \begin{subequations}
      \begin{align}
	[v_{\phi,r} ] &= \left. \frac{i\alpha}{2} \left( R v_{\phi, zz} + {3 v_{z, \phi z}} + \frac{4 (v_{r,\phi} + v_{\phi, \phi\phi})}{R} \right) \right|_{r=R} \, , \\
	[v_{z,r} ] &= \left. \frac{i\alpha}{2} \left( 4 R v_{z,zz} + {2 v_{r,z} + 3 v_{\phi, z\phi}} + \frac{v_{z, \phi\phi}}{R} \right) \right|_{r=R} \, , \\
	\left[-\frac{p}{\eta} \right] &= \left. -i\alpha \left( \frac{2 (v_r + v_{\phi, \phi} )}{R} + v_{z,z} \right) \right|_{r=R} \, , 
      \end{align}
    \end{subequations}
    where $\alpha := 2\kS/(3 \eta R \omega) $ is a dimensionless number characteristic for shear.
    Note that it follows from the incompressibility equation 
    \begin{equation}
    \frac{v_r + v_{\phi, \phi}}{r} + v_{r,r} + v_{z,z} = 0 \, , 
    \end{equation}
    that $[v_{r,r}] = 0$.
    Hereafter, we shall derive the traction jump equations across a membrane possessing a bending rigidity.

    %%%%%%%%%%%%%%%%%%%%%%%%%%%%%%%%%%%%%%%%%%%%%%%%%%%%%%%%%%%%

    \subsection{Bending}

    Here we use the full Helfrich model for the bending energy. For small deformations and planar membranes, this is equivalent to the ''linear bending model'' used in our earlier works~\cite{daddi16, daddi16b, daddi16c, daddi17}, see ref.~\cite{Guckenberger_preprint} for details. For a curved surface that we consider here, however, the latter leads to unphysical tangential components.
    The traction jump equations across the membranes are given by~\cite{zhong87, Guckenberger_preprint}
    \begin{equation}
    \Delta \vect{f}        = -2\kB \left( 2(H^2-K+H_0 H) + \Delta_\parallel \right) (H-H_0) \, \vect{n} \, ,  \label{helfrich_tractionJump}
    \end{equation}
    where $\kB$ is the bending modulus, $H$ and $K$ are the mean and Gaussian curvatures, respectively given by
    \begin{equation}
    H = \frac{1}{2} \, b_\alpha^\alpha \, , \qquad 
    K = \mathrm{det~} b_\alpha^\beta \, , 
    \end{equation}
    with $b_\alpha^\beta$ being the mixed version of the curvature tensor related to the covariant representation of the curvature tensor by $b_{\alpha}^{\beta} = b_{\alpha\delta} g^{\delta\beta}$.
    Continuing, $\Delta_\parallel$ is the horizontal Laplace operator and $H_0$ is the spontaneous curvature for which we take the initial undisturbed shape here. 
    The linearized traction jumps are therefore given by 
    \begin{equation}
    \begin{split}
      -\kB \Big( & R^3 u_{r,zzzz} + 2R(u_{r,zz} + u_{r,zz\phi\phi})  
	+ \frac{u_r+2u_{r,\phi\phi}+u_{r, \phi\phi\phi\phi}}{R} \Big)  + \Delta f_r = 0 \, .
    \end{split}
    \end{equation}
    and $\Delta f_\phi = \Delta f_z = 0$.

    Interestingly, bending does not introduce at leading order a jump in the tangential traction~\cite{guckenberger16}.
    The traction jump equations take the following final from
    \begin{subequations}
    \begin{align}
      [v_{\phi, r}] &= 0 \, , \\ 
      [v_{z,r} ] &= 0 \, , \\
      \left[-\frac{p}{\eta} \right] &= - i\alphaB^3 \Big( R^3 v_{r,zzzz} + 2R(v_{r,zz} + v_{r,zz\phi\phi}) 
	    + \left. \frac{v_r+2v_{r,\phi\phi}+v_{r, \phi\phi\phi\phi}}{R} \Big) \right|_{r=R} \, , 
    \end{align}
    \end{subequations}
    where $\alphaB = (\kB /(\eta \omega))^{1/3}/R$ is the dimensionless number characteristic for bending.

    %%%%%%%%%%%%%%%%%%%%%%%%%%%%%%%%%%%

    \section{Determination of the unknown functions for axial motion} \label{appendix:determinationConstantsAxial}
    
    In this appendix, we derive the expressions of the two functions $\psi^*_\parallel$ and $\varphi^*_\parallel$ associated to the solution of the Stokes equations due to a point force directed along a cylindrical membrane possessing pure shear or pure bending rigidities.
    
    \subsection{Pure shear}\label{shear_Axial}

    \quad As a first model, we consider an idealized membrane with a finite shear resistance and no bending resistance, such as an artificial capsule~\cite{rao94, eggleton98, navot98, sui08POF, clausen10}. The tangential traction jump given by Eq.~\eqref{BC:sigma_r_z} is in leading order independent of bending resistance and readily leads to
    \begin{equation}
     \begin{split}
      &- s^2 I_1\psi^*_\parallel  - s^2 \big( I_1+sI_0 \big)\varphi^*_\parallel  +s^2 \big( (i\alpha - 1)K_1 + 2i\alpha s K_0 \big) {\psi_2}_\parallel  \\ 
      &\phanSp-\big( (1+i\alpha  + 2 i\alpha s^2)K_1 - (1+i\alpha)s K_0 \big) s^2 {\varphi_2}_\parallel  =   R s \big( sK_0 - 2K_1 \big) \, , \label{Traction_Tan}
     \end{split}
    \end{equation}
    where $\alpha={2\kS}/{(3\eta R \omega)}$ is the shear parameter. 
    Neglecting the bending contribution $\Delta f^\mathrm{B}_{r}$ in the radial traction jump in  Eq.~\eqref{BC:sigma_r_r} yields
    \begin{equation}
      \begin{split}
      & 2s^2 I_0 \varphi^*_\parallel -i\alpha s \big( sK_0+2K_1 \big) {\psi_2}_\parallel + s\big( i\alpha (2+s^2)K_1 + s(i\alpha-2) K_0 \big) {\varphi_2}_\parallel  = - 2  R s K_0 \, . \label{Traction_Nor}
      \end{split}
    \end{equation}

    Eqs.~\eqref{psi_2_phi_2_from_psi_1_phi_1} together with \eqref{Traction_Tan} and \eqref{Traction_Nor} form a linear system of equations for the four unknown functions, amenable to immediate resolution via the standard substitution method.
    We obtain
    \begin{equation}
    \psi^*_\parallel  =  R \, \frac{\MSpara}{\NSpara} \, , \quad \varphi^*_\parallel  =   R \, \frac{\LSpara}{\NSpara} \, ,  \label{psi_1_AND_phi_1}
    \end{equation}
    where the numerators read
    %\begin{subequations}
    \begin{align}
    \MSpara &= \alpha \Big( (I_0 K_1+I_1 K_0) \big( 3i\alpha K_0^2-(4+3i\alpha) K_1^2 \big) s^3 + \big( -3i\alpha I_0 K_0^3+(8+3i\alpha)I_1 K_0^2 K_1 \notag \\
            &+ (8+9i\alpha) I_0 K_0 K_1^2 + 3i\alpha I_1 K_1^3 \big)s^2 + \big( 6(i\alpha-1)I_1 K_0^3-6(i\alpha+1)I_0 K_0^2 K_1 \notag \\
	&-2(1+6i\alpha)I_1 K_0 K_1^2 - 2I_0 K_1^3 \big)s +12i\alpha  K_0^2 K_1 I_1 \Big)   \, , \notag \\
    \LSpara &=  \Big(
      \big( -3i\alpha I_0 K_0^3 + (4-3i\alpha) I_1 K_0^2 K_1 +(4+3i\alpha) I_0 K_0 K_1^2 + 3i\alpha I_1 K_1^3  \big) s^2 \notag \\
      &+ \big( 6(i\alpha-1) I_1 K_0^3 - 6(1+i\alpha) I_0 K_0^2 K_1 + 2(1-6i\alpha) I_1 K_0 K_1^2 + 2I_0 K_1^3 \big) s \notag \\
      &+12i\alpha I_1 K_0^2 K_1 \Big) \alpha \, , \notag
    \end{align}
    %\end{subequations}
    and the denominator
    \begin{equation}
    \begin{split}
      \NSpara &= \big( 3i(K_0^2-K_1^2)(I_0^2-I_1^2)\alpha+4(I_1^2 K_0^2-I_0^2K_1^2) \big)\alpha s^3 +2\alpha s^2 (I_0 K_0+I_1 K_1) \notag \\
      &\times \big( 3i\alpha(I_0 K_1-I_1 K_0) +2(I_0 K_1 + I_1 K_0) \big)  +4 \big( -3i I_0 I_1 K_0 K_1 \alpha^2+\alpha(I_1^2 K_0^2-I_0^2 K_1^2) \notag  \\
	  & +i(I_0 K_1 + I_1 K_0)^2 \big) s +8 \alpha I_1 K_1 (I_0 K_1 + I_1 K_0) \, . 
    \end{split}
    \end{equation}
    Taking $\alpha\to\infty$, which is achieved either by considering an infinite shear modulus $\kS$ or a vanishing actuation frequency,
    we recover the known solution for a hard cylinder with stick boundary conditions, namely
    \begin{subequations}\label{LironandShahar}
    \begin{align}
    \lim_{\alpha \to \infty} \frac{\psi^*_\parallel }{R} &=   \, \frac{(I_0 K_1+I_1 K_0)s^2 - (I_0 K_0+I_1 K_1) s+2 I_1 K_0}{s \big( s I_0^2 - s I_1^2 - 2 I_0 I_1 \big)} \, , \label{psi_1_alphaInfty} \\
    \lim_{\alpha \to \infty} \frac{\varphi^*_\parallel}{R} &=   \, \frac{ 2 I_1 K_0 - (I_0 K_0+I_1 K_1) s }{s \big( s I_0^2 - s I_1^2 - 2 I_0 I_1 \big)} \, , \label{phi_1_alphaInfty}
    \end{align}
    \end{subequations}
    in agreement with the results of Liron and Shahar~\cite{liron78}.
    Note that both ${\psi_2}_\parallel$ and ${\varphi_2}_\parallel$ vanish in this limit, meaning that the fluid outside the cylinder is stagnant.

    \subsection{Pure bending}\label{bending_Axial}

    \quad A complimentary model membrane involves only a finite bending resistance, as considered previously to model a typical fluid vesicle~\cite{bukman96, luo13, dupont15, kaoui16, kaoui17, aouane17}. The effects of bending are determined by the dimensionless number $\alphaB=(\kB/(\eta\omega))^{1/3}/R$. 
    We now set $\Delta f_z^\mathrm{S} = \Delta f_r^\mathrm{S} = 0$ in Eqs.~\eqref{BC:sigma_r_z} and \eqref{BC:sigma_r_r}. The tangential-normal stress component is therefore continuous, leading to
    \begin{equation}
	- s^2 I_1\psi^*_\parallel  - s^2 \big( I_1+sI_0 \big)\varphi^*_\parallel  -s^2 K_1 {\psi_2}_\parallel -\big( K_1 - s K_0 \big) s^2 {\varphi_2}_\parallel =   R s \big( sK_0 - 2K_1 \big) \, , \notag 
    \end{equation}
    while the discontinuity in the normal traction jump leads to
    \begin{equation}
      \begin{split}
	&s \left( 2sI_0+i\alphaB^3 (sI_0-I_1)(s^2-1)^2 \right) \varphi^*_\parallel + i\alphaB^3 s(s^2-1)^2 I_1 \psi^*_\parallel \\
	&\phanSp- 2s^2 K_0 {\varphi_2}_\parallel = R s \left( 2+i\alphaB^3 (s^2-1)^2 \right) K_0 \, . \notag
      \end{split}
    \end{equation}
    The functions $\psi^*_\parallel$ and $\varphi^*_\parallel$ can be cast in a form similar to Eq.~\eqref{psi_1_AND_phi_1} as
    \begin{equation}
    \psi^*_\parallel  =  R \, \frac{\MBpara}{\NBpara} \, , \quad \varphi^*_\parallel  =   R \, \frac{\LBpara}{\NBpara} \, ,  \label{psi_1_AND_phi_1_bending}
    \end{equation}
    with the numerators
    \begin{align}
    \MBpara &=  \alphaB^3 (s^2-1)^2 K_0 (K_1+sK_0) \, , \notag  \\
    \LBpara &=  -\alphaB^3 (s^2-1)^2 K_0 K_1 \, ,   \notag
    \end{align}
    and the denominator 
    \begin{equation}
    \begin{split}
      \NBpara &= (s^2-1)^2 \left( s(I_0K_1-I_1K_0)-2I_1K_1 \right)\alphaB^3 -2is (I_0K_1+I_1K_0) \, . \notag 
    \end{split}
    \end{equation}

    Importantly, by considering the limit $\alphaB\to\infty$ (corresponding to an infinite bending modulus of a vanishing actuation frequency) we obtain
    \begin{align}
    \lim_{\alphaB\to\infty} \frac{\psi^*_\parallel}{R} &=   \, \frac{K_0 \big( s K_0 +K_1 \big)}{\big( s I_0 -2I_1 \big)K_1 - s K_0 I_1 } \, , \notag  \\
    \lim_{\alphaB\to\infty} \frac{\varphi^*_\parallel}{R} &= - \frac{K_0 K_1}{\big( s I_0 -2I_1 \big)K_1 - s K_0 I_1} \, , \notag 
    \end{align}
    which is found to be different from the solution for a hard cylinder given by Eqs.~\eqref{LironandShahar}. 
    This difference will be explained later on, as it is characteristic for many elastohydrodynamic systems.

    \section{Determination of the unknown functions for radial motion} \label{appendix:determinationConstantsRadial}
    
    In this appendix, we provide analytical expressions of the three functions $\varphi^*_\perp$, $\psi^*_\perp$ and $\gamma^*_\perp$ associated to a point force acting perpendicular to a cylindrical membrane with either shear or bending rigidities.

    \subsection{Pure shear}\label{shear_Radial}

    \quad We first consider an idealized membrane with a finite shear resistance and no bending resistance.
    The tangential traction jump along the $z$ direction given by Eq.~\eqref{BC:sigma_r_z} is independent of bending leading to
    \begin{align} \nonumber
    & s^2(I_0+sI_1) \varphi^*_\perp + s(sI_0 - I_1) \psi^*_\perp + s \left( s \left( 1+i\alpha(3+2s^2) \right)K_0 +  \left( i\alpha(5+s^2)-s^2 \right) K_1 \right) {\varphi_2}_\perp \notag \\
    &+ \frac{i\alpha s}{2} \left( 3sK_0 + 5K_1 \right) {\gamma_2}_\perp  +s \left( s(1-i\alpha) K_0 + \left( 1-i\alpha (3+2s^2) \right) K_1 \right) {\psi_2}_\perp = R s (K_0 - s K_1) \, ,  \label{Eq_4}
    \end{align}
    and the tangential traction jump along the $\phi$ direction given by Eq.~\eqref{BC:sigma_r_phi} leads to 
    \begin{align} 
    &\left( (4+s^2) I_1 -2sI_0 \right)\varphi^*_\perp + (sI_0-2I_1)\psi^*_\perp + \left( (2+s^2)I_1-sI_0 \right) \gamma^*_\perp + \tfrac{1}{2} \Big( \big( i\alpha \left(8+s^2 \right)\notag \\
    &-(4+2s^2) \big) K_1 + s \left( i\alpha \left(4+s^2 \right)-2 \right) K_0 \Big) {\gamma_2}_\perp +  \big( \left( i\alpha (8+3s^2)-(4+s^2) \right)K_1 \notag \\
    &+ 2s \left( i\alpha (2+s^2) - 1 \right) K_0 \big) {\varphi_2}_\perp + \left( 2\left(1-i\alpha (2+s^2)\right) K_1 + s(1-2i\alpha) K_0 \right) {\psi_2}_\perp = Rs K_1 \, . \label{Eq_5}
    \end{align}
    Continuing, the shear related part in the normal traction jump given by Eq.~\eqref{BC:sigma_r_r} yields
    \begin{align} \nonumber
      & 2s^2 I_1 \varphi^*_\perp + \left( i\alpha s(4+s^2)K_0 + 2 \left( i\alpha (4+s^2)-s^2 \right) K_1 \right) {\varphi_2}_\perp  - i\alpha \left( 2s K_0 + (4+s^2) K_1 \right) {\psi_2}_\perp \\ \nonumber
      &\phanSp + 2i\alpha (sK_0 + 2K_1) {\gamma_2}_\perp = -2Rs K_1 \, . \label{Eq_6}
    \end{align}

    Inserting the expressions of ${\varphi_2}_\perp$, ${\psi_2}_\perp$ and ${\gamma_2}_\perp$ given by Eqs.~\eqref{Eq_1_forPhi2} through \eqref{Eq_3_forOmega2} into Eqs.~\eqref{Eq_4} through \eqref{Eq_6}, we obtain the unknown functions $\varphi^*_\perp$, $\psi^*_\perp$ and $\gamma^*_\perp$ inside the channel.
    Explicit analytical expressions are not listed here due to their complexity and lengthiness.
    Particularly, by taking $\alpha \to \infty$, we recover the solution for a no-slip cylinder, namely
    \begin{subequations}\label{LironandShahar_radial}
    \begin{align}
    \lim_{\alpha\to\infty} \frac{\varphi^*_\perp}{R} &=  \, \frac{s (s I_0 - I_1)(I_0K_0+I_1K_1)-2 I_1^2 K_0}
	  {s \left(s(sI_0 - I_1) (I_0^2-I_1^2) - 2I_0 I_1^2 \right)} \, ,  \\
    \lim_{\alpha\to\infty} \frac{ \psi^*_\perp}{R} &= \, \frac{s (I_1 - sI_0) (I_0 K_1 + I_1 K_0)}{s(sI_0-I_1) (I_0^2 - I_1^2) - 2I_0 I_1^2} \, ,  \\
      \lim_{\alpha\to\infty} \frac{\gamma^*_\perp}{R} &= 2 \, \frac{s I_1 (I_0K_0+I_1K_1)+2I_1^2 K_0 -s^2 K_0(I_0^2-I_1^2)}{s \left( s(sI_0-I_1)(I_0^2-I_1^2)-2I_0I_1^2 \right)} \, , 
    \end{align}
    \end{subequations}
    and ${\varphi_2}_\perp = {\psi_2}_\perp = {\gamma_2}_\perp = 0$, in complete agreement with the results by Liron and Shahar~\cite{liron78}.

    \subsection{Pure bending}\label{bending_Radial}

    \quad Neglecting the shear contribution in the tangential traction jump along the $z$ direction given by Eq.~\eqref{BC:sigma_r_z}, we obtain
    \begin{equation}
      s^2(I_0+sI_1)\varphi^*_\perp + s(sI_0-I_1)\psi^*_\perp + s^2(K_0-sK_1) {\varphi_2}_\perp  
      + s(K_1+sK_0) {\psi_2}_\perp = Rs (K_0-sK_1) \, . \label{Eq_4_bending}
    \end{equation}

    The traction jump along the $\phi$ direction stated by Eq.~\eqref{BC:sigma_r_phi} is continuous, leading to
    \begin{align} \nonumber
      &\left( (4+s^2)I_1 - 2sI_0 \right) \varphi^*_\perp + (sI_0 - 2I_1) \psi^*_\perp + \left( (2+s^2)I_1 - sI_0 \right) \gamma^*_\perp + (sK_0+2K_1)  {\psi_2}_\perp \\ \nonumber
      &\phanSp-\left( 2sK_0+(4+s^2)K_1 \right)  {\varphi_2}_\perp  - \left( sK_0 + \left( s^2  + 2 \right)K_1 \right) {\gamma_2}_\perp = RsK_1 \, . \label{Eq_5_bending}
    \end{align}
    while the discontinuity of the normal traction jump due to pure bending leads to
    \begin{equation}
    \begin{split}
      &2s I_1 \psi^*_\perp + \left( i\alphaB^3 s^3 \left( (s^2+2)K_1+s K_0 \right)-2s K_1 \right) {\varphi_2}_\perp  \\
      &\phanSp-i\alphaB^3 s^3 (sK_0+K1) {\psi_2}_\perp + i\alphaB^3 s^3 K_1 {\gamma_2}_\perp = -2R K_1 \, .
    \end{split} \label{Eq_6_bending}
    \end{equation}

    The unknown functions $\varphi^*_\perp$, $\psi^*_\perp$ and $\gamma^*_\perp$ are readily obtained after plugging  the expressions of ${\varphi_2}_\perp$, ${\psi_2}_\perp$ and ${\gamma_2}_\perp$ given by Eqs.~\eqref{Eq_1_forPhi2} through \eqref{Eq_3_forOmega2} into Eqs.~\eqref{Eq_4_bending} through \eqref{Eq_6_bending}.
    % Similar, analytical expression cannot be listed here due to their complexity and lengthiness.
    Further, by taking $\alphaB \to \infty$, we obtain
    \begin{subequations}
     \begin{align}
    \lim_{\alphaB\to\infty} \frac{\varphi^*_\perp}{R} &=  \frac{(K_0+sK_1)(sK_0+K_1)}{sK_0 \left( (3+s^2)I_1-2sI_0 \right)-(3+s^2)(2I_1-sI_0) K_1} \, , \\
    \lim_{\alphaB\to\infty} \frac{\psi^*_\perp}{R} &=  \frac{(K_0+sK_1) \left( sK_0+(2+s^2)K_1 \right)}{sK_0 \left( (3+s^2)I_1-2sI_0 \right)-(3+s^2)(2I_1-sI_0) K_1} \, ,  \\
      \lim_{\alphaB\to\infty} \frac{\gamma^*_\perp}{R}    &=  \frac{2 K_1 (K_0+sK_1)}{sK_0 \left( (3+s^2)I_1-2sI_0 \right)-(3+s^2)(2I_1-sI_0) K_1} \, , 
      \label{eqn:flowHardBending} 
    \end{align}
    \end{subequations}
    which is not identical to the solution for a no-slip cylinder given by Eqs.~\eqref{LironandShahar_radial}, i.e.\@ in the same way as observed for the axial motion. 
    This feature is justified by the fact that bending does not introduce a discontinuity in the tangential traction jumps and that the normal traction jumps due to bending resistance as prescribed by Helfrich law in Eq.~\eqref{bendingTraction} depends only on the normal displacement~$u_r$.
    Therefore, even when considering an infinite bending modulus, the tangential components of the membrane displacement $u_\phi$ and $u_z$ are still completely free.
    As a result, this behavior cannot represent the hard cylinder where all membrane displacements should be restricted.
    A similar feature has been found for spherical membranes~\cite{daddi17b, daddi17c}.
    
    \clearpage

    % \bibliography{biblio}

\begin{thebibliography}{129}%
\makeatletter
\providecommand \@ifxundefined [1]{%
 \@ifx{#1\undefined}
}%
\providecommand \@ifnum [1]{%
 \ifnum #1\expandafter \@firstoftwo
 \else \expandafter \@secondoftwo
 \fi
}%
\providecommand \@ifx [1]{%
 \ifx #1\expandafter \@firstoftwo
 \else \expandafter \@secondoftwo
 \fi
}%
\providecommand \natexlab [1]{#1}%
\providecommand \enquote  [1]{``#1''}%
\providecommand \bibnamefont  [1]{#1}%
\providecommand \bibfnamefont [1]{#1}%
\providecommand \citenamefont [1]{#1}%
\providecommand \href@noop [0]{\@secondoftwo}%
\providecommand \href [0]{\begingroup \@sanitize@url \@href}%
\providecommand \@href[1]{\@@startlink{#1}\@@href}%
\providecommand \@@href[1]{\endgroup#1\@@endlink}%
\providecommand \@sanitize@url [0]{\catcode `\\12\catcode `\$12\catcode
  `\&12\catcode `\#12\catcode `\^12\catcode `\_12\catcode `\%12\relax}%
\providecommand \@@startlink[1]{}%
\providecommand \@@endlink[0]{}%
\providecommand \url  [0]{\begingroup\@sanitize@url \@url }%
\providecommand \@url [1]{\endgroup\@href {#1}{\urlprefix }}%
\providecommand \urlprefix  [0]{URL }%
\providecommand \Eprint [0]{\href }%
\providecommand \doibase [0]{http://dx.doi.org/}%
\providecommand \selectlanguage [0]{\@gobble}%
\providecommand \bibinfo  [0]{\@secondoftwo}%
\providecommand \bibfield  [0]{\@secondoftwo}%
\providecommand \translation [1]{[#1]}%
\providecommand \BibitemOpen [0]{}%
\providecommand \bibitemStop [0]{}%
\providecommand \bibitemNoStop [0]{.\EOS\space}%
\providecommand \EOS [0]{\spacefactor3000\relax}%
\providecommand \BibitemShut  [1]{\csname bibitem#1\endcsname}%
\let\auto@bib@innerbib\@empty
%</preamble>
\bibitem [{\citenamefont {Brady}\ and\ \citenamefont {Bossis}(1988)}]{brady88}%
  \BibitemOpen
  \bibfield  {author} {\bibinfo {author} {\bibfnamefont {J.~F.}\ \bibnamefont
  {Brady}}\ and\ \bibinfo {author} {\bibfnamefont {G.}~\bibnamefont {Bossis}},\
  }\bibfield  {title} {\enquote {\bibinfo {title} {Stokesian dynamics},}\
  }\href@noop {} {\bibfield  {journal} {\bibinfo  {journal} {Ann. Rev. Fluid
  Mech.}\ }\textbf {\bibinfo {volume} {20}},\ \bibinfo {pages} {111--157}
  (\bibinfo {year} {1988})}\BibitemShut {NoStop}%
\bibitem [{\citenamefont {Bleibel}\ \emph {et~al.}(2014)\citenamefont
  {Bleibel}, \citenamefont {Dom{\'\i}nguez}, \citenamefont {G{\"u}nther},
  \citenamefont {Harting},\ and\ \citenamefont {Oettel}}]{bleibel14}%
  \BibitemOpen
  \bibfield  {author} {\bibinfo {author} {\bibfnamefont {J.}~\bibnamefont
  {Bleibel}}, \bibinfo {author} {\bibfnamefont {A.}~\bibnamefont
  {Dom{\'\i}nguez}}, \bibinfo {author} {\bibfnamefont {F.}~\bibnamefont
  {G{\"u}nther}}, \bibinfo {author} {\bibfnamefont {J.}~\bibnamefont
  {Harting}}, \ and\ \bibinfo {author} {\bibfnamefont {M.}~\bibnamefont
  {Oettel}},\ }\bibfield  {title} {\enquote {\bibinfo {title} {Hydrodynamic
  interactions induce anomalous diffusion under partial confinement},}\
  }\href@noop {} {\bibfield  {journal} {\bibinfo  {journal} {Soft Matter}\
  }\textbf {\bibinfo {volume} {10}},\ \bibinfo {pages} {2945--2948} (\bibinfo
  {year} {2014})}\BibitemShut {NoStop}%
\bibitem [{\citenamefont {Wei}, \citenamefont {Bechinger},\ and\ \citenamefont
  {Leiderer}(2000)}]{wei00}%
  \BibitemOpen
  \bibfield  {author} {\bibinfo {author} {\bibfnamefont {Q.-H.}\ \bibnamefont
  {Wei}}, \bibinfo {author} {\bibfnamefont {C.}~\bibnamefont {Bechinger}}, \
  and\ \bibinfo {author} {\bibfnamefont {P.}~\bibnamefont {Leiderer}},\
  }\bibfield  {title} {\enquote {\bibinfo {title} {Single-file diffusion of
  colloids in one-dimensional channels},}\ }\href@noop {} {\bibfield  {journal}
  {\bibinfo  {journal} {Science}\ }\textbf {\bibinfo {volume} {287}},\ \bibinfo
  {pages} {625--627} (\bibinfo {year} {2000})}\BibitemShut {NoStop}%
\bibitem [{\citenamefont {Lutz}, \citenamefont {Kollmann},\ and\ \citenamefont
  {Bechinger}(2004)}]{lutz04}%
  \BibitemOpen
  \bibfield  {author} {\bibinfo {author} {\bibfnamefont {C.}~\bibnamefont
  {Lutz}}, \bibinfo {author} {\bibfnamefont {M.}~\bibnamefont {Kollmann}}, \
  and\ \bibinfo {author} {\bibfnamefont {C.}~\bibnamefont {Bechinger}},\
  }\bibfield  {title} {\enquote {\bibinfo {title} {Single-file diffusion of
  colloids in one-dimensional channels},}\ }\href@noop {} {\bibfield  {journal}
  {\bibinfo  {journal} {Phys. Rev. Lett.}\ }\textbf {\bibinfo {volume} {93}},\
  \bibinfo {pages} {026001} (\bibinfo {year} {2004})}\BibitemShut {NoStop}%
\bibitem [{\citenamefont {Janssen}\ \emph {et~al.}(2012)\citenamefont
  {Janssen}, \citenamefont {Baron}, \citenamefont {Anderson}, \citenamefont
  {Blawzdziewicz}, \citenamefont {Loewenberg},\ and\ \citenamefont
  {Wajnryb}}]{Janssen_2012}%
  \BibitemOpen
  \bibfield  {author} {\bibinfo {author} {\bibfnamefont {P.~J.~A.}\
  \bibnamefont {Janssen}}, \bibinfo {author} {\bibfnamefont {M.~D.}\
  \bibnamefont {Baron}}, \bibinfo {author} {\bibfnamefont {P.~D.}\ \bibnamefont
  {Anderson}}, \bibinfo {author} {\bibfnamefont {J.}~\bibnamefont
  {Blawzdziewicz}}, \bibinfo {author} {\bibfnamefont {M.}~\bibnamefont
  {Loewenberg}}, \ and\ \bibinfo {author} {\bibfnamefont {E.}~\bibnamefont
  {Wajnryb}},\ }\bibfield  {title} {\enquote {\bibinfo {title} {{Collective
  dynamics of confined rigid spheres and deformable drops}},}\ }\href@noop {}
  {\bibfield  {journal} {\bibinfo  {journal} {Soft Matter}\ }\textbf {\bibinfo
  {volume} {8}},\ \bibinfo {pages} {7495--13} (\bibinfo {year}
  {2012})}\BibitemShut {NoStop}%
\bibitem [{\citenamefont {Lele}\ \emph {et~al.}(2011)\citenamefont {Lele},
  \citenamefont {Swan}, \citenamefont {Brady}, \citenamefont {Wagner},\ and\
  \citenamefont {Furst}}]{lele11}%
  \BibitemOpen
  \bibfield  {author} {\bibinfo {author} {\bibfnamefont {P.~P.}\ \bibnamefont
  {Lele}}, \bibinfo {author} {\bibfnamefont {J.~W.}\ \bibnamefont {Swan}},
  \bibinfo {author} {\bibfnamefont {J.~F.}\ \bibnamefont {Brady}}, \bibinfo
  {author} {\bibfnamefont {N.~J.}\ \bibnamefont {Wagner}}, \ and\ \bibinfo
  {author} {\bibfnamefont {E.~M.}\ \bibnamefont {Furst}},\ }\bibfield  {title}
  {\enquote {\bibinfo {title} {Colloidal diffusion and hydrodynamic screening
  near boundaries},}\ }\href@noop {} {\bibfield  {journal} {\bibinfo  {journal}
  {Soft Matter}\ }\textbf {\bibinfo {volume} {7}},\ \bibinfo {pages}
  {6844--6852} (\bibinfo {year} {2011})}\BibitemShut {NoStop}%
\bibitem [{\citenamefont {Gross}, \citenamefont {Kr{\"u}ger},\ and\
  \citenamefont {Varnik}(2014)}]{gross14}%
  \BibitemOpen
  \bibfield  {author} {\bibinfo {author} {\bibfnamefont {M.}~\bibnamefont
  {Gross}}, \bibinfo {author} {\bibfnamefont {T.}~\bibnamefont {Kr{\"u}ger}}, \
  and\ \bibinfo {author} {\bibfnamefont {F.}~\bibnamefont {Varnik}},\
  }\bibfield  {title} {\enquote {\bibinfo {title} {Rheology of dense
  suspensions of elastic capsules: normal stresses, yield stress, jamming and
  confinement effects},}\ }\href@noop {} {\bibfield  {journal} {\bibinfo
  {journal} {Soft Matter}\ }\textbf {\bibinfo {volume} {10}},\ \bibinfo {pages}
  {4360--4372} (\bibinfo {year} {2014})}\BibitemShut {NoStop}%
\bibitem [{\citenamefont {Frey-Wyssling}(1952)}]{Frey1952}%
  \BibitemOpen
  \bibinfo {editor} {\bibfnamefont {A.}~\bibnamefont {Frey-Wyssling}},\ ed.,\
  \href@noop {} {\emph {\bibinfo {title} {Deformation and flow in biological
  systems}}}\ (\bibinfo  {publisher} {North-Holland Publishing Co.,
  Amsterdam},\ \bibinfo {year} {1952})\BibitemShut {NoStop}%
\bibitem [{\citenamefont {Shadwick}(1999)}]{Shadwick1999}%
  \BibitemOpen
  \bibfield  {author} {\bibinfo {author} {\bibfnamefont {R.}~\bibnamefont
  {Shadwick}},\ }\bibfield  {title} {\enquote {\bibinfo {title} {Mechanical
  design in arteries},}\ }\href@noop {} {\bibfield  {journal} {\bibinfo
  {journal} {J. Exp. Biol.}\ }\textbf {\bibinfo {volume} {202}},\ \bibinfo
  {pages} {3305--3313} (\bibinfo {year} {1999})}\BibitemShut {NoStop}%
\bibitem [{\citenamefont {Caro}\ \emph {et~al.}(2011)\citenamefont {Caro},
  \citenamefont {Pedley}, \citenamefont {Schroter},\ and\ \citenamefont
  {Seed}}]{Caro2011}%
  \BibitemOpen
  \bibfield  {author} {\bibinfo {author} {\bibfnamefont {C.~G.}\ \bibnamefont
  {Caro}}, \bibinfo {author} {\bibfnamefont {T.~J.}\ \bibnamefont {Pedley}},
  \bibinfo {author} {\bibfnamefont {R.~C.}\ \bibnamefont {Schroter}}, \ and\
  \bibinfo {author} {\bibfnamefont {W.~A.}\ \bibnamefont {Seed}},\ }\href@noop
  {} {\emph {\bibinfo {title} {The Mechanics of the Circulation}}},\ \bibinfo
  {edition} {2nd}\ ed.\ (\bibinfo  {publisher} {Cambridge University Press},\
  \bibinfo {year} {2011})\BibitemShut {NoStop}%
\bibitem [{\citenamefont {Stone}, \citenamefont {Stroock},\ and\ \citenamefont
  {Ajdari}(2004)}]{stone04}%
  \BibitemOpen
  \bibfield  {author} {\bibinfo {author} {\bibfnamefont {H.~A.}\ \bibnamefont
  {Stone}}, \bibinfo {author} {\bibfnamefont {A.~D.}\ \bibnamefont {Stroock}},
  \ and\ \bibinfo {author} {\bibfnamefont {A.}~\bibnamefont {Ajdari}},\
  }\bibfield  {title} {\enquote {\bibinfo {title} {Engineering flows in small
  devices: microfluidics toward a lab-on-a-chip},}\ }\href@noop {} {\bibfield
  {journal} {\bibinfo  {journal} {Annu. Rev. Fluid Mech.}\ }\textbf {\bibinfo
  {volume} {36}},\ \bibinfo {pages} {381--411} (\bibinfo {year}
  {2004})}\BibitemShut {NoStop}%
\bibitem [{\citenamefont {Holmes}\ \emph {et~al.}(2013)\citenamefont {Holmes},
  \citenamefont {Tavakol}, \citenamefont {Froehlicher},\ and\ \citenamefont
  {Stone}}]{holmes13}%
  \BibitemOpen
  \bibfield  {author} {\bibinfo {author} {\bibfnamefont {D.~P.}\ \bibnamefont
  {Holmes}}, \bibinfo {author} {\bibfnamefont {B.}~\bibnamefont {Tavakol}},
  \bibinfo {author} {\bibfnamefont {G.}~\bibnamefont {Froehlicher}}, \ and\
  \bibinfo {author} {\bibfnamefont {H.~A.}\ \bibnamefont {Stone}},\ }\bibfield
  {title} {\enquote {\bibinfo {title} {Control and manipulation of microfluidic
  flow via elastic deformations},}\ }\href@noop {} {\bibfield  {journal}
  {\bibinfo  {journal} {Soft Matter}\ }\textbf {\bibinfo {volume} {9}},\
  \bibinfo {pages} {7049--7053} (\bibinfo {year} {2013})}\BibitemShut {NoStop}%
\bibitem [{\citenamefont {Tavakol}\ \emph {et~al.}(2014)\citenamefont
  {Tavakol}, \citenamefont {Bozlar}, \citenamefont {Punckt}, \citenamefont
  {Froehlicher}, \citenamefont {Stone}, \citenamefont {Aksay},\ and\
  \citenamefont {Holmes}}]{tavakol14}%
  \BibitemOpen
  \bibfield  {author} {\bibinfo {author} {\bibfnamefont {B.}~\bibnamefont
  {Tavakol}}, \bibinfo {author} {\bibfnamefont {M.}~\bibnamefont {Bozlar}},
  \bibinfo {author} {\bibfnamefont {C.}~\bibnamefont {Punckt}}, \bibinfo
  {author} {\bibfnamefont {G.}~\bibnamefont {Froehlicher}}, \bibinfo {author}
  {\bibfnamefont {H.~A.}\ \bibnamefont {Stone}}, \bibinfo {author}
  {\bibfnamefont {I.~A.}\ \bibnamefont {Aksay}}, \ and\ \bibinfo {author}
  {\bibfnamefont {D.~P.}\ \bibnamefont {Holmes}},\ }\bibfield  {title}
  {\enquote {\bibinfo {title} {Buckling of dielectric elastomeric plates for
  soft, electrically active microfluidic pumps},}\ }\href@noop {} {\bibfield
  {journal} {\bibinfo  {journal} {Soft Matter}\ }\textbf {\bibinfo {volume}
  {10}},\ \bibinfo {pages} {4789--4794} (\bibinfo {year} {2014})}\BibitemShut
  {NoStop}%
\bibitem [{\citenamefont {Happel}\ and\ \citenamefont
  {Brenner}(2012)}]{happel12}%
  \BibitemOpen
  \bibfield  {author} {\bibinfo {author} {\bibfnamefont {J.}~\bibnamefont
  {Happel}}\ and\ \bibinfo {author} {\bibfnamefont {H.}~\bibnamefont
  {Brenner}},\ }\href@noop {} {\emph {\bibinfo {title} {Low Reynolds number
  hydrodynamics: with special applications to particulate media}}},\
  Vol.~\bibinfo {volume} {1}\ (\bibinfo  {publisher} {Springer Science \&
  Business Media},\ \bibinfo {year} {2012})\BibitemShut {NoStop}%
\bibitem [{\citenamefont {Fax{\'e}n}(1959)}]{faxen59}%
  \BibitemOpen
  \bibfield  {author} {\bibinfo {author} {\bibfnamefont {H.}~\bibnamefont
  {Fax{\'e}n}},\ }\bibfield  {title} {\enquote {\bibinfo {title} {About t.
  bohlin's paper: On the drag on rigid spheres, moving in a viscous liquid
  inside cylindrical tubes},}\ }\href@noop {} {\bibfield  {journal} {\bibinfo
  {journal} {Colloid. Polym. Sci.}\ }\textbf {\bibinfo {volume} {167}},\
  \bibinfo {pages} {146--146} (\bibinfo {year} {1959})}\BibitemShut {NoStop}%
\bibitem [{\citenamefont {Fax{\'e}n}(1922)}]{faxen22}%
  \BibitemOpen
  \bibfield  {author} {\bibinfo {author} {\bibfnamefont {H.}~\bibnamefont
  {Fax{\'e}n}},\ }\bibfield  {title} {\enquote {\bibinfo {title} {Der
  {Widerstand} gegen die {Bewegung} einer starren {Kugel} in einer z{\"a}hen
  {Fl{\"u}ssigkeit}, die zwischen zwei parallelen ebenen {W{\"a}nden}
  eingeschlossen ist},}\ }\href@noop {} {\bibfield  {journal} {\bibinfo
  {journal} {Ann. Phys.}\ }\textbf {\bibinfo {volume} {373}},\ \bibinfo {pages}
  {89--119} (\bibinfo {year} {1922})}\BibitemShut {NoStop}%
\bibitem [{\citenamefont {Wakiya}(1953)}]{wakiya53}%
  \BibitemOpen
  \bibfield  {author} {\bibinfo {author} {\bibfnamefont {S.}~\bibnamefont
  {Wakiya}},\ }\bibfield  {title} {\enquote {\bibinfo {title} {A spherical
  obstacle in the flow of a viscous fluid through a tube},}\ }\href@noop {}
  {\bibfield  {journal} {\bibinfo  {journal} {J. Phys. Soc. Japan}\ }\textbf
  {\bibinfo {volume} {8}},\ \bibinfo {pages} {254--256} (\bibinfo {year}
  {1953})}\BibitemShut {NoStop}%
\bibitem [{\citenamefont {Bohlin}(1960)}]{bohlin60}%
  \BibitemOpen
  \bibfield  {author} {\bibinfo {author} {\bibfnamefont {T.}~\bibnamefont
  {Bohlin}},\ }\bibfield  {title} {\enquote {\bibinfo {title} {On the drag on a
  rigid sphere moving in a viscous liquid inside a cylindrical tube},}\
  }\href@noop {} {\bibfield  {journal} {\bibinfo  {journal} {Trans. Roy. Inst.
  Technol. Stockholm}\ }\textbf {\bibinfo {volume} {155}},\ \bibinfo {pages}
  {64} (\bibinfo {year} {1960})}\BibitemShut {NoStop}%
\bibitem [{\citenamefont {Zimmerman}(2004)}]{zimmerman04}%
  \BibitemOpen
  \bibfield  {author} {\bibinfo {author} {\bibfnamefont {W.~B.}\ \bibnamefont
  {Zimmerman}},\ }\bibfield  {title} {\enquote {\bibinfo {title} {On the
  resistance of a spherical particle settling in a tube of viscous fluid},}\
  }\href@noop {} {\bibfield  {journal} {\bibinfo  {journal} {Int. J. Eng.
  Sci.}\ }\textbf {\bibinfo {volume} {42}},\ \bibinfo {pages} {1753--1778}
  (\bibinfo {year} {2004})}\BibitemShut {NoStop}%
\bibitem [{\citenamefont {Leichtberg}, \citenamefont {Pfeffer},\ and\
  \citenamefont {Weinbaum}(1976)}]{Leichtberg_1976}%
  \BibitemOpen
  \bibfield  {author} {\bibinfo {author} {\bibfnamefont {S.}~\bibnamefont
  {Leichtberg}}, \bibinfo {author} {\bibfnamefont {R.}~\bibnamefont {Pfeffer}},
  \ and\ \bibinfo {author} {\bibfnamefont {S.}~\bibnamefont {Weinbaum}},\
  }\bibfield  {title} {\enquote {\bibinfo {title} {{Stokes flow past finite
  coaxial clusters of spheres in a circular cylinder}},}\ }\href@noop {}
  {\bibfield  {journal} {\bibinfo  {journal} {Int. J. Multiphase Flow}\
  }\textbf {\bibinfo {volume} {3}},\ \bibinfo {pages} {147} (\bibinfo {year}
  {1976})}\BibitemShut {NoStop}%
\bibitem [{\citenamefont {Kedzierski}\ and\ \citenamefont
  {Wajnryb}(2010)}]{kkedzierski10}%
  \BibitemOpen
  \bibfield  {author} {\bibinfo {author} {\bibfnamefont {M.}~\bibnamefont
  {Kedzierski}}\ and\ \bibinfo {author} {\bibfnamefont {E.}~\bibnamefont
  {Wajnryb}},\ }\bibfield  {title} {\enquote {\bibinfo {title} {Precise
  multipole method for calculating many-body hydrodynamic interactions in a
  microchannel},}\ }\href@noop {} {\bibfield  {journal} {\bibinfo  {journal}
  {J. Chem. Phys.}\ }\textbf {\bibinfo {volume} {133}},\ \bibinfo {pages}
  {154105} (\bibinfo {year} {2010})}\BibitemShut {NoStop}%
\bibitem [{\citenamefont {Cui}, \citenamefont {Diamant},\ and\ \citenamefont
  {Lin}(2002)}]{cui02}%
  \BibitemOpen
  \bibfield  {author} {\bibinfo {author} {\bibfnamefont {B.}~\bibnamefont
  {Cui}}, \bibinfo {author} {\bibfnamefont {H.}~\bibnamefont {Diamant}}, \ and\
  \bibinfo {author} {\bibfnamefont {B.}~\bibnamefont {Lin}},\ }\bibfield
  {title} {\enquote {\bibinfo {title} {Screened hydrodynamic interaction in a
  narrow channel},}\ }\href@noop {} {\bibfield  {journal} {\bibinfo  {journal}
  {Phys. Rev. Lett.}\ }\textbf {\bibinfo {volume} {89}},\ \bibinfo {pages}
  {188302} (\bibinfo {year} {2002})}\BibitemShut {NoStop}%
\bibitem [{\citenamefont {Cui}\ \emph {et~al.}(2002)\citenamefont {Cui},
  \citenamefont {Lin}, \citenamefont {Sharma},\ and\ \citenamefont
  {Rice}}]{cui02b}%
  \BibitemOpen
  \bibfield  {author} {\bibinfo {author} {\bibfnamefont {B.}~\bibnamefont
  {Cui}}, \bibinfo {author} {\bibfnamefont {B.}~\bibnamefont {Lin}}, \bibinfo
  {author} {\bibfnamefont {S.}~\bibnamefont {Sharma}}, \ and\ \bibinfo {author}
  {\bibfnamefont {S.~A.}\ \bibnamefont {Rice}},\ }\bibfield  {title} {\enquote
  {\bibinfo {title} {Equilibrium structure and effective pair interaction in a
  quasi-one-dimensional colloid liquid},}\ }\href@noop {} {\bibfield  {journal}
  {\bibinfo  {journal} {J. Chem. Phys.}\ }\textbf {\bibinfo {volume} {116}},\
  \bibinfo {pages} {3119--3127} (\bibinfo {year} {2002})}\BibitemShut {NoStop}%
\bibitem [{\citenamefont {Misiunas}\ \emph {et~al.}(2015)\citenamefont
  {Misiunas}, \citenamefont {Pagliara}, \citenamefont {Lauga}, \citenamefont
  {Lister},\ and\ \citenamefont {Keyser}}]{misiunas15}%
  \BibitemOpen
  \bibfield  {author} {\bibinfo {author} {\bibfnamefont {K.}~\bibnamefont
  {Misiunas}}, \bibinfo {author} {\bibfnamefont {S.}~\bibnamefont {Pagliara}},
  \bibinfo {author} {\bibfnamefont {E.}~\bibnamefont {Lauga}}, \bibinfo
  {author} {\bibfnamefont {J.~R.}\ \bibnamefont {Lister}}, \ and\ \bibinfo
  {author} {\bibfnamefont {U.~F.}\ \bibnamefont {Keyser}},\ }\bibfield  {title}
  {\enquote {\bibinfo {title} {Nondecaying hydrodynamic interactions along
  narrow channels},}\ }\href@noop {} {\bibfield  {journal} {\bibinfo  {journal}
  {Phys. Rev. Lett.}\ }\textbf {\bibinfo {volume} {115}},\ \bibinfo {pages}
  {038301} (\bibinfo {year} {2015})}\BibitemShut {NoStop}%
\bibitem [{\citenamefont {Yeh}\ and\ \citenamefont {Keh}(2013)}]{Yeh_2013}%
  \BibitemOpen
  \bibfield  {author} {\bibinfo {author} {\bibfnamefont {H.~Y.}\ \bibnamefont
  {Yeh}}\ and\ \bibinfo {author} {\bibfnamefont {H.~J.}\ \bibnamefont {Keh}},\
  }\bibfield  {title} {\enquote {\bibinfo {title} {{Axisymmetric creeping
  motion of a prolate particle in a cylindrical pore}},}\ }\href@noop {}
  {\bibfield  {journal} {\bibinfo  {journal} {Eur. J. Mech. B Fluid}\ }\textbf
  {\bibinfo {volume} {39}},\ \bibinfo {pages} {52--58} (\bibinfo {year}
  {2013})}\BibitemShut {NoStop}%
\bibitem [{\citenamefont {Happel}\ and\ \citenamefont
  {Bryne}(1954)}]{happel54}%
  \BibitemOpen
  \bibfield  {author} {\bibinfo {author} {\bibfnamefont {J.}~\bibnamefont
  {Happel}}\ and\ \bibinfo {author} {\bibfnamefont {B.~J.}\ \bibnamefont
  {Bryne}},\ }\bibfield  {title} {\enquote {\bibinfo {title} {Viscous flow in
  multiparticle systems. motion of a sphere in a cylindrical tube},}\
  }\href@noop {} {\bibfield  {journal} {\bibinfo  {journal} {Ind. Eng. Chem.}\
  }\textbf {\bibinfo {volume} {46}},\ \bibinfo {pages} {1181--1186} (\bibinfo
  {year} {1954})}\BibitemShut {NoStop}%
\bibitem [{\citenamefont {Brenner}\ and\ \citenamefont
  {Happel}(1958)}]{brenner58}%
  \BibitemOpen
  \bibfield  {author} {\bibinfo {author} {\bibfnamefont {H.}~\bibnamefont
  {Brenner}}\ and\ \bibinfo {author} {\bibfnamefont {J.}~\bibnamefont
  {Happel}},\ }\bibfield  {title} {\enquote {\bibinfo {title} {Slow viscous
  flow past a sphere in a cylindrical tube},}\ }\href@noop {} {\bibfield
  {journal} {\bibinfo  {journal} {J. Fluid Mech.}\ }\textbf {\bibinfo {volume}
  {4}},\ \bibinfo {pages} {195--213} (\bibinfo {year} {1958})}\BibitemShut
  {NoStop}%
\bibitem [{\citenamefont {Greenstein}\ and\ \citenamefont
  {Happel}(1968)}]{greenstein68}%
  \BibitemOpen
  \bibfield  {author} {\bibinfo {author} {\bibfnamefont {T.}~\bibnamefont
  {Greenstein}}\ and\ \bibinfo {author} {\bibfnamefont {J.}~\bibnamefont
  {Happel}},\ }\bibfield  {title} {\enquote {\bibinfo {title} {Theoretical
  study of the slow motion of a sphere and a fluid in a cylindrical tube},}\
  }\href@noop {} {\bibfield  {journal} {\bibinfo  {journal} {J. Fluid Mech.}\
  }\textbf {\bibinfo {volume} {34}},\ \bibinfo {pages} {705--710} (\bibinfo
  {year} {1968})}\BibitemShut {NoStop}%
\bibitem [{\citenamefont {Bungay}\ and\ \citenamefont
  {Brenner}(1973)}]{bungay73}%
  \BibitemOpen
  \bibfield  {author} {\bibinfo {author} {\bibfnamefont {P.~M.}\ \bibnamefont
  {Bungay}}\ and\ \bibinfo {author} {\bibfnamefont {H.}~\bibnamefont
  {Brenner}},\ }\bibfield  {title} {\enquote {\bibinfo {title} {The motion of a
  closely-fitting sphere in a fluid-filled tube},}\ }\href@noop {} {\bibfield
  {journal} {\bibinfo  {journal} {Int. J. Multiphase Flow}\ }\textbf {\bibinfo
  {volume} {1}},\ \bibinfo {pages} {25--56} (\bibinfo {year}
  {1973})}\BibitemShut {NoStop}%
\bibitem [{\citenamefont {Liron}\ and\ \citenamefont {Shahar}(1978)}]{liron78}%
  \BibitemOpen
  \bibfield  {author} {\bibinfo {author} {\bibfnamefont {N.}~\bibnamefont
  {Liron}}\ and\ \bibinfo {author} {\bibfnamefont {R.}~\bibnamefont {Shahar}},\
  }\bibfield  {title} {\enquote {\bibinfo {title} {Stokes flow due to a
  stokeslet in a pipe},}\ }\href@noop {} {\bibfield  {journal} {\bibinfo
  {journal} {J. Fluid Mech.}\ }\textbf {\bibinfo {volume} {86}},\ \bibinfo
  {pages} {727--744} (\bibinfo {year} {1978})}\BibitemShut {NoStop}%
\bibitem [{\citenamefont {Greenstein}\ and\ \citenamefont
  {Happel}(1970)}]{greenstein70}%
  \BibitemOpen
  \bibfield  {author} {\bibinfo {author} {\bibfnamefont {T.}~\bibnamefont
  {Greenstein}}\ and\ \bibinfo {author} {\bibfnamefont {J.}~\bibnamefont
  {Happel}},\ }\bibfield  {title} {\enquote {\bibinfo {title} {The slow motion
  of two particles symmetrically placed about the axis of a circular cylinder
  in a direction perpendicular to their line of centers},}\ }\href@noop {}
  {\bibfield  {journal} {\bibinfo  {journal} {Applied Scientific Research}\
  }\textbf {\bibinfo {volume} {22}},\ \bibinfo {pages} {345--359} (\bibinfo
  {year} {1970})}\BibitemShut {NoStop}%
\bibitem [{\citenamefont {Lecoq}\ \emph {et~al.}(1993)\citenamefont {Lecoq},
  \citenamefont {Feuillebois}, \citenamefont {Anthore}, \citenamefont
  {Anthore}, \citenamefont {Bostel},\ and\ \citenamefont
  {Petipas}}]{Lecoq1993}%
  \BibitemOpen
  \bibfield  {author} {\bibinfo {author} {\bibfnamefont {N.}~\bibnamefont
  {Lecoq}}, \bibinfo {author} {\bibfnamefont {F.}~\bibnamefont {Feuillebois}},
  \bibinfo {author} {\bibfnamefont {N.}~\bibnamefont {Anthore}}, \bibinfo
  {author} {\bibfnamefont {R.}~\bibnamefont {Anthore}}, \bibinfo {author}
  {\bibfnamefont {F.}~\bibnamefont {Bostel}}, \ and\ \bibinfo {author}
  {\bibfnamefont {C.}~\bibnamefont {Petipas}},\ }\bibfield  {title} {\enquote
  {\bibinfo {title} {Precise measurement of particle--wall hydrodynamic
  interactions at low reynolds number using laser interferometry},}\
  }\href@noop {} {\bibfield  {journal} {\bibinfo  {journal} {Phys. Fluids A}\
  }\textbf {\bibinfo {volume} {5}},\ \bibinfo {pages} {3--12} (\bibinfo {year}
  {1993})}\BibitemShut {NoStop}%
\bibitem [{\citenamefont {Higdon}\ and\ \citenamefont
  {Muldowney}(1995)}]{higdon95}%
  \BibitemOpen
  \bibfield  {author} {\bibinfo {author} {\bibfnamefont {J.~J.~L.}\
  \bibnamefont {Higdon}}\ and\ \bibinfo {author} {\bibfnamefont {G.~P.}\
  \bibnamefont {Muldowney}},\ }\bibfield  {title} {\enquote {\bibinfo {title}
  {Resistance functions for spherical particles, droplets and bubbles in
  cylindrical tubes},}\ }\href@noop {} {\bibfield  {journal} {\bibinfo
  {journal} {J. Fluid Mech.}\ }\textbf {\bibinfo {volume} {298}},\ \bibinfo
  {pages} {193--210} (\bibinfo {year} {1995})}\BibitemShut {NoStop}%
\bibitem [{\citenamefont {Pozrikidis}(2005)}]{pozrikidis05}%
  \BibitemOpen
  \bibfield  {author} {\bibinfo {author} {\bibfnamefont {C.}~\bibnamefont
  {Pozrikidis}},\ }\bibfield  {title} {\enquote {\bibinfo {title} {Computation
  of stokes flow due to the motion or presence of a particle in a tube},}\
  }\href@noop {} {\bibfield  {journal} {\bibinfo  {journal} {J. Eng. Math.}\
  }\textbf {\bibinfo {volume} {53}},\ \bibinfo {pages} {1--20} (\bibinfo {year}
  {2005})}\BibitemShut {NoStop}%
\bibitem [{\citenamefont {Keh}\ and\ \citenamefont {Chang}(2007)}]{keh07}%
  \BibitemOpen
  \bibfield  {author} {\bibinfo {author} {\bibfnamefont {H.~J.}\ \bibnamefont
  {Keh}}\ and\ \bibinfo {author} {\bibfnamefont {Y.~C.}\ \bibnamefont
  {Chang}},\ }\bibfield  {title} {\enquote {\bibinfo {title} {Creeping motion
  of a slip spherical particle in a circular cylindrical pore},}\ }\href@noop
  {} {\bibfield  {journal} {\bibinfo  {journal} {Int. J. Multiph. Flow}\
  }\textbf {\bibinfo {volume} {33}},\ \bibinfo {pages} {726--741} (\bibinfo
  {year} {2007})}\BibitemShut {NoStop}%
\bibitem [{\citenamefont {Bhattacharya}, \citenamefont {Mishra},\ and\
  \citenamefont {Bhattacharya}(2010)}]{Bhattacharya_2010}%
  \BibitemOpen
  \bibfield  {author} {\bibinfo {author} {\bibfnamefont {S.}~\bibnamefont
  {Bhattacharya}}, \bibinfo {author} {\bibfnamefont {C.}~\bibnamefont
  {Mishra}}, \ and\ \bibinfo {author} {\bibfnamefont {S.}~\bibnamefont
  {Bhattacharya}},\ }\bibfield  {title} {\enquote {\bibinfo {title} {{Analysis
  of general creeping motion of a sphere inside a cylinder}},}\ }\href@noop {}
  {\bibfield  {journal} {\bibinfo  {journal} {J. Fluid Mech.}\ }\textbf
  {\bibinfo {volume} {642}},\ \bibinfo {pages} {295} (\bibinfo {year}
  {2010})}\BibitemShut {NoStop}%
\bibitem [{\citenamefont {Imperio}, \citenamefont {Padding},\ and\
  \citenamefont {Briels}(2011)}]{imperio11}%
  \BibitemOpen
  \bibfield  {author} {\bibinfo {author} {\bibfnamefont {A.}~\bibnamefont
  {Imperio}}, \bibinfo {author} {\bibfnamefont {J.~T.}\ \bibnamefont
  {Padding}}, \ and\ \bibinfo {author} {\bibfnamefont {W.~J.}\ \bibnamefont
  {Briels}},\ }\bibfield  {title} {\enquote {\bibinfo {title} {Diffusion of
  spherical particles in microcavities},}\ }\href@noop {} {\bibfield  {journal}
  {\bibinfo  {journal} {J. Chem. Phys.}\ }\textbf {\bibinfo {volume} {134}},\
  \bibinfo {pages} {154904} (\bibinfo {year} {2011})}\BibitemShut {NoStop}%
\bibitem [{\citenamefont {Navardi}, \citenamefont {Bhattacharya},\ and\
  \citenamefont {Wu}(2015)}]{Navardi_2015}%
  \BibitemOpen
  \bibfield  {author} {\bibinfo {author} {\bibfnamefont {S.}~\bibnamefont
  {Navardi}}, \bibinfo {author} {\bibfnamefont {S.}~\bibnamefont
  {Bhattacharya}}, \ and\ \bibinfo {author} {\bibfnamefont {H.}~\bibnamefont
  {Wu}},\ }\bibfield  {title} {\enquote {\bibinfo {title} {{Stokesian
  simulation of two unequal spheres in a pressure-driven creeping flow through
  a cylinder}},}\ }\href@noop {} {\bibfield  {journal} {\bibinfo  {journal}
  {Comput. Fluids}\ }\textbf {\bibinfo {volume} {121}},\ \bibinfo {pages} {145}
  (\bibinfo {year} {2015})}\BibitemShut {NoStop}%
\bibitem [{\citenamefont {Hasimoto}(1976)}]{hasimoto76}%
  \BibitemOpen
  \bibfield  {author} {\bibinfo {author} {\bibfnamefont {H.}~\bibnamefont
  {Hasimoto}},\ }\bibfield  {title} {\enquote {\bibinfo {title} {Slow motion of
  a small sphere in a cylindrical domain},}\ }\href@noop {} {\bibfield
  {journal} {\bibinfo  {journal} {J. Phys. Soc. Japan}\ }\textbf {\bibinfo
  {volume} {41}},\ \bibinfo {pages} {2143--2144} (\bibinfo {year}
  {1976})}\BibitemShut {NoStop}%
\bibitem [{\citenamefont {Sano}(1987)}]{sano87}%
  \BibitemOpen
  \bibfield  {author} {\bibinfo {author} {\bibfnamefont {O.}~\bibnamefont
  {Sano}},\ }\bibfield  {title} {\enquote {\bibinfo {title} {Mobility of a
  small sphere in a viscous fluid confined in a rigid circular cylinder of
  finite length},}\ }\href@noop {} {\bibfield  {journal} {\bibinfo  {journal}
  {J. Phys. Soc. Japan}\ }\textbf {\bibinfo {volume} {56}},\ \bibinfo {pages}
  {2713--2720} (\bibinfo {year} {1987})}\BibitemShut {NoStop}%
\bibitem [{\citenamefont {Sheard}\ and\ \citenamefont {Ryan}(2007)}]{sheard07}%
  \BibitemOpen
  \bibfield  {author} {\bibinfo {author} {\bibfnamefont {G.~J.}\ \bibnamefont
  {Sheard}}\ and\ \bibinfo {author} {\bibfnamefont {K.}~\bibnamefont {Ryan}},\
  }\bibfield  {title} {\enquote {\bibinfo {title} {Pressure-driven flow past
  spheres moving in a circular tube},}\ }\href@noop {} {\bibfield  {journal}
  {\bibinfo  {journal} {J. Fluid Mech.}\ }\textbf {\bibinfo {volume} {592}},\
  \bibinfo {pages} {233--262} (\bibinfo {year} {2007})}\BibitemShut {NoStop}%
\bibitem [{\citenamefont {Felderhof}(2009)}]{felderhof09}%
  \BibitemOpen
  \bibfield  {author} {\bibinfo {author} {\bibfnamefont {B.~U.}\ \bibnamefont
  {Felderhof}},\ }\bibfield  {title} {\enquote {\bibinfo {title} {Transient
  flow of a viscous incompressible fluid in a circular tube after a sudden
  point impulse},}\ }\href@noop {} {\bibfield  {journal} {\bibinfo  {journal}
  {J. Fluid Mech.}\ }\textbf {\bibinfo {volume} {637}},\ \bibinfo {pages}
  {285--303} (\bibinfo {year} {2009})}\BibitemShut {NoStop}%
\bibitem [{\citenamefont {Felderhof}(2010{\natexlab{a}})}]{felderhof10}%
  \BibitemOpen
  \bibfield  {author} {\bibinfo {author} {\bibfnamefont {B.~U.}\ \bibnamefont
  {Felderhof}},\ }\bibfield  {title} {\enquote {\bibinfo {title} {Transient
  flow of a viscous compressible fluid in a circular tube after a sudden point
  impulse},}\ }\href@noop {} {\bibfield  {journal} {\bibinfo  {journal} {J.
  Fluid Mech.}\ }\textbf {\bibinfo {volume} {644}},\ \bibinfo {pages} {97--106}
  (\bibinfo {year} {2010}{\natexlab{a}})}\BibitemShut {NoStop}%
\bibitem [{\citenamefont {Felderhof}(2010{\natexlab{b}})}]{felderhof10b}%
  \BibitemOpen
  \bibfield  {author} {\bibinfo {author} {\bibfnamefont {B.~U.}\ \bibnamefont
  {Felderhof}},\ }\bibfield  {title} {\enquote {\bibinfo {title} {Transient
  flow of a viscous compressible fluid in a circular tube after a sudden point
  impulse transverse to the axis},}\ }\href@noop {} {\bibfield  {journal}
  {\bibinfo  {journal} {J. Fluid Mech.}\ }\textbf {\bibinfo {volume} {649}},\
  \bibinfo {pages} {329--340} (\bibinfo {year}
  {2010}{\natexlab{b}})}\BibitemShut {NoStop}%
\bibitem [{\citenamefont {Felderhof}\ and\ \citenamefont
  {Ooms}(2011)}]{felderhof11}%
  \BibitemOpen
  \bibfield  {author} {\bibinfo {author} {\bibfnamefont {B.~U.}\ \bibnamefont
  {Felderhof}}\ and\ \bibinfo {author} {\bibfnamefont {G.}~\bibnamefont
  {Ooms}},\ }\bibfield  {title} {\enquote {\bibinfo {title} {Flow of a viscous
  compressible fluid produced in a circular tube by an impulsive point
  source},}\ }\href@noop {} {\bibfield  {journal} {\bibinfo  {journal} {J.
  Fluid Mech.}\ }\textbf {\bibinfo {volume} {668}},\ \bibinfo {pages}
  {100--112} (\bibinfo {year} {2011})}\BibitemShut {NoStop}%
\bibitem [{\citenamefont {Rubinow}\ and\ \citenamefont
  {Keller}(1972)}]{Rubinow1972}%
  \BibitemOpen
  \bibfield  {author} {\bibinfo {author} {\bibfnamefont {S.}~\bibnamefont
  {Rubinow}}\ and\ \bibinfo {author} {\bibfnamefont {J.~B.}\ \bibnamefont
  {Keller}},\ }\bibfield  {title} {\enquote {\bibinfo {title} {Flow of a
  viscous fluid through an elastic tube with applications to blood flow},}\
  }\href@noop {} {\bibfield  {journal} {\bibinfo  {journal} {J. Theo. Biol.}\
  }\textbf {\bibinfo {volume} {35}},\ \bibinfo {pages} {299 -- 313} (\bibinfo
  {year} {1972})}\BibitemShut {NoStop}%
\bibitem [{\citenamefont {Fung}(2013)}]{fung13}%
  \BibitemOpen
  \bibfield  {author} {\bibinfo {author} {\bibfnamefont {Y.-C.}\ \bibnamefont
  {Fung}},\ }\href@noop {} {\emph {\bibinfo {title} {Biomechanics:
  circulation}}}\ (\bibinfo  {publisher} {Springer Science \& Business Media},\
  \bibinfo {year} {2013})\BibitemShut {NoStop}%
\bibitem [{\citenamefont {Bertram}, \citenamefont {Raymond},\ and\
  \citenamefont {Butcher}(1989)}]{bertram89}%
  \BibitemOpen
  \bibfield  {author} {\bibinfo {author} {\bibfnamefont {C.~D.}\ \bibnamefont
  {Bertram}}, \bibinfo {author} {\bibfnamefont {C.~J.}\ \bibnamefont
  {Raymond}}, \ and\ \bibinfo {author} {\bibfnamefont {K.~S.~A.}\ \bibnamefont
  {Butcher}},\ }\bibfield  {title} {\enquote {\bibinfo {title} {Oscillations in
  a collapsed-tube analog of the brachial artery under a sphygmomanometer
  cuff},}\ }\href@noop {} {\bibfield  {journal} {\bibinfo  {journal} {J.
  Biomech. Eng.}\ }\textbf {\bibinfo {volume} {111}},\ \bibinfo {pages}
  {185--191} (\bibinfo {year} {1989})}\BibitemShut {NoStop}%
\bibitem [{\citenamefont {Grotberg}(2001)}]{grotberg01}%
  \BibitemOpen
  \bibfield  {author} {\bibinfo {author} {\bibfnamefont {J.~B.}\ \bibnamefont
  {Grotberg}},\ }\bibfield  {title} {\enquote {\bibinfo {title} {Respiratory
  fluid mechanics and transport processes},}\ }\href@noop {} {\bibfield
  {journal} {\bibinfo  {journal} {Annu. Rev. Biomed. Eng.}\ }\textbf {\bibinfo
  {volume} {3}},\ \bibinfo {pages} {421--457} (\bibinfo {year}
  {2001})}\BibitemShut {NoStop}%
\bibitem [{\citenamefont {Grotberg}\ and\ \citenamefont
  {Jensen}(2004)}]{grotberg04}%
  \BibitemOpen
  \bibfield  {author} {\bibinfo {author} {\bibfnamefont {J.~B.}\ \bibnamefont
  {Grotberg}}\ and\ \bibinfo {author} {\bibfnamefont {O.~E.}\ \bibnamefont
  {Jensen}},\ }\bibfield  {title} {\enquote {\bibinfo {title} {Biofluid
  mechanics in flexible tubes},}\ }\href@noop {} {\bibfield  {journal}
  {\bibinfo  {journal} {Ann. Rev. Fluid Mech.}\ }\textbf {\bibinfo {volume}
  {36}},\ \bibinfo {pages} {121} (\bibinfo {year} {2004})}\BibitemShut
  {NoStop}%
\bibitem [{\citenamefont {Canic}\ \emph {et~al.}(2006)\citenamefont {Canic},
  \citenamefont {Tambaca}, \citenamefont {Guidoboni}, \citenamefont {Mikelic},
  \citenamefont {Hartley},\ and\ \citenamefont {Rosenstrauch}}]{canic06}%
  \BibitemOpen
  \bibfield  {author} {\bibinfo {author} {\bibfnamefont {S.}~\bibnamefont
  {Canic}}, \bibinfo {author} {\bibfnamefont {J.}~\bibnamefont {Tambaca}},
  \bibinfo {author} {\bibfnamefont {G.}~\bibnamefont {Guidoboni}}, \bibinfo
  {author} {\bibfnamefont {A.}~\bibnamefont {Mikelic}}, \bibinfo {author}
  {\bibfnamefont {C.~J.}\ \bibnamefont {Hartley}}, \ and\ \bibinfo {author}
  {\bibfnamefont {D.}~\bibnamefont {Rosenstrauch}},\ }\bibfield  {title}
  {\enquote {\bibinfo {title} {Modeling viscoelastic behavior of arterial walls
  and their interaction with pulsatile blood flow},}\ }\href@noop {} {\bibfield
   {journal} {\bibinfo  {journal} {SIAM J. Appl. Math.}\ }\textbf {\bibinfo
  {volume} {67}},\ \bibinfo {pages} {164--193} (\bibinfo {year}
  {2006})}\BibitemShut {NoStop}%
\bibitem [{\citenamefont {Takagi}\ and\ \citenamefont
  {Balmforth}(2011)}]{Takagi2011}%
  \BibitemOpen
  \bibfield  {author} {\bibinfo {author} {\bibfnamefont {D.}~\bibnamefont
  {Takagi}}\ and\ \bibinfo {author} {\bibfnamefont {N.~J.}\ \bibnamefont
  {Balmforth}},\ }\bibfield  {title} {\enquote {\bibinfo {title} {Peristaltic
  pumping of viscous fluid in an elastic tube},}\ }\href@noop {} {\bibfield
  {journal} {\bibinfo  {journal} {J. Fluid Mech.}\ }\textbf {\bibinfo {volume}
  {672}},\ \bibinfo {pages} {196–218} (\bibinfo {year} {2011})}\BibitemShut
  {NoStop}%
\bibitem [{\citenamefont {Heap}\ and\ \citenamefont {Juel}(2008)}]{heap08}%
  \BibitemOpen
  \bibfield  {author} {\bibinfo {author} {\bibfnamefont {A.}~\bibnamefont
  {Heap}}\ and\ \bibinfo {author} {\bibfnamefont {A.}~\bibnamefont {Juel}},\
  }\bibfield  {title} {\enquote {\bibinfo {title} {Anomalous bubble propagation
  in elastic tubes},}\ }\href@noop {} {\bibfield  {journal} {\bibinfo
  {journal} {Phys. Fluids}\ }\textbf {\bibinfo {volume} {20}},\ \bibinfo
  {pages} {081702} (\bibinfo {year} {2008})}\BibitemShut {NoStop}%
\bibitem [{\citenamefont {Heap}\ and\ \citenamefont {Juel}(2009)}]{heap09}%
  \BibitemOpen
  \bibfield  {author} {\bibinfo {author} {\bibfnamefont {A.}~\bibnamefont
  {Heap}}\ and\ \bibinfo {author} {\bibfnamefont {A.}~\bibnamefont {Juel}},\
  }\bibfield  {title} {\enquote {\bibinfo {title} {Bubble transitions in
  strongly collapsed elastic tubes},}\ }\href@noop {} {\bibfield  {journal}
  {\bibinfo  {journal} {J. Fluid Mech.}\ }\textbf {\bibinfo {volume} {633}},\
  \bibinfo {pages} {485--507} (\bibinfo {year} {2009})}\BibitemShut {NoStop}%
\bibitem [{\citenamefont {Zheng}, \citenamefont {Phan-Thien},\ and\
  \citenamefont {Tanner}(1991)}]{zheng91}%
  \BibitemOpen
  \bibfield  {author} {\bibinfo {author} {\bibfnamefont {R.}~\bibnamefont
  {Zheng}}, \bibinfo {author} {\bibfnamefont {N.}~\bibnamefont {Phan-Thien}}, \
  and\ \bibinfo {author} {\bibfnamefont {R.~I.}\ \bibnamefont {Tanner}},\
  }\bibfield  {title} {\enquote {\bibinfo {title} {The flow past a sphere in a
  cylindrical tube: effects of intertia, shear-thinning and elasticity},}\
  }\href@noop {} {\bibfield  {journal} {\bibinfo  {journal} {Rheol. Acta}\
  }\textbf {\bibinfo {volume} {30}},\ \bibinfo {pages} {499--510} (\bibinfo
  {year} {1991})}\BibitemShut {NoStop}%
\bibitem [{\citenamefont {Nahar}, \citenamefont {Jeelani},\ and\ \citenamefont
  {Windhab}(2012)}]{nahar12}%
  \BibitemOpen
  \bibfield  {author} {\bibinfo {author} {\bibfnamefont {S.}~\bibnamefont
  {Nahar}}, \bibinfo {author} {\bibfnamefont {S.~A.~K.}\ \bibnamefont
  {Jeelani}}, \ and\ \bibinfo {author} {\bibfnamefont {E.~J.}\ \bibnamefont
  {Windhab}},\ }\bibfield  {title} {\enquote {\bibinfo {title} {Influence of
  elastic tube deformation on flow behavior of a shear thinning fluid},}\
  }\href@noop {} {\bibfield  {journal} {\bibinfo  {journal} {Chem. Eng. Sci.}\
  }\textbf {\bibinfo {volume} {75}},\ \bibinfo {pages} {445--455} (\bibinfo
  {year} {2012})}\BibitemShut {NoStop}%
\bibitem [{\citenamefont {Nahar}, \citenamefont {Jeelani},\ and\ \citenamefont
  {Windhab}(2013)}]{nahar13}%
  \BibitemOpen
  \bibfield  {author} {\bibinfo {author} {\bibfnamefont {S.}~\bibnamefont
  {Nahar}}, \bibinfo {author} {\bibfnamefont {S.~A.~K.}\ \bibnamefont
  {Jeelani}}, \ and\ \bibinfo {author} {\bibfnamefont {E.~J.}\ \bibnamefont
  {Windhab}},\ }\bibfield  {title} {\enquote {\bibinfo {title} {Prediction of
  velocity profiles of shear thinning fluids flowing in elastic tubes},}\
  }\href@noop {} {\bibfield  {journal} {\bibinfo  {journal} {Chem. Eng. Comm.}\
  }\textbf {\bibinfo {volume} {200}},\ \bibinfo {pages} {820--835} (\bibinfo
  {year} {2013})}\BibitemShut {NoStop}%
\bibitem [{\citenamefont {Mikelic}, \citenamefont {Guidoboni},\ and\
  \citenamefont {Canic}(2007)}]{mikelic07}%
  \BibitemOpen
  \bibfield  {author} {\bibinfo {author} {\bibfnamefont {A.}~\bibnamefont
  {Mikelic}}, \bibinfo {author} {\bibfnamefont {G.}~\bibnamefont {Guidoboni}},
  \ and\ \bibinfo {author} {\bibfnamefont {S.}~\bibnamefont {Canic}},\
  }\bibfield  {title} {\enquote {\bibinfo {title} {Fluid-structure interaction
  in a pre-stressed tube with thick elastic walls i: the stationary stokes
  problem},}\ }\href@noop {} {\bibfield  {journal} {\bibinfo  {journal} {Netw.
  Heterog. Media}\ }\textbf {\bibinfo {volume} {2}},\ \bibinfo {pages} {397}
  (\bibinfo {year} {2007})}\BibitemShut {NoStop}%
\bibitem [{\citenamefont {Marzo}, \citenamefont {Luo},\ and\ \citenamefont
  {Bertram}(2005)}]{marzo05}%
  \BibitemOpen
  \bibfield  {author} {\bibinfo {author} {\bibfnamefont {A.}~\bibnamefont
  {Marzo}}, \bibinfo {author} {\bibfnamefont {X.~Y.}\ \bibnamefont {Luo}}, \
  and\ \bibinfo {author} {\bibfnamefont {C.~D.}\ \bibnamefont {Bertram}},\
  }\bibfield  {title} {\enquote {\bibinfo {title} {Three-dimensional collapse
  and steady flow in thick-walled flexible tubes},}\ }\href@noop {} {\bibfield
  {journal} {\bibinfo  {journal} {J. Fluid Struct.}\ }\textbf {\bibinfo
  {volume} {20}},\ \bibinfo {pages} {817--835} (\bibinfo {year}
  {2005})}\BibitemShut {NoStop}%
\bibitem [{\citenamefont {Alexander}(1971)}]{alexander71}%
  \BibitemOpen
  \bibfield  {author} {\bibinfo {author} {\bibfnamefont {H.}~\bibnamefont
  {Alexander}},\ }\bibfield  {title} {\enquote {\bibinfo {title} {The tensile
  instability of an inflated cylindrical membrane as affected by an axial
  load},}\ }\href@noop {} {\bibfield  {journal} {\bibinfo  {journal} {Int. J.
  Mech. Sci.}\ }\textbf {\bibinfo {volume} {13}},\ \bibinfo {pages} {87--94}
  (\bibinfo {year} {1971})}\BibitemShut {NoStop}%
\bibitem [{\citenamefont {Shan}\ \emph {et~al.}(2006)\citenamefont {Shan},
  \citenamefont {Philen}, \citenamefont {Bakis}, \citenamefont {Wang},\ and\
  \citenamefont {Rahn}}]{shan06}%
  \BibitemOpen
  \bibfield  {author} {\bibinfo {author} {\bibfnamefont {Y.}~\bibnamefont
  {Shan}}, \bibinfo {author} {\bibfnamefont {M.~P.}\ \bibnamefont {Philen}},
  \bibinfo {author} {\bibfnamefont {C.~E.}\ \bibnamefont {Bakis}}, \bibinfo
  {author} {\bibfnamefont {K.-W.}\ \bibnamefont {Wang}}, \ and\ \bibinfo
  {author} {\bibfnamefont {C.~D.}\ \bibnamefont {Rahn}},\ }\bibfield  {title}
  {\enquote {\bibinfo {title} {Nonlinear-elastic finite axisymmetric
  deformation of flexible matrix composite membranes under internal pressure
  and axial force},}\ }\href@noop {} {\bibfield  {journal} {\bibinfo  {journal}
  {Compos. Sci. Tech.}\ }\textbf {\bibinfo {volume} {66}},\ \bibinfo {pages}
  {3053--3063} (\bibinfo {year} {2006})}\BibitemShut {NoStop}%
\bibitem [{\citenamefont {Rahimi}, \citenamefont {DeSimone},\ and\
  \citenamefont {Arroyo}(2013)}]{rahimi13}%
  \BibitemOpen
  \bibfield  {author} {\bibinfo {author} {\bibfnamefont {M.}~\bibnamefont
  {Rahimi}}, \bibinfo {author} {\bibfnamefont {A.}~\bibnamefont {DeSimone}}, \
  and\ \bibinfo {author} {\bibfnamefont {M.}~\bibnamefont {Arroyo}},\
  }\bibfield  {title} {\enquote {\bibinfo {title} {Curved fluid membranes
  behave laterally as effective viscoelastic media},}\ }\href@noop {}
  {\bibfield  {journal} {\bibinfo  {journal} {Soft Matter}\ }\textbf {\bibinfo
  {volume} {9}},\ \bibinfo {pages} {11033--11045} (\bibinfo {year}
  {2013})}\BibitemShut {NoStop}%
\bibitem [{\citenamefont {Rahimi}(2013)}]{rahimi13thesis}%
  \BibitemOpen
  \bibfield  {author} {\bibinfo {author} {\bibfnamefont {M.}~\bibnamefont
  {Rahimi}},\ }\emph {\bibinfo {title} {Shape dynamics and lipid hydrodynamics
  of bilayer membranes: modeling, simulation and experiments}},\ \href@noop {}
  {Ph.D. thesis} (\bibinfo {year} {2013})\BibitemShut {NoStop}%
\bibitem [{\citenamefont {Lighthill}(1968)}]{Lighthill_1968}%
  \BibitemOpen
  \bibfield  {author} {\bibinfo {author} {\bibfnamefont {M.~J.}\ \bibnamefont
  {Lighthill}},\ }\bibfield  {title} {\enquote {\bibinfo {title}
  {{Pressure-forcing of tightly fitting pellets along fluid-filled elastic
  tubes}},}\ }\href@noop {} {\bibfield  {journal} {\bibinfo  {journal} {J Fluid
  Mech}\ }\textbf {\bibinfo {volume} {34}},\ \bibinfo {pages} {113--143}
  (\bibinfo {year} {1968})}\BibitemShut {NoStop}%
\bibitem [{\citenamefont {T{\"o}zeren}, \citenamefont {{\"O}zkaya},\ and\
  \citenamefont {T{\"o}zeren}(1982)}]{Tozeren_1982}%
  \BibitemOpen
  \bibfield  {author} {\bibinfo {author} {\bibfnamefont {A.}~\bibnamefont
  {T{\"o}zeren}}, \bibinfo {author} {\bibfnamefont {N.}~\bibnamefont
  {{\"O}zkaya}}, \ and\ \bibinfo {author} {\bibfnamefont {H.}~\bibnamefont
  {T{\"o}zeren}},\ }\bibfield  {title} {\enquote {\bibinfo {title} {{Flow of
  Particles Along a Deformable Tube}},}\ }\href@noop {} {\bibfield  {journal}
  {\bibinfo  {journal} {J Biomechanics}\ }\textbf {\bibinfo {volume} {15}},\
  \bibinfo {pages} {517--527} (\bibinfo {year} {1982})}\BibitemShut {NoStop}%
\bibitem [{\citenamefont {Daddi-Moussa-Ider}, \citenamefont {Lisicki},\ and\
  \citenamefont {Gekle}(2017{\natexlab{a}})}]{daddi17d}%
  \BibitemOpen
  \bibfield  {author} {\bibinfo {author} {\bibfnamefont {A.}~\bibnamefont
  {Daddi-Moussa-Ider}}, \bibinfo {author} {\bibfnamefont {M.}~\bibnamefont
  {Lisicki}}, \ and\ \bibinfo {author} {\bibfnamefont {S.}~\bibnamefont
  {Gekle}},\ }\bibfield  {title} {\enquote {\bibinfo {title} {Slow rotation of
  a spherical particle inside an elastic tube},}\ }\href@noop {} {\bibfield
  {journal} {\bibinfo  {journal} {Acta Mech. (accepted)}\ } (\bibinfo {year}
  {2017}{\natexlab{a}})}\BibitemShut {NoStop}%
\bibitem [{\citenamefont {Leal}(1980)}]{leal80}%
  \BibitemOpen
  \bibfield  {author} {\bibinfo {author} {\bibfnamefont {L.~G.}\ \bibnamefont
  {Leal}},\ }\bibfield  {title} {\enquote {\bibinfo {title} {Particle motions
  in a viscous fluid},}\ }\href@noop {} {\bibfield  {journal} {\bibinfo
  {journal} {Ann. Rev. Fluid Mech.}\ }\textbf {\bibinfo {volume} {12}},\
  \bibinfo {pages} {435--476} (\bibinfo {year} {1980})}\BibitemShut {NoStop}%
\bibitem [{\citenamefont {Tan}, \citenamefont {Thomas},\ and\ \citenamefont
  {Liu}(2012)}]{tan12}%
  \BibitemOpen
  \bibfield  {author} {\bibinfo {author} {\bibfnamefont {J.}~\bibnamefont
  {Tan}}, \bibinfo {author} {\bibfnamefont {A.}~\bibnamefont {Thomas}}, \ and\
  \bibinfo {author} {\bibfnamefont {Y.}~\bibnamefont {Liu}},\ }\bibfield
  {title} {\enquote {\bibinfo {title} {Influence of red blood cells on
  nanoparticle targeted delivery in microcirculation},}\ }\href@noop {}
  {\bibfield  {journal} {\bibinfo  {journal} {Soft Matter}\ }\textbf {\bibinfo
  {volume} {8}},\ \bibinfo {pages} {1934--1946} (\bibinfo {year}
  {2012})}\BibitemShut {NoStop}%
\bibitem [{\citenamefont {Daddi-Moussa-Ider}, \citenamefont {Guckenberger},\
  and\ \citenamefont {Gekle}(2016{\natexlab{a}})}]{daddi16}%
  \BibitemOpen
  \bibfield  {author} {\bibinfo {author} {\bibfnamefont {A.}~\bibnamefont
  {Daddi-Moussa-Ider}}, \bibinfo {author} {\bibfnamefont {A.}~\bibnamefont
  {Guckenberger}}, \ and\ \bibinfo {author} {\bibfnamefont {S.}~\bibnamefont
  {Gekle}},\ }\bibfield  {title} {\enquote {\bibinfo {title} {Long-lived
  anomalous thermal diffusion induced by elastic cell membranes on nearby
  particles},}\ }\href@noop {} {\bibfield  {journal} {\bibinfo  {journal}
  {Phys. Rev. E}\ }\textbf {\bibinfo {volume} {93}},\ \bibinfo {pages} {012612}
  (\bibinfo {year} {2016}{\natexlab{a}})}\BibitemShut {NoStop}%
\bibitem [{\citenamefont {Daddi-Moussa-Ider}, \citenamefont {Guckenberger},\
  and\ \citenamefont {Gekle}(2016{\natexlab{b}})}]{daddi16b}%
  \BibitemOpen
  \bibfield  {author} {\bibinfo {author} {\bibfnamefont {A.}~\bibnamefont
  {Daddi-Moussa-Ider}}, \bibinfo {author} {\bibfnamefont {A.}~\bibnamefont
  {Guckenberger}}, \ and\ \bibinfo {author} {\bibfnamefont {S.}~\bibnamefont
  {Gekle}},\ }\bibfield  {title} {\enquote {\bibinfo {title} {Particle mobility
  between two planar elastic membranes: {B}rownian motion and membrane
  deformation},}\ }\href@noop {} {\bibfield  {journal} {\bibinfo  {journal}
  {Phys. Fluids}\ }\textbf {\bibinfo {volume} {28}},\ \bibinfo {pages} {071903}
  (\bibinfo {year} {2016}{\natexlab{b}})}\BibitemShut {NoStop}%
\bibitem [{\citenamefont {Daddi-Moussa-Ider}\ and\ \citenamefont
  {Gekle}(2016)}]{daddi16c}%
  \BibitemOpen
  \bibfield  {author} {\bibinfo {author} {\bibfnamefont {A.}~\bibnamefont
  {Daddi-Moussa-Ider}}\ and\ \bibinfo {author} {\bibfnamefont {S.}~\bibnamefont
  {Gekle}},\ }\bibfield  {title} {\enquote {\bibinfo {title} {Hydrodynamic
  interaction between particles near elastic interfaces},}\ }\href@noop {}
  {\bibfield  {journal} {\bibinfo  {journal} {J. Chem. Phys.}\ }\textbf
  {\bibinfo {volume} {145}},\ \bibinfo {pages} {014905} (\bibinfo {year}
  {2016})}\BibitemShut {NoStop}%
\bibitem [{\citenamefont {Ramanujan}\ and\ \citenamefont
  {Pozrikidis}(1998)}]{ramanujan98}%
  \BibitemOpen
  \bibfield  {author} {\bibinfo {author} {\bibfnamefont {S.}~\bibnamefont
  {Ramanujan}}\ and\ \bibinfo {author} {\bibfnamefont {C.}~\bibnamefont
  {Pozrikidis}},\ }\bibfield  {title} {\enquote {\bibinfo {title} {Deformation
  of liquid capsules enclosed by elastic membranes in simple shear flow: large
  deformations and the effect of fluid viscosities},}\ }\href@noop {}
  {\bibfield  {journal} {\bibinfo  {journal} {J. Fluid Mech.}\ }\textbf
  {\bibinfo {volume} {361}},\ \bibinfo {pages} {117--143} (\bibinfo {year}
  {1998})}\BibitemShut {NoStop}%
\bibitem [{\citenamefont {Lac}\ \emph {et~al.}(2004)\citenamefont {Lac},
  \citenamefont {Barth\`{e}s-Biesel}, \citenamefont {Pelekasis},\ and\
  \citenamefont {Tsamopoulos}}]{lac04}%
  \BibitemOpen
  \bibfield  {author} {\bibinfo {author} {\bibfnamefont {E.}~\bibnamefont
  {Lac}}, \bibinfo {author} {\bibfnamefont {D.}~\bibnamefont
  {Barth\`{e}s-Biesel}}, \bibinfo {author} {\bibfnamefont {N.~A.}\ \bibnamefont
  {Pelekasis}}, \ and\ \bibinfo {author} {\bibfnamefont {J.}~\bibnamefont
  {Tsamopoulos}},\ }\bibfield  {title} {\enquote {\bibinfo {title} {Spherical
  capsules in three-dimensional unbounded {S}tokes flows: effect of the
  membrane constitutive law and onset of buckling},}\ }\href@noop {} {\bibfield
   {journal} {\bibinfo  {journal} {J. Fluid Mech.}\ }\textbf {\bibinfo {volume}
  {516}},\ \bibinfo {pages} {303--334} (\bibinfo {year} {2004})}\BibitemShut
  {NoStop}%
\bibitem [{\citenamefont {Freund}(2014)}]{Freund_2014}%
  \BibitemOpen
  \bibfield  {author} {\bibinfo {author} {\bibfnamefont {J.~B.}\ \bibnamefont
  {Freund}},\ }\bibfield  {title} {\enquote {\bibinfo {title} {Numerical
  simulation of flowing blood cells},}\ }\href@noop {} {\bibfield  {journal}
  {\bibinfo  {journal} {Annu. Rev. Fluid Mech.}\ }\textbf {\bibinfo {volume}
  {46}},\ \bibinfo {pages} {67--95} (\bibinfo {year} {2014})}\BibitemShut
  {NoStop}%
\bibitem [{\citenamefont {Barth{\`e}s-Biesel}(2016)}]{BarthesBiesel_2016}%
  \BibitemOpen
  \bibfield  {author} {\bibinfo {author} {\bibfnamefont {D.}~\bibnamefont
  {Barth{\`e}s-Biesel}},\ }\bibfield  {title} {\enquote {\bibinfo {title}
  {{Motion and Deformation of Elastic Capsules and Vesicles in Flow}},}\
  }\href@noop {} {\bibfield  {journal} {\bibinfo  {journal} {Annu. Rev. Fluid
  Mech.}\ }\textbf {\bibinfo {volume} {48}},\ \bibinfo {pages} {25--52}
  (\bibinfo {year} {2016})}\BibitemShut {NoStop}%
\bibitem [{\citenamefont {Helfrich}(1973)}]{helfrich73}%
  \BibitemOpen
  \bibfield  {author} {\bibinfo {author} {\bibfnamefont {W.}~\bibnamefont
  {Helfrich}},\ }\bibfield  {title} {\enquote {\bibinfo {title} {Elastic
  properties of lipid bilayers - theory and possible experiments},}\
  }\href@noop {} {\bibfield  {journal} {\bibinfo  {journal} {Z. Naturf. C.}\
  }\textbf {\bibinfo {volume} {28}},\ \bibinfo {pages} {693} (\bibinfo {year}
  {1973})}\BibitemShut {NoStop}%
\bibitem [{\citenamefont {Guckenberger}\ and\ \citenamefont
  {Gekle}(2017)}]{Guckenberger_preprint}%
  \BibitemOpen
  \bibfield  {author} {\bibinfo {author} {\bibfnamefont {A.}~\bibnamefont
  {Guckenberger}}\ and\ \bibinfo {author} {\bibfnamefont {S.}~\bibnamefont
  {Gekle}},\ }\bibfield  {title} {\enquote {\bibinfo {title} {Theory and
  algorithms to compute helfrich bending forces: a review},}\ }\href@noop {}
  {\bibfield  {journal} {\bibinfo  {journal} {J. Phys. Condens. Matter}\
  }\textbf {\bibinfo {volume} {29}},\ \bibinfo {pages} {203001} (\bibinfo
  {year} {2017})}\BibitemShut {NoStop}%
\bibitem [{\citenamefont {Kim}\ and\ \citenamefont {Karrila}(2013)}]{kim13}%
  \BibitemOpen
  \bibfield  {author} {\bibinfo {author} {\bibfnamefont {S.}~\bibnamefont
  {Kim}}\ and\ \bibinfo {author} {\bibfnamefont {S.~J.}\ \bibnamefont
  {Karrila}},\ }\href@noop {} {\emph {\bibinfo {title} {Microhydrodynamics:
  principles and selected applications}}}\ (\bibinfo  {publisher} {Courier
  Corporation},\ \bibinfo {year} {2013})\BibitemShut {NoStop}%
\bibitem [{\citenamefont {Bickel}(2006)}]{bickel06}%
  \BibitemOpen
  \bibfield  {author} {\bibinfo {author} {\bibfnamefont {T.}~\bibnamefont
  {Bickel}},\ }\bibfield  {title} {\enquote {\bibinfo {title} {Brownian motion
  near a liquid-like membrane},}\ }\href@noop {} {\bibfield  {journal}
  {\bibinfo  {journal} {Eur. Phys. J. E}\ }\textbf {\bibinfo {volume} {20}},\
  \bibinfo {pages} {379--385} (\bibinfo {year} {2006})}\BibitemShut {NoStop}%
\bibitem [{\citenamefont {Weekley}, \citenamefont {Waters},\ and\ \citenamefont
  {Jensen}(2006)}]{weekley06}%
  \BibitemOpen
  \bibfield  {author} {\bibinfo {author} {\bibfnamefont {S.~J.}\ \bibnamefont
  {Weekley}}, \bibinfo {author} {\bibfnamefont {S.~L.}\ \bibnamefont {Waters}},
  \ and\ \bibinfo {author} {\bibfnamefont {O.~E.}\ \bibnamefont {Jensen}},\
  }\bibfield  {title} {\enquote {\bibinfo {title} {Transient elastohydrodynamic
  drag on a particle moving near a deformable wall},}\ }\href@noop {}
  {\bibfield  {journal} {\bibinfo  {journal} {Q. J. Mech. Appl. Math.}\
  }\textbf {\bibinfo {volume} {59}},\ \bibinfo {pages} {277--300} (\bibinfo
  {year} {2006})}\BibitemShut {NoStop}%
\bibitem [{\citenamefont {Salez}\ and\ \citenamefont
  {Mahadevan}(2015)}]{salez15}%
  \BibitemOpen
  \bibfield  {author} {\bibinfo {author} {\bibfnamefont {T.}~\bibnamefont
  {Salez}}\ and\ \bibinfo {author} {\bibfnamefont {L.}~\bibnamefont
  {Mahadevan}},\ }\bibfield  {title} {\enquote {\bibinfo {title}
  {Elastohydrodynamics of a sliding, spinning and sedimenting cylinder near a
  soft wall},}\ }\href@noop {} {\bibfield  {journal} {\bibinfo  {journal} {J.
  Fluid Mech.}\ }\textbf {\bibinfo {volume} {779}},\ \bibinfo {pages}
  {181--196} (\bibinfo {year} {2015})}\BibitemShut {NoStop}%
\bibitem [{\citenamefont {Saintyves}\ \emph {et~al.}(2016)\citenamefont
  {Saintyves}, \citenamefont {Jules}, \citenamefont {Salez},\ and\
  \citenamefont {Mahadevan}}]{saintyves16}%
  \BibitemOpen
  \bibfield  {author} {\bibinfo {author} {\bibfnamefont {B.}~\bibnamefont
  {Saintyves}}, \bibinfo {author} {\bibfnamefont {T.}~\bibnamefont {Jules}},
  \bibinfo {author} {\bibfnamefont {T.}~\bibnamefont {Salez}}, \ and\ \bibinfo
  {author} {\bibfnamefont {L.}~\bibnamefont {Mahadevan}},\ }\bibfield  {title}
  {\enquote {\bibinfo {title} {Self-sustained lift and low friction via soft
  lubrication},}\ }\href@noop {} {\bibfield  {journal} {\bibinfo  {journal}
  {Proc. Nat. Acad. Sci.}\ }\textbf {\bibinfo {volume} {113}},\ \bibinfo
  {pages} {5847--5849} (\bibinfo {year} {2016})}\BibitemShut {NoStop}%
\bibitem [{\citenamefont {Rallabandi}\ \emph {et~al.}(2017)\citenamefont
  {Rallabandi}, \citenamefont {Saintyves}, \citenamefont {Jules}, \citenamefont
  {Salez}, \citenamefont {Sch{\"o}necker}, \citenamefont {Mahadevan},\ and\
  \citenamefont {Stone}}]{rallabandi17}%
  \BibitemOpen
  \bibfield  {author} {\bibinfo {author} {\bibfnamefont {B.}~\bibnamefont
  {Rallabandi}}, \bibinfo {author} {\bibfnamefont {B.}~\bibnamefont
  {Saintyves}}, \bibinfo {author} {\bibfnamefont {T.}~\bibnamefont {Jules}},
  \bibinfo {author} {\bibfnamefont {T.}~\bibnamefont {Salez}}, \bibinfo
  {author} {\bibfnamefont {C.}~\bibnamefont {Sch{\"o}necker}}, \bibinfo
  {author} {\bibfnamefont {L.}~\bibnamefont {Mahadevan}}, \ and\ \bibinfo
  {author} {\bibfnamefont {H.~A.}\ \bibnamefont {Stone}},\ }\bibfield  {title}
  {\enquote {\bibinfo {title} {Rotation of an immersed cylinder sliding near a
  thin elastic coating},}\ }\href@noop {} {\bibfield  {journal} {\bibinfo
  {journal} {Phys. Rev. Fluids}\ }\textbf {\bibinfo {volume} {2}},\ \bibinfo
  {pages} {074102} (\bibinfo {year} {2017})}\BibitemShut {NoStop}%
\bibitem [{\citenamefont {Fuentes}, \citenamefont {Kim},\ and\ \citenamefont
  {Jeffrey}(1988)}]{fuentes88}%
  \BibitemOpen
  \bibfield  {author} {\bibinfo {author} {\bibfnamefont {Y.~O.}\ \bibnamefont
  {Fuentes}}, \bibinfo {author} {\bibfnamefont {S.}~\bibnamefont {Kim}}, \ and\
  \bibinfo {author} {\bibfnamefont {D.~J.}\ \bibnamefont {Jeffrey}},\
  }\bibfield  {title} {\enquote {\bibinfo {title} {Mobility functions for two
  unequal viscous drops in {S}tokes flow. i. axisymmetric motions},}\
  }\href@noop {} {\bibfield  {journal} {\bibinfo  {journal} {Phys. Fluids}\
  }\textbf {\bibinfo {volume} {31}},\ \bibinfo {pages} {2445--2455} (\bibinfo
  {year} {1988})}\BibitemShut {NoStop}%
\bibitem [{\citenamefont {Fuentes}, \citenamefont {Kim},\ and\ \citenamefont
  {Jeffrey}(1989)}]{fuentes89}%
  \BibitemOpen
  \bibfield  {author} {\bibinfo {author} {\bibfnamefont {Y.~O.}\ \bibnamefont
  {Fuentes}}, \bibinfo {author} {\bibfnamefont {S.}~\bibnamefont {Kim}}, \ and\
  \bibinfo {author} {\bibfnamefont {D.~J.}\ \bibnamefont {Jeffrey}},\
  }\bibfield  {title} {\enquote {\bibinfo {title} {Mobility functions for two
  unequal viscous drops in {S}tokes flow. ii. asymmetric motions},}\
  }\href@noop {} {\bibfield  {journal} {\bibinfo  {journal} {Phys. Fluids A}\
  }\textbf {\bibinfo {volume} {1}},\ \bibinfo {pages} {61--76} (\bibinfo {year}
  {1989})}\BibitemShut {NoStop}%
\bibitem [{\citenamefont {Watson}(1995)}]{watson95}%
  \BibitemOpen
  \bibfield  {author} {\bibinfo {author} {\bibfnamefont {G.~N.}\ \bibnamefont
  {Watson}},\ }\href@noop {} {\emph {\bibinfo {title} {A treatise on the theory
  of Bessel functions}}}\ (\bibinfo  {publisher} {Cambridge university press},\
  \bibinfo {year} {1995})\BibitemShut {NoStop}%
\bibitem [{\citenamefont {Abramowitz}, \citenamefont {Stegun}\ \emph
  {et~al.}(1972)\citenamefont {Abramowitz}, \citenamefont {Stegun} \emph
  {et~al.}}]{abramowitz72}%
  \BibitemOpen
  \bibfield  {author} {\bibinfo {author} {\bibfnamefont {M.}~\bibnamefont
  {Abramowitz}}, \bibinfo {author} {\bibfnamefont {I.~A.}\ \bibnamefont
  {Stegun}},  \emph {et~al.},\ }\href@noop {} {\emph {\bibinfo {title}
  {Handbook of mathematical functions}}},\ Vol.~\bibinfo {volume} {1}\
  (\bibinfo  {publisher} {Dover New York},\ \bibinfo {year} {1972})\BibitemShut
  {NoStop}%
\bibitem [{\citenamefont {Daddi-Moussa-Ider}\ and\ \citenamefont
  {Gekle}(2017)}]{daddi17b}%
  \BibitemOpen
  \bibfield  {author} {\bibinfo {author} {\bibfnamefont {A.}~\bibnamefont
  {Daddi-Moussa-Ider}}\ and\ \bibinfo {author} {\bibfnamefont {S.}~\bibnamefont
  {Gekle}},\ }\bibfield  {title} {\enquote {\bibinfo {title} {Hydrodynamic
  mobility of a solid particle near a spherical elastic membrane: Axisymmetric
  motion},}\ }\href@noop {} {\bibfield  {journal} {\bibinfo  {journal} {Phys.
  Rev. E}\ }\textbf {\bibinfo {volume} {95}},\ \bibinfo {pages} {013108}
  (\bibinfo {year} {2017})}\BibitemShut {NoStop}%
\bibitem [{\citenamefont {Daddi-Moussa-Ider}, \citenamefont {Lisicki},\ and\
  \citenamefont {Gekle}(2017{\natexlab{b}})}]{daddi17c}%
  \BibitemOpen
  \bibfield  {author} {\bibinfo {author} {\bibfnamefont {A.}~\bibnamefont
  {Daddi-Moussa-Ider}}, \bibinfo {author} {\bibfnamefont {M.}~\bibnamefont
  {Lisicki}}, \ and\ \bibinfo {author} {\bibfnamefont {S.}~\bibnamefont
  {Gekle}},\ }\bibfield  {title} {\enquote {\bibinfo {title} {Hydrodynamic
  mobility of a solid particle near a spherical elastic membrane. {II.}
  {A}symmetric motion},}\ }\href@noop {} {\bibfield  {journal} {\bibinfo
  {journal} {Phys. Rev. E}\ }\textbf {\bibinfo {volume} {95}},\ \bibinfo
  {pages} {053117} (\bibinfo {year} {2017}{\natexlab{b}})}\BibitemShut
  {NoStop}%
\bibitem [{\citenamefont {Kim}\ and\ \citenamefont {Mifflin}(1984)}]{kim84}%
  \BibitemOpen
  \bibfield  {author} {\bibinfo {author} {\bibfnamefont {S.}~\bibnamefont
  {Kim}}\ and\ \bibinfo {author} {\bibfnamefont {R.~T.}\ \bibnamefont
  {Mifflin}},\ }\href@noop {} {\enquote {\bibinfo {title} {The resistance and
  mobility functions of two equal spheres in low reynolds number flow.}}\
  }\bibinfo {type} {Tech. Rep.}\ (\bibinfo {year} {1984})\BibitemShut {NoStop}%
\bibitem [{\citenamefont {Dufresne}\ \emph {et~al.}(2000)\citenamefont
  {Dufresne}, \citenamefont {Squires}, \citenamefont {Brenner},\ and\
  \citenamefont {Grier}}]{dufresne00}%
  \BibitemOpen
  \bibfield  {author} {\bibinfo {author} {\bibfnamefont {E.~R.}\ \bibnamefont
  {Dufresne}}, \bibinfo {author} {\bibfnamefont {T.~M.}\ \bibnamefont
  {Squires}}, \bibinfo {author} {\bibfnamefont {M.~P.}\ \bibnamefont
  {Brenner}}, \ and\ \bibinfo {author} {\bibfnamefont {D.~G.}\ \bibnamefont
  {Grier}},\ }\bibfield  {title} {\enquote {\bibinfo {title} {Hydrodynamic
  coupling of two {B}rownian spheres to a planar surface},}\ }\href@noop {}
  {\bibfield  {journal} {\bibinfo  {journal} {Phys. Rev. Lett.}\ }\textbf
  {\bibinfo {volume} {85}},\ \bibinfo {pages} {3317} (\bibinfo {year}
  {2000})}\BibitemShut {NoStop}%
\bibitem [{\citenamefont {Swan}\ and\ \citenamefont {Brady}(2007)}]{swan07}%
  \BibitemOpen
  \bibfield  {author} {\bibinfo {author} {\bibfnamefont {J.~W.}\ \bibnamefont
  {Swan}}\ and\ \bibinfo {author} {\bibfnamefont {J.~F.}\ \bibnamefont
  {Brady}},\ }\bibfield  {title} {\enquote {\bibinfo {title} {Simulation of
  hydrodynamically interacting particles near a no-slip boundary},}\
  }\href@noop {} {\bibfield  {journal} {\bibinfo  {journal} {Phys. Fluids}\
  }\textbf {\bibinfo {volume} {19}},\ \bibinfo {pages} {113306} (\bibinfo
  {year} {2007})}\BibitemShut {NoStop}%
\bibitem [{\citenamefont {Tr{\"a}nkle}, \citenamefont {Ruh},\ and\
  \citenamefont {Rohrbach}(2016)}]{traenkle16}%
  \BibitemOpen
  \bibfield  {author} {\bibinfo {author} {\bibfnamefont {B.}~\bibnamefont
  {Tr{\"a}nkle}}, \bibinfo {author} {\bibfnamefont {D.}~\bibnamefont {Ruh}}, \
  and\ \bibinfo {author} {\bibfnamefont {A.}~\bibnamefont {Rohrbach}},\
  }\bibfield  {title} {\enquote {\bibinfo {title} {Interaction dynamics of two
  diffusing particles: contact times and influence of nearby surfaces},}\
  }\href@noop {} {\bibfield  {journal} {\bibinfo  {journal} {Soft Matter}\
  }\textbf {\bibinfo {volume} {12}},\ \bibinfo {pages} {2729--2736} (\bibinfo
  {year} {2016})}\BibitemShut {NoStop}%
\bibitem [{\citenamefont {Blake}(1979)}]{blake79}%
  \BibitemOpen
  \bibfield  {author} {\bibinfo {author} {\bibfnamefont {J.~R.}\ \bibnamefont
  {Blake}},\ }\bibfield  {title} {\enquote {\bibinfo {title} {On the generation
  of viscous toroidal eddies in a cylinder},}\ }\href@noop {} {\bibfield
  {journal} {\bibinfo  {journal} {J. Fluid Mech.}\ }\textbf {\bibinfo {volume}
  {95}},\ \bibinfo {pages} {209--222} (\bibinfo {year} {1979})}\BibitemShut
  {NoStop}%
\bibitem [{\citenamefont {Bracewell}(1999)}]{bracewell99}%
  \BibitemOpen
  \bibfield  {author} {\bibinfo {author} {\bibfnamefont {R.}~\bibnamefont
  {Bracewell}},\ }\href@noop {} {\emph {\bibinfo {title} {The Fourier Transform
  and Its Applications}}}\ (\bibinfo  {publisher} {McGraw-Hill},\ \bibinfo
  {year} {1999})\BibitemShut {NoStop}%
\bibitem [{\citenamefont {Hahn}(2005)}]{hahn05}%
  \BibitemOpen
  \bibfield  {author} {\bibinfo {author} {\bibfnamefont {T.}~\bibnamefont
  {Hahn}},\ }\bibfield  {title} {\enquote {\bibinfo {title} {Cuba—a library
  for multidimensional numerical integration},}\ }\href@noop {} {\bibfield
  {journal} {\bibinfo  {journal} {Comp. Phys. Comm.}\ }\textbf {\bibinfo
  {volume} {168}},\ \bibinfo {pages} {78--95} (\bibinfo {year}
  {2005})}\BibitemShut {NoStop}%
\bibitem [{\citenamefont {Hahn}(2016)}]{hahn16}%
  \BibitemOpen
  \bibfield  {author} {\bibinfo {author} {\bibfnamefont {T.}~\bibnamefont
  {Hahn}},\ }\bibfield  {title} {\enquote {\bibinfo {title} {Concurrent
  cuba},}\ }\href@noop {} {\bibfield  {journal} {\bibinfo  {journal} {Comp.
  Phys. Comm.}\ }\textbf {\bibinfo {volume} {207}},\ \bibinfo {pages}
  {341--349} (\bibinfo {year} {2016})}\BibitemShut {NoStop}%
\bibitem [{\citenamefont {Bickel}(2007)}]{bickel07}%
  \BibitemOpen
  \bibfield  {author} {\bibinfo {author} {\bibfnamefont {T.}~\bibnamefont
  {Bickel}},\ }\bibfield  {title} {\enquote {\bibinfo {title} {Hindered
  mobility of a particle near a soft interface},}\ }\href@noop {} {\bibfield
  {journal} {\bibinfo  {journal} {Phys. Rev. E}\ }\textbf {\bibinfo {volume}
  {75}},\ \bibinfo {pages} {041403} (\bibinfo {year} {2007})}\BibitemShut
  {NoStop}%
\bibitem [{\citenamefont {Phan-Thien}\ and\ \citenamefont
  {Tullock}(1993)}]{phan93}%
  \BibitemOpen
  \bibfield  {author} {\bibinfo {author} {\bibfnamefont {N.}~\bibnamefont
  {Phan-Thien}}\ and\ \bibinfo {author} {\bibfnamefont {D.}~\bibnamefont
  {Tullock}},\ }\bibfield  {title} {\enquote {\bibinfo {title} {Completed
  double layer boundary element method in elasticity},}\ }\href@noop {}
  {\bibfield  {journal} {\bibinfo  {journal} {J. Mech. Phys. Solids}\ }\textbf
  {\bibinfo {volume} {41}},\ \bibinfo {pages} {1067 -- 1086} (\bibinfo {year}
  {1993})}\BibitemShut {NoStop}%
\bibitem [{\citenamefont {Kohr}\ and\ \citenamefont {Pop}(2004)}]{kohr04}%
  \BibitemOpen
  \bibfield  {author} {\bibinfo {author} {\bibfnamefont {M.}~\bibnamefont
  {Kohr}}\ and\ \bibinfo {author} {\bibfnamefont {I.~I.}\ \bibnamefont {Pop}},\
  }\href@noop {} {\emph {\bibinfo {title} {Viscous incompressible flow for low
  Reynolds numbers}}},\ Vol.~\bibinfo {volume} {16}\ (\bibinfo  {publisher}
  {Wit Pr/Comp. Mech.},\ \bibinfo {year} {2004})\BibitemShut {NoStop}%
\bibitem [{\citenamefont {Zhao}\ and\ \citenamefont {Shaqfeh}(2011)}]{zhao11}%
  \BibitemOpen
  \bibfield  {author} {\bibinfo {author} {\bibfnamefont {H.}~\bibnamefont
  {Zhao}}\ and\ \bibinfo {author} {\bibfnamefont {E.~S.~G.}\ \bibnamefont
  {Shaqfeh}},\ }\bibfield  {title} {\enquote {\bibinfo {title} {Shear-induced
  platelet margination in a microchannel},}\ }\href@noop {} {\bibfield
  {journal} {\bibinfo  {journal} {Phys. Rev. E}\ }\textbf {\bibinfo {volume}
  {83}},\ \bibinfo {pages} {061924} (\bibinfo {year} {2011})}\BibitemShut
  {NoStop}%
\bibitem [{\citenamefont {Zhao}, \citenamefont {Shaqfeh},\ and\ \citenamefont
  {Narsimhan}(2012)}]{zhao12}%
  \BibitemOpen
  \bibfield  {author} {\bibinfo {author} {\bibfnamefont {H.}~\bibnamefont
  {Zhao}}, \bibinfo {author} {\bibfnamefont {E.~S.~G.}\ \bibnamefont
  {Shaqfeh}}, \ and\ \bibinfo {author} {\bibfnamefont {V.}~\bibnamefont
  {Narsimhan}},\ }\bibfield  {title} {\enquote {\bibinfo {title} {Shear-induced
  particle migration and margination in a cellular suspension},}\ }\href@noop
  {} {\bibfield  {journal} {\bibinfo  {journal} {Phys. Fluids}\ }\textbf
  {\bibinfo {volume} {24}},\ \bibinfo {pages} {011902} (\bibinfo {year}
  {2012})}\BibitemShut {NoStop}%
\bibitem [{\citenamefont {Guckenberger}\ \emph {et~al.}(2016)\citenamefont
  {Guckenberger}, \citenamefont {Schraml}, \citenamefont {Chen}, \citenamefont
  {Leonetti},\ and\ \citenamefont {Gekle}}]{guckenberger16}%
  \BibitemOpen
  \bibfield  {author} {\bibinfo {author} {\bibfnamefont {A.}~\bibnamefont
  {Guckenberger}}, \bibinfo {author} {\bibfnamefont {M.~P.}\ \bibnamefont
  {Schraml}}, \bibinfo {author} {\bibfnamefont {P.~G.}\ \bibnamefont {Chen}},
  \bibinfo {author} {\bibfnamefont {M.}~\bibnamefont {Leonetti}}, \ and\
  \bibinfo {author} {\bibfnamefont {S.}~\bibnamefont {Gekle}},\ }\bibfield
  {title} {\enquote {\bibinfo {title} {On the bending algorithms for soft
  objects in flows},}\ }\href@noop {} {\bibfield  {journal} {\bibinfo
  {journal} {Comp. Phys. Comm.}\ }\textbf {\bibinfo {volume} {207}},\ \bibinfo
  {pages} {1--23} (\bibinfo {year} {2016})}\BibitemShut {NoStop}%
\bibitem [{\citenamefont {Marquardt}(1963)}]{marquardt63}%
  \BibitemOpen
  \bibfield  {author} {\bibinfo {author} {\bibfnamefont {D.~W.}\ \bibnamefont
  {Marquardt}},\ }\bibfield  {title} {\enquote {\bibinfo {title} {An algorithm
  for least-squares estimation of nonlinear parameters},}\ }\href@noop {}
  {\bibfield  {journal} {\bibinfo  {journal} {SIAM J Appl Math.}\ }\textbf
  {\bibinfo {volume} {11}},\ \bibinfo {pages} {431--441} (\bibinfo {year}
  {1963})}\BibitemShut {NoStop}%
\bibitem [{\citenamefont {Conn}, \citenamefont {Gould},\ and\ \citenamefont
  {Toint}(2000)}]{conn00}%
  \BibitemOpen
  \bibfield  {author} {\bibinfo {author} {\bibfnamefont {A.~R.}\ \bibnamefont
  {Conn}}, \bibinfo {author} {\bibfnamefont {N.~I.~M.}\ \bibnamefont {Gould}},
  \ and\ \bibinfo {author} {\bibfnamefont {P.~L.}\ \bibnamefont {Toint}},\
  }\href@noop {} {\emph {\bibinfo {title} {Trust region methods}}},\
  Vol.~\bibinfo {volume} {1}\ (\bibinfo  {publisher} {Siam},\ \bibinfo {year}
  {2000})\BibitemShut {NoStop}%
\bibitem [{\citenamefont {Pozrikidis}(2001)}]{pozrikidis01jfm}%
  \BibitemOpen
  \bibfield  {author} {\bibinfo {author} {\bibfnamefont {C.}~\bibnamefont
  {Pozrikidis}},\ }\bibfield  {title} {\enquote {\bibinfo {title} {Effect of
  membrane bending stiffness on the deformation of capsules in simple shear
  flow},}\ }\href@noop {} {\bibfield  {journal} {\bibinfo  {journal} {J. Fluid
  Mech.}\ }\textbf {\bibinfo {volume} {440}},\ \bibinfo {pages} {269--291}
  (\bibinfo {year} {2001})}\BibitemShut {NoStop}%
\bibitem [{\citenamefont {Cipparrone}\ \emph {et~al.}(2010)\citenamefont
  {Cipparrone}, \citenamefont {Ricardez-Vargas}, \citenamefont {Pagliusi},\
  and\ \citenamefont {Provenzano}}]{cipparrone10}%
  \BibitemOpen
  \bibfield  {author} {\bibinfo {author} {\bibfnamefont {G.}~\bibnamefont
  {Cipparrone}}, \bibinfo {author} {\bibfnamefont {I.}~\bibnamefont
  {Ricardez-Vargas}}, \bibinfo {author} {\bibfnamefont {P.}~\bibnamefont
  {Pagliusi}}, \ and\ \bibinfo {author} {\bibfnamefont {C.}~\bibnamefont
  {Provenzano}},\ }\bibfield  {title} {\enquote {\bibinfo {title} {Polarization
  gradient: exploring an original route for optical trapping and
  manipulation},}\ }\href@noop {} {\bibfield  {journal} {\bibinfo  {journal}
  {Optics express}\ }\textbf {\bibinfo {volume} {18}},\ \bibinfo {pages}
  {6008--6013} (\bibinfo {year} {2010})}\BibitemShut {NoStop}%
\bibitem [{\citenamefont {Green}\ and\ \citenamefont {Adkins}(1960)}]{green60}%
  \BibitemOpen
  \bibfield  {author} {\bibinfo {author} {\bibfnamefont {A.~E.}\ \bibnamefont
  {Green}}\ and\ \bibinfo {author} {\bibfnamefont {J.~C.}\ \bibnamefont
  {Adkins}},\ }\href@noop {} {\emph {\bibinfo {title} {Large Elastic
  Deformations and Non-linear Continuum Mechanics}}}\ (\bibinfo  {publisher}
  {Oxford University Press},\ \bibinfo {year} {1960})\BibitemShut {NoStop}%
\bibitem [{\citenamefont {Zhu}(2014)}]{zhu14}%
  \BibitemOpen
  \bibfield  {author} {\bibinfo {author} {\bibfnamefont {L.}~\bibnamefont
  {Zhu}},\ }\emph {\bibinfo {title} {Simulation of individual cells in flow}},\
  \href@noop {} {Ph.D. thesis},\ \bibinfo  {school} {KTH Royal Institute of
  Technology} (\bibinfo {year} {2014})\BibitemShut {NoStop}%
\bibitem [{\citenamefont {Daddi-Moussa-Ider}(2017)}]{daddi-thesis}%
  \BibitemOpen
  \bibfield  {author} {\bibinfo {author} {\bibfnamefont {A.}~\bibnamefont
  {Daddi-Moussa-Ider}},\ }\emph {\bibinfo {title} {Diffusion of nanoparticles
  nearby elastic cell membranes : A theoretical study}},\ \href@noop {} {Ph.D.
  thesis},\ \bibinfo  {school} {University of Bayreuth, Germany} (\bibinfo
  {year} {2017})\BibitemShut {NoStop}%
\bibitem [{\citenamefont {Zhu}\ and\ \citenamefont {Brandt}(2015)}]{zhu15}%
  \BibitemOpen
  \bibfield  {author} {\bibinfo {author} {\bibfnamefont {L.}~\bibnamefont
  {Zhu}}\ and\ \bibinfo {author} {\bibfnamefont {L.}~\bibnamefont {Brandt}},\
  }\bibfield  {title} {\enquote {\bibinfo {title} {The motion of a deforming
  capsule through a corner},}\ }\href@noop {} {\bibfield  {journal} {\bibinfo
  {journal} {J. Fluid Mech.}\ }\textbf {\bibinfo {volume} {770}},\ \bibinfo
  {pages} {374--397} (\bibinfo {year} {2015})}\BibitemShut {NoStop}%
\bibitem [{\citenamefont {Sadd}(2009)}]{sadd09}%
  \BibitemOpen
  \bibfield  {author} {\bibinfo {author} {\bibfnamefont {M.~H.}\ \bibnamefont
  {Sadd}},\ }\href@noop {} {\emph {\bibinfo {title} {Elasticity: theory,
  applications, and numerics}}}\ (\bibinfo  {publisher} {Academic Press},\
  \bibinfo {year} {2009})\BibitemShut {NoStop}%
\bibitem [{\citenamefont {Kr{\"u}ger}(2012)}]{krueger12}%
  \BibitemOpen
  \bibfield  {author} {\bibinfo {author} {\bibfnamefont {T.}~\bibnamefont
  {Kr{\"u}ger}},\ }\href@noop {} {\emph {\bibinfo {title} {Computer simulation
  study of collective phenomena in dense suspensions of red blood cells under
  shear}}}\ (\bibinfo  {publisher} {Springer Science \& Business Media},\
  \bibinfo {year} {2012})\BibitemShut {NoStop}%
\bibitem [{\citenamefont {Lac}\ and\ \citenamefont
  {Barth{\`e}s-Biesel}(2005)}]{lac05}%
  \BibitemOpen
  \bibfield  {author} {\bibinfo {author} {\bibfnamefont {E.}~\bibnamefont
  {Lac}}\ and\ \bibinfo {author} {\bibfnamefont {D.}~\bibnamefont
  {Barth{\`e}s-Biesel}},\ }\bibfield  {title} {\enquote {\bibinfo {title}
  {Deformation of a capsule in simple shear flow: effect of membrane
  prestress},}\ }\href@noop {} {\bibfield  {journal} {\bibinfo  {journal}
  {Phys. Fluids}\ }\textbf {\bibinfo {volume} {17}},\ \bibinfo {pages} {072105}
  (\bibinfo {year} {2005})}\BibitemShut {NoStop}%
\bibitem [{\citenamefont {Deserno}(2015)}]{deserno15}%
  \BibitemOpen
  \bibfield  {author} {\bibinfo {author} {\bibfnamefont {M.}~\bibnamefont
  {Deserno}},\ }\bibfield  {title} {\enquote {\bibinfo {title} {Fluid lipid
  membranes: From differential geometry to curvature stresses},}\ }\href@noop
  {} {\bibfield  {journal} {\bibinfo  {journal} {Chem. Phys. Lipids}\ }\textbf
  {\bibinfo {volume} {185}},\ \bibinfo {pages} {11--45} (\bibinfo {year}
  {2015})}\BibitemShut {NoStop}%
\bibitem [{\citenamefont {Synge}\ and\ \citenamefont {Schild}(1969)}]{synge69}%
  \BibitemOpen
  \bibfield  {author} {\bibinfo {author} {\bibfnamefont {J.~L.}\ \bibnamefont
  {Synge}}\ and\ \bibinfo {author} {\bibfnamefont {A.}~\bibnamefont {Schild}},\
  }\href@noop {} {\emph {\bibinfo {title} {Tensor calculus}}},\ Vol.~\bibinfo
  {volume} {5}\ (\bibinfo  {publisher} {Courier Corporation},\ \bibinfo {year}
  {1969})\BibitemShut {NoStop}%
\bibitem [{\citenamefont {Daddi-Moussa-Ider}, \citenamefont {Lisicki},\ and\
  \citenamefont {Gekle}(2017{\natexlab{c}})}]{daddi17}%
  \BibitemOpen
  \bibfield  {author} {\bibinfo {author} {\bibfnamefont {A.}~\bibnamefont
  {Daddi-Moussa-Ider}}, \bibinfo {author} {\bibfnamefont {M.}~\bibnamefont
  {Lisicki}}, \ and\ \bibinfo {author} {\bibfnamefont {S.}~\bibnamefont
  {Gekle}},\ }\bibfield  {title} {\enquote {\bibinfo {title} {Mobility of an
  axisymmetric particle near an elastic interface},}\ }\href@noop {} {\bibfield
   {journal} {\bibinfo  {journal} {J. Fluid Mech.}\ }\textbf {\bibinfo {volume}
  {811}},\ \bibinfo {pages} {210--233} (\bibinfo {year}
  {2017}{\natexlab{c}})}\BibitemShut {NoStop}%
\bibitem [{\citenamefont {Zhong-Can}\ and\ \citenamefont
  {Helfrich}(1987)}]{zhong87}%
  \BibitemOpen
  \bibfield  {author} {\bibinfo {author} {\bibfnamefont {O.-Y.}\ \bibnamefont
  {Zhong-Can}}\ and\ \bibinfo {author} {\bibfnamefont {W.}~\bibnamefont
  {Helfrich}},\ }\bibfield  {title} {\enquote {\bibinfo {title} {Instability
  and deformation of a spherical vesicle by pressure},}\ }\href@noop {}
  {\bibfield  {journal} {\bibinfo  {journal} {Phys. Rev. Lett.}\ }\textbf
  {\bibinfo {volume} {59}},\ \bibinfo {pages} {2486} (\bibinfo {year}
  {1987})}\BibitemShut {NoStop}%
\bibitem [{\citenamefont {Rao}, \citenamefont {Zahalak},\ and\ \citenamefont
  {Sutera}(1994)}]{rao94}%
  \BibitemOpen
  \bibfield  {author} {\bibinfo {author} {\bibfnamefont {P.~R.}\ \bibnamefont
  {Rao}}, \bibinfo {author} {\bibfnamefont {G.~I.}\ \bibnamefont {Zahalak}}, \
  and\ \bibinfo {author} {\bibfnamefont {S.~P.}\ \bibnamefont {Sutera}},\
  }\bibfield  {title} {\enquote {\bibinfo {title} {Large deformations of
  elastic cylindrical capsules in shear flows},}\ }\href@noop {} {\bibfield
  {journal} {\bibinfo  {journal} {J. Fluid Mech.}\ }\textbf {\bibinfo {volume}
  {270}},\ \bibinfo {pages} {73--90} (\bibinfo {year} {1994})}\BibitemShut
  {NoStop}%
\bibitem [{\citenamefont {Eggleton}\ and\ \citenamefont
  {Popel}(1998)}]{eggleton98}%
  \BibitemOpen
  \bibfield  {author} {\bibinfo {author} {\bibfnamefont {C.~D.}\ \bibnamefont
  {Eggleton}}\ and\ \bibinfo {author} {\bibfnamefont {A.~S.}\ \bibnamefont
  {Popel}},\ }\bibfield  {title} {\enquote {\bibinfo {title} {Large deformation
  of red blood cell ghosts in a simple shear flow},}\ }\href@noop {} {\bibfield
   {journal} {\bibinfo  {journal} {Phys. Fluids}\ }\textbf {\bibinfo {volume}
  {10}},\ \bibinfo {pages} {1834--1845} (\bibinfo {year} {1998})}\BibitemShut
  {NoStop}%
\bibitem [{\citenamefont {Navot}(1998)}]{navot98}%
  \BibitemOpen
  \bibfield  {author} {\bibinfo {author} {\bibfnamefont {Y.}~\bibnamefont
  {Navot}},\ }\bibfield  {title} {\enquote {\bibinfo {title} {Elastic membranes
  in viscous shear flow},}\ }\href@noop {} {\bibfield  {journal} {\bibinfo
  {journal} {Phys. Fluids}\ }\textbf {\bibinfo {volume} {10}},\ \bibinfo
  {pages} {1819--1833} (\bibinfo {year} {1998})}\BibitemShut {NoStop}%
\bibitem [{\citenamefont {Sui}\ \emph {et~al.}(2008)\citenamefont {Sui},
  \citenamefont {Chew}, \citenamefont {Roy}, \citenamefont {Cheng},\ and\
  \citenamefont {Low}}]{sui08POF}%
  \BibitemOpen
  \bibfield  {author} {\bibinfo {author} {\bibfnamefont {Y.}~\bibnamefont
  {Sui}}, \bibinfo {author} {\bibfnamefont {Y.~T.}\ \bibnamefont {Chew}},
  \bibinfo {author} {\bibfnamefont {P.}~\bibnamefont {Roy}}, \bibinfo {author}
  {\bibfnamefont {Y.~P.}\ \bibnamefont {Cheng}}, \ and\ \bibinfo {author}
  {\bibfnamefont {H.~T.}\ \bibnamefont {Low}},\ }\bibfield  {title} {\enquote
  {\bibinfo {title} {Dynamic motion of red blood cells in simple shear flow},}\
  }\href@noop {} {\bibfield  {journal} {\bibinfo  {journal} {Phys. Fluids}\
  }\textbf {\bibinfo {volume} {20}} (\bibinfo {year} {2008})}\BibitemShut
  {NoStop}%
\bibitem [{\citenamefont {Clausen}\ and\ \citenamefont
  {Aidun}(2010)}]{clausen10}%
  \BibitemOpen
  \bibfield  {author} {\bibinfo {author} {\bibfnamefont {J.~R.}\ \bibnamefont
  {Clausen}}\ and\ \bibinfo {author} {\bibfnamefont {C.~K.}\ \bibnamefont
  {Aidun}},\ }\bibfield  {title} {\enquote {\bibinfo {title} {Capsule dynamics
  and rheology in shear flow: Particle pressure and normal stress},}\
  }\href@noop {} {\bibfield  {journal} {\bibinfo  {journal} {Phys. Fluids}\
  }\textbf {\bibinfo {volume} {22}},\ \bibinfo {pages} {123302} (\bibinfo
  {year} {2010})}\BibitemShut {NoStop}%
\bibitem [{\citenamefont {Bukman}, \citenamefont {Yao},\ and\ \citenamefont
  {Wortis}(1996)}]{bukman96}%
  \BibitemOpen
  \bibfield  {author} {\bibinfo {author} {\bibfnamefont {D.~J.}\ \bibnamefont
  {Bukman}}, \bibinfo {author} {\bibfnamefont {J.~H.}\ \bibnamefont {Yao}}, \
  and\ \bibinfo {author} {\bibfnamefont {M.}~\bibnamefont {Wortis}},\
  }\bibfield  {title} {\enquote {\bibinfo {title} {Stability of cylindrical
  vesicles under axial tension},}\ }\href@noop {} {\bibfield  {journal}
  {\bibinfo  {journal} {Phys. Rev. E}\ }\textbf {\bibinfo {volume} {54}},\
  \bibinfo {pages} {5463} (\bibinfo {year} {1996})}\BibitemShut {NoStop}%
\bibitem [{\citenamefont {Luo}\ \emph {et~al.}(2013)\citenamefont {Luo},
  \citenamefont {Wang}, \citenamefont {He}, \citenamefont {Xu},\ and\
  \citenamefont {Bai}}]{luo13}%
  \BibitemOpen
  \bibfield  {author} {\bibinfo {author} {\bibfnamefont {Z.~Y.}\ \bibnamefont
  {Luo}}, \bibinfo {author} {\bibfnamefont {S.~Q.}\ \bibnamefont {Wang}},
  \bibinfo {author} {\bibfnamefont {L.}~\bibnamefont {He}}, \bibinfo {author}
  {\bibfnamefont {F.}~\bibnamefont {Xu}}, \ and\ \bibinfo {author}
  {\bibfnamefont {B.~F.}\ \bibnamefont {Bai}},\ }\bibfield  {title} {\enquote
  {\bibinfo {title} {Inertia-dependent dynamics of three-dimensional vesicles
  and red blood cells in shear flow},}\ }\href@noop {} {\bibfield  {journal}
  {\bibinfo  {journal} {Soft Matter}\ }\textbf {\bibinfo {volume} {9}},\
  \bibinfo {pages} {9651--9660} (\bibinfo {year} {2013})}\BibitemShut {NoStop}%
\bibitem [{\citenamefont {Dupont}\ \emph {et~al.}(2015)\citenamefont {Dupont},
  \citenamefont {Salsac}, \citenamefont {Barth{\`e}s-Biesel}, \citenamefont
  {Vidrascu},\ and\ \citenamefont {Le~Tallec}}]{dupont15}%
  \BibitemOpen
  \bibfield  {author} {\bibinfo {author} {\bibfnamefont {C.}~\bibnamefont
  {Dupont}}, \bibinfo {author} {\bibfnamefont {A.-V.}\ \bibnamefont {Salsac}},
  \bibinfo {author} {\bibfnamefont {D.}~\bibnamefont {Barth{\`e}s-Biesel}},
  \bibinfo {author} {\bibfnamefont {M.}~\bibnamefont {Vidrascu}}, \ and\
  \bibinfo {author} {\bibfnamefont {P.}~\bibnamefont {Le~Tallec}},\ }\bibfield
  {title} {\enquote {\bibinfo {title} {Influence of bending resistance on the
  dynamics of a spherical capsule in shear flow},}\ }\href@noop {} {\bibfield
  {journal} {\bibinfo  {journal} {Phys. Fluids}\ }\textbf {\bibinfo {volume}
  {27}},\ \bibinfo {pages} {051902} (\bibinfo {year} {2015})}\BibitemShut
  {NoStop}%
\bibitem [{\citenamefont {Kaoui}\ and\ \citenamefont
  {Harting}(2016)}]{kaoui16}%
  \BibitemOpen
  \bibfield  {author} {\bibinfo {author} {\bibfnamefont {B.}~\bibnamefont
  {Kaoui}}\ and\ \bibinfo {author} {\bibfnamefont {J.}~\bibnamefont
  {Harting}},\ }\bibfield  {title} {\enquote {\bibinfo {title} {Two-dimensional
  lattice boltzmann simulations of vesicles with viscosity contrast},}\
  }\href@noop {} {\bibfield  {journal} {\bibinfo  {journal} {Rheol. Acta}\
  }\textbf {\bibinfo {volume} {55}},\ \bibinfo {pages} {465--475} (\bibinfo
  {year} {2016})}\BibitemShut {NoStop}%
\bibitem [{\citenamefont {Kaoui}(2017)}]{kaoui17}%
  \BibitemOpen
  \bibfield  {author} {\bibinfo {author} {\bibfnamefont {B.}~\bibnamefont
  {Kaoui}},\ }\bibfield  {title} {\enquote {\bibinfo {title} {Flow and mass
  transfer around a core-shell reservoir},}\ }\href@noop {} {\bibfield
  {journal} {\bibinfo  {journal} {Phys. Rev. E}\ }\textbf {\bibinfo {volume}
  {95}},\ \bibinfo {pages} {063310} (\bibinfo {year} {2017})}\BibitemShut
  {NoStop}%
\bibitem [{\citenamefont {Aouane}\ \emph {et~al.}(2017)\citenamefont {Aouane},
  \citenamefont {Farutin}, \citenamefont {Thi\'ebaud}, \citenamefont
  {Benyoussef}, \citenamefont {Wagner},\ and\ \citenamefont
  {Misbah}}]{aouane17}%
  \BibitemOpen
  \bibfield  {author} {\bibinfo {author} {\bibfnamefont {O.}~\bibnamefont
  {Aouane}}, \bibinfo {author} {\bibfnamefont {A.}~\bibnamefont {Farutin}},
  \bibinfo {author} {\bibfnamefont {M.}~\bibnamefont {Thi\'ebaud}}, \bibinfo
  {author} {\bibfnamefont {A.}~\bibnamefont {Benyoussef}}, \bibinfo {author}
  {\bibfnamefont {C.}~\bibnamefont {Wagner}}, \ and\ \bibinfo {author}
  {\bibfnamefont {C.}~\bibnamefont {Misbah}},\ }\bibfield  {title} {\enquote
  {\bibinfo {title} {Hydrodynamic pairing of soft particles in a confined
  flow},}\ }\href@noop {} {\bibfield  {journal} {\bibinfo  {journal} {Phys.
  Rev. Fluids}\ }\textbf {\bibinfo {volume} {2}},\ \bibinfo {pages} {063102}
  (\bibinfo {year} {2017})}\BibitemShut {NoStop}%
\end{thebibliography}

    \end{document}